\documentclass[12pt]{article}
\usepackage[letterpaper,left=2.5cm,right=2.5cm,top=3cm,bottom=3cm]{geometry}
\usepackage{setspace}
\usepackage{color}
\usepackage[numbers, sort&compress]{natbib}
\usepackage{authblk}
\usepackage[utf8]{inputenc}
\usepackage{lmodern}
\usepackage[T1]{fontenc}
\usepackage{graphicx}
\usepackage{amsmath, amssymb, mathabx}
\usepackage{bm}
\usepackage{mathtools}
\usepackage{relsize}
\usepackage{caption}

\newcommand\sfigSingleLevel{S1}
\newcommand\sfigWithinMoment{S2}
\newcommand\sfigDisequilibrium{S3}
\newcommand\sfigPhaseDiagramDisc{S4}

\title{A scaling law of multilevel evolution: how the balance between
  within- and among-collective evolution is determined}

\author[1,4]{\normalsize Nobuto Takeuchi}

\author[2,4]{\normalsize Namiko Mitarai}
\author[3,4]{\normalsize Kunihiko Kaneko}

\affil[1]{\normalsize School of Biological Sciences, University of Auckland, Private Bag 92019, Auckland 1142, New Zealand}

\affil[2]{\normalsize The Niels Bohr Institute, University of Copenhagen, Blegdamsvej 17, Copenhagen, 2100-DK, Denmark}

\affil[3]{\normalsize Graduate School of Arts and Sciences, University of Tokyo, Komaba 3-8-1, Meguro-ku, Tokyo 153-8902, Japan}

\affil[4]{\normalsize Research Center for Complex Systems Biology, Universal Biology Institute, University of Tokyo, Komaba 3-8-1, Meguro-ku, Tokyo 153-8902, Japan}

\date{}
\begin{document}

\maketitle

\doublespacing

\begin{abstract}
  Numerous living systems are hierarchically organised, whereby
  replicating components are grouped into reproducing
  collectives---e.g., organelles are grouped into cells, and cells are
  grouped into multicellular organisms. In such systems, evolution can
  operate at two levels: evolution among collectives, which tends to
  promote selfless cooperation among components within collectives
  (called altruism), and evolution within collectives, which tends to
  promote cheating among components within collectives. The balance
  between within- and among-collective evolution thus exerts profound
  impacts on the fitness of these systems. Here, we investigate how this
  balance depends on the size of a collective (denoted by $N$) and the
  mutation rate of components ($m$) through mathematical analyses and
  computer simulations of multiple population genetics models. We first
  confirm a previous result that increasing $N$ or $m$ accelerates
  within-collective evolution relative to among-collective evolution,
  thus promoting the evolution of cheating. Moreover, we show that when
  within- and among-collective evolution exactly balance each other out,
  the following scaling relation generally holds: $Nm^{\alpha}$ is a
  constant, where scaling exponent $\alpha$ depends on multiple
  parameters, such as the strength of selection and whether altruism is
  a binary or quantitative trait. This relation indicates that although
  $N$ and $m$ have quantitatively distinct impacts on the balance
  between within- and among-collective evolution, their impacts become
  identical if $m$ is scaled with a proper exponent. Our results thus
  provide a novel insight into conditions under which cheating or
  altruism evolves in hierarchically-organised replicating systems.
\end{abstract}


\section*{Introduction}

A fundamental feature of living systems is hierarchical organization, in
which replicating components are grouped into reproducing collectives
\cite{maynard_smith_major_1995}. For example, replicating molecules are
grouped into protocells \cite{Joyce2018}, organelles such as
mitochondria are grouped into cells \cite{burt_genes_2006}, cells are
grouped into multicellular organisms \cite{buss_evolution_1987}, and
multicellular organisms are grouped into eusocial colonies
\cite{Davies2012,Gershwin2014}.

Such hierarchical organization hinges on altruism among replicating
components \cite{Bourke2011}, the selfless action that increases
collective-level fitness at the cost of self-replication of individual
components \cite{West2007}. For example, molecules in a protocell
catalyze chemical reactions to facilitate the growth of the protocell at
the cost of self-replication of the molecules, a cost that arises from a
trade-off between serving as catalysts and serving as templates
\cite{Durand2010,Ivica2013}. Likewise, cells in a multicellular organism
perform somatic functions beneficial to the whole organism, such as
defence and locomotion, at the cost of cell proliferation due to
different trade-offs \cite{Bell1985,buss_evolution_1987,Kirk1998}.

Altruism, however, entails the risk of invasion by cheaters---selfish
components that avoid altruism and instead replicate themselves to the
detriment of a collective. For example, parasitic templates replicate to
the detriment of a protocell \cite{MaynardSmith1979,Bansho2016}, selfish
organelles multiply to the detriment of a cell \cite{burt_genes_2006},
and cancer cells proliferate to the detriment of a multicellular
organism \cite{Greaves2012,Aktipis2015}. Since cheaters replicate faster
than altruists within a collective, they can out-compete the altruists,
causing the decline of collective-level fitness---within-collective
evolution, for short. However, collectives containing many altruists can
reproduce faster than those containing many cheaters, so that altruists
can be selected through competition among collectives---among-collective
evolution. Evolution thus operates at multiple levels of the biological
hierarchy in conflicting directions---conflicting multilevel
evolution. Whether within- or among-collective evolution predominates
exerts profound impacts on the stability and evolution of
hierarchically-organised replicating systems
\cite{Wilson1975,Slatkin1978,Aoki1982,Crow1982,Leigh1983,Kimura1984,Kimura1986,Frank1994,Rispe2000,Goodnight2005,Traulsen2006,Bijma2007,Leigh2010,Chuang2009,Frank2012,Simon2013,Fontanari2014,Luo2014,Tarnita2013,Blokhuis2018,VanVliet2019,Cooney2019,Takeuchi2016,Takeuchi2017,Takeuchi2019b},
which abound in nature
\cite{buss_evolution_1987,maynard_smith_major_1995,burt_genes_2006,Davies2012,Gershwin2014,Joyce2018}. Therefore,
how the balance between within- and among-collective evolution is
determined is an important question in biology.

Previously, we have demonstrated that the balance between within- and
among-collective evolution involves a simple scaling relation between
parameters of population dynamics
\cite{Takeuchi2016,Takeuchi2017,Takeuchi2019b}. These parameters are the
mutation rate of components (denoted by $m$) and the number of
replicating components per collective (denoted by $N$)---in general, $N$
represents the `size' of a collective, such as the number of replicating
molecules per protocell, organelles per cell, cells per multicellular
organism, and organisms per colony. As $m$ or $N$ increases,
within-collective evolution accelerates relative to among-collective
evolution (i.e., promoting the evolution of cheating), and $m$ and $N$
display the following scaling relation when within- and among-collective
evolution exactly balance each other out (i.e., no bias towards the
evolution of cheating or altruism): $Nm^{\alpha}$ is a constant (i.e.,
$N\,\propto\, m^{-\alpha}$), where scaling exponent $\alpha$ is
approximately one half
\cite{Takeuchi2016,Takeuchi2017,Takeuchi2019b}. This scaling relation
indicates that although $m$ and $N$ have quantitatively different
impacts on the balance between within- and among-collective evolution,
their impacts are identical if $m$ is scaled with exponent $\alpha$
(e.g., doubling $N$ and quartering $m$ approximately cancel each other
out, keeping the balance of multilevel evolution).

While the above scaling relation provides a novel insight into how the
balance between within- and among-collective evolution is determined,
the generality of this relation is unknown because the relation has
originally been demonstrated in specific models of protocells through
computer simulations \cite{Takeuchi2016,Takeuchi2017,Takeuchi2019b}. To
shed light on the generality of the scaling relation, here we adapt a
standard model of population genetics, the Wright-Fisher model
\cite{Ewens2004}, to investigate the balance between within- and
among-collective evolution. Combining computer simulations and
mathematical analyses, we establish the following generalized scaling
relation under the assumption that selection strengths are stationary in
time: $N\,\propto\,m^{-\alpha}$, where $\alpha$ decreases to zero as
selection strength $s$ decreases to zero. To examine further the
generality of the scaling relation, we analyse another simple model of
multilevel evolution, which approaches the model studied by Kimura
\cite{Kimura1984,Kimura1986} as $s\to0$. Interestingly, our results show
that this model displays a distinct scaling relation:
$N\,\propto\,m^{-\alpha}$, where $\alpha$ \textit{increases} to one as
$s$ decreases to zero. We show that this difference stems from the fact
that our first model considers a quantitative trait, whereas our second
model and Kimura's consider a binary trait. Taken together, our results
suggest that the existence of scaling relation $N\,\propto\,m^{-\alpha}$
is a general feature of conflicting multilevel evolution, but scaling
exponent $\alpha$ depends on multiple factors in a non-trivial manner.

\section*{Model}
\begin{figure}[t]
\centerline{\includegraphics{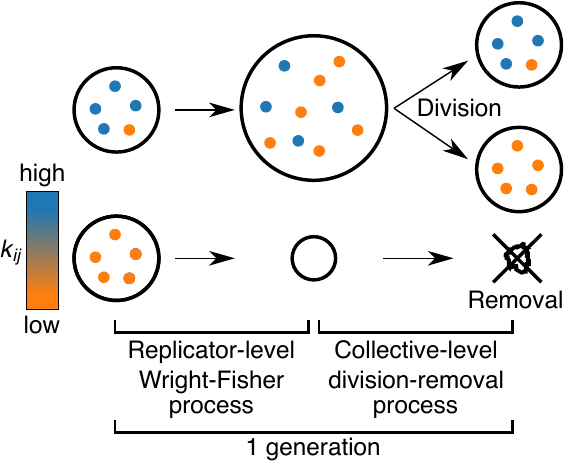}}
\caption{\label{fig:model}Schematic of model. Replicators (dot) are
  grouped into collectives (circles). $k_{ij}$ represents degree of
  altruism performed by replicators within collectives.}
\end{figure}

Our model is an extension of the Wright-Fisher model to incorporate
conflicting multilevel evolution \cite{Ewens2004}. The model consists of
a population of $M$ replicators grouped into collectives, each
consisting of at most $N$ replicators (Fig.\,\ref{fig:model} and
Table\,\ref{tbl:symbols}). The number of replicators in a collective can
increase or decrease, and if this number exceeds $N$, the collective
randomly divides into two.

Replicator $j$ in collective $i$ is assigned a heritable quantitative
trait (denoted by $k_{ij}$) representing the degree of altruism it
performs within collective $i$ (e.g., $k_{ij}$ represents the amount of
chemical catalysis a replicating molecule provides in a protocell or the
amount of somatic work a cell performs in a multicellular
organism). Replicators are assumed to face a trade-off between
performing altruism and undergoing self-replication. Thus, the fitness
of individual replicators (denoted by $w_{ij}$) decreases with
individual trait $k_{ij}$, whereas the collective-level fitness of
replicators $\langle w_{i\tilde{j}}\rangle$ increases with
collective-level trait $\langle k_{i\tilde{j}}\rangle$, where
$\langle x_{i\tilde{j}}\rangle$ is $x_{ij}$ averaged over replicators in
collective $i$ (i.e., $x_{ij}$ is averaged over the index marked with a
tilde; see also Table\,\ref{tbl:symbols}). For simplicity, we assume
that the strengths of selection within and among collectives, defined as
\begin{align}
    &s_\mathrm{w}=-\frac{\partial\ln w_{ij}}{\partial k_{ij}} &\text{and}&
    &s_\mathrm{a}=\frac{\partial\ln \langle w_{i\tilde{j}}\rangle}{\partial\langle k_{i\tilde{j}}\rangle},
\end{align}
respectively, depend only very weakly on $k_{ij}$ and
$\langle k_{i\tilde{j}}\rangle$ (i.e.,
$\partial s_\mathrm{w}/\partial k_{ij}\approx0$ and
$\partial s_\mathrm{a}/\partial\langle
k_{i\tilde{j}}\rangle\approx0$). This assumption implies that the
relative fitness of replicators and collectives is translationally
invariant with respect to $k_{ij}$ and $\langle k_{i\tilde{j}}\rangle$,
respectively---i.e.,
$(w_{ij}+ \Delta w_{ij})/w_{ij}\approx 1+s_\mathrm{w} \Delta k_{ij}$,
and
$(\langle w_{i\tilde{j}}\rangle +\Delta\langle
w_{i\tilde{j}}\rangle)/\langle
w_{i\tilde{j}}\rangle\approx1+s_\mathrm{a} \Delta\langle
k_{i\tilde{j}}\rangle$. Owing to this assumption, our model informs only
about short-term evolution. For computer simulations, we used the
following fitness function:
\begin{equation}
\label{eq:fitness_def}
    w_{ij} = e^{s_\mathrm{a}\langle k_{i\tilde{j}}\rangle}\frac{e^{-s_\mathrm{w} k_{ij}}}{\langle e^{-s_\mathrm{w} k_{i\tilde{j}}}\rangle},
\end{equation}
where $s_\mathrm{w}$ and $s_\mathrm{a}$ are constant so that
Eq.\,(\ref{eq:fitness_def}) satisfies the above assumption. This
particular form of fitness function, however, does not affect our main
conclusion, as will be seen from the fact that the mathematical analysis
presented below is independent of it.

The state of the model is updated in discrete time
(Fig.\,\ref{fig:model}). In each generation, $M$ replicators are sampled
with replacement from replicators of the previous generation with
probabilities proportional to $w_{ij}$, as in the Wright-Fisher process
\cite{Ewens2004}.

During the above sampling, a replicator inherits group identity $i$ and
trait $k_{ij}$ from its parental replicator with potential mutation (no
migration among collectives is allowed). More precisely, the $k_{ij}$
value of a replicator is set to $k_{ij_p} + \epsilon$, where $k_{ij_p}$
is the trait of the parental replicator, and $\epsilon$ takes a value of
zero with a probability of $1-m$ or a value sampled from a Gaussian
distribution with mean zero and variance $\sigma$ (determining mutation
step size) with a probability of $m$ (representing a genetic or
epigenetic mutation rate). The assumption that the mean of $\epsilon$ is
zero is made on the following premise: evolution is mainly driven by
selection or random genetic drift, and the direction of evolution is not
directly determined by mutation. Although this premise often
approximates reality, it can be wrong if a mutation rate is so high as
to dictate the direction of evolution as in the error catastrophe
\cite{Domingo2005}, a situation that is ignored in this study. The
assumption that $\sigma$ is independent of $k_{ij}$ is made for
simplicity and implies that our model informs only about short-term
evolution. Although we could reduce the number of parameters by
aggregating $m$ and $\sigma$ into $m\sigma$ (which is the variance of
$\epsilon$), we keep them separate so that the mutation rate as usually
defined is discernible.

After the above sampling, collectives containing more than $N$
replicators are randomly divided into two, and those with no replicators
removed (Fig.\,\ref{fig:model}).

\begin{center}
  \renewcommand{\arraystretch}{0.7}
  \begin{tabular}{c l}
    symbol & description\\\hline
    $M$ & total number of replicators\\
    $N$ & maximum number of replicators per collective\\
    $L$ & total number of collectives\\
    $n_i$ & number of replicators in collective $i$\\
    $k_{ij}$ & degree of altruism performed by replicator $j$ in collective $i$\\
    $w_{ij}$ & fitness of replicator $j$ in collective $i$\\
    $s_\mathrm{w}$ & within-collective selection strength: $-(\partial/\partial k_{ij}) \ln w_{ij}$\\
    $s_\mathrm{a}$ & among-collective selection strength: $(\partial/\partial\langle k_{i\tilde{j}}\rangle)\ln\langle w_{i\tilde{j}}\rangle$\\
    $m$ & probability of mutation of $k_{ij}$ per generation\\
    $\epsilon$ & effect of mutation on $k_{ij}$\\
    $\sigma$ & variance of $\epsilon$ (only in continuous-trait model)\\
    $\langle x_{i\tilde{j}}\rangle$ & within-collective average: $n_i^{-1}\sum_{j=1}^{n_i}x_{ij}$\\
    $\mathrm{ave}_{\tilde{i}}[x_i]$ & among-collective average: $M^{-1}\sum_{i=1}^{L}n_ix_{i}$\\
    $\langle\!\langle x_{\tilde{i}\tilde{j}}\rangle\!\rangle$ & global average: $M^{-1}\sum_{i=1}^{L}\sum_{j=1}^{n_i}x_{ij}(=\mathrm{ave}_{\tilde{i}}[\langle x_{i\tilde{j}}\rangle])$\\
    $\mathbb{E}[X]$ & expected $X$ after one
                                iteration of Wright-Fisher process\\
    $\mathrm{cov}_{i\tilde{j}}[x_{ij},y_{ij}]$ & within-collective covariance: $n_i^{-1}\sum_{j=1}^{n_i}(x_{ij}-\langle x_{i\tilde{j}}\rangle)(y_{ij}-\langle y_{i\tilde{j}}\rangle)$\\
    $\mathrm{cov}_{\tilde{i}}[x_i,y_i]$ & among-collective covariance: $M^{-1}\sum_{i=1}^{L}n_i(x_{i}-\mathrm{ave}_{\tilde{i}}[x_i])(y_{i}-\mathrm{ave}_{\tilde{i}}[y_i])$\\
    $v_\mathrm{w}$ & within-collective variance of $k_{ij}$: $\mathrm{ave}_{\tilde{i}}[\langle(k_{i\tilde{j}}-\langle k_{i\tilde{j}}\rangle)^2\rangle]$\\
    $v_\mathrm{a}$ & among-collective variance of $k_{ij}$:  $\mathrm{ave}_{\tilde{i}}[(\langle k_{i\tilde{j}}\rangle-\langle\!\langle k_{\tilde{i}\tilde{j}}\rangle\!\rangle)^2]$\\
    $v_\mathrm{t}$ & total variance of  $k_{ij}$:  $\langle\!\langle(k_{\tilde{i}\tilde{j}}-\langle\!\langle k_{\tilde{i}\tilde{j}}\rangle\!\rangle)^2\rangle\!\rangle$ ($=v_\mathrm{a}+v_\mathrm{w}$)\\
    $c_\mathrm{w}$ & within-collective 3rd central moment of $k_{ij}$: $\mathrm{ave}_{\tilde{i}}[\langle(k_{i\tilde{j}}-\langle k_{i\tilde{j}}\rangle)^3\rangle]$\\
    $c_\mathrm{a}$ & among-collective 3rd central moment of $k_{ij}$: $\mathrm{ave}_{\tilde{i}}[(\langle k_{i\tilde{j}}\rangle-\langle\!\langle k_{\tilde{i}\tilde{j}}\rangle\!\rangle)^3]$\\
  \end{tabular}
  \captionof{table}{\label{tbl:symbols}Symbol list.}
\end{center}

\section*{Results}
\subsection*{Demonstration of a scaling relation by computer
  simulations}
\begin{figure}[t]
\centerline{\includegraphics{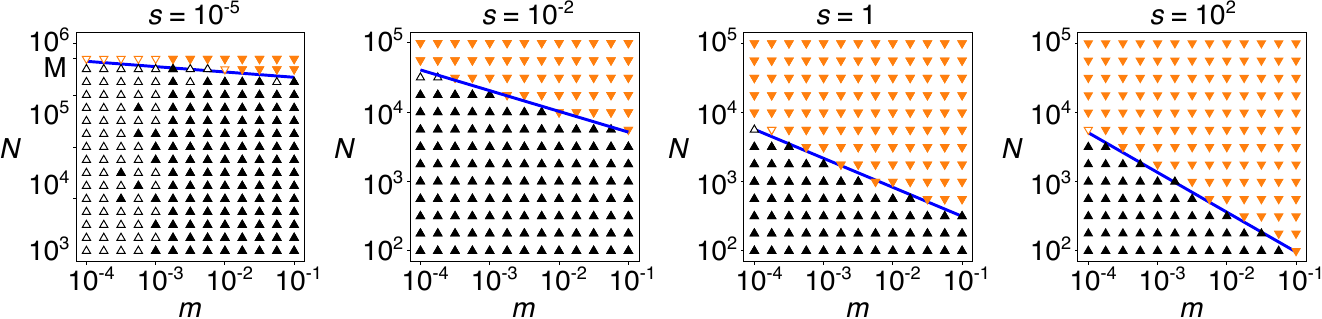}}
\caption{\label{fig:pase-diagram}Parameter-sweep diagrams
  ($s_\mathrm{w}=s_\mathrm{a}=s$, $M=5\times 10^5$, and
  $\sigma=10^{-4}$). Symbols have following meaning:
  $\Delta\langle\!\langle
  k_{\tilde{i}\tilde{j}}\rangle\!\rangle>3\times10^{-7}$ (black filled
  triangle up);
  $\Delta\langle\!\langle
  k_{\tilde{i}\tilde{j}}\rangle\!\rangle<-3\times10^{-7}$ (orange filled
  triangle down); RHS of Eq.\,(\ref{eq:price-mean}) measured in
  simulations is positive (black open triangle up) or negative (orange
  open triangle down), where
  $|\Delta\langle\!\langle
  k_{\tilde{i}\tilde{j}}\rangle\!\rangle|<3\times10^{-7}$. Lines are
  estimated boundaries where
  $\Delta\langle\!\langle k_{\tilde{i}\tilde{j}}\rangle\!\rangle$
  changes sign (see Methods under `Parameter-sweep diagram').}
\end{figure}

By simulating the above model, we measured the rate of change of
$\langle\!\langle k_{\tilde{i}\tilde{j}}\rangle\!\rangle$, where
$\langle\!\langle x_{\tilde{i}\tilde{j}}\rangle\!\rangle$ is $x_{ij}$
averaged over all replicators, at steady states as a function of $m$ and
$N$, assuming $s_\mathrm{w}=s_\mathrm{a}$. The result indicates the
existence of two distinct parameter regions, where
$\langle\!\langle k_{\tilde{i}\tilde{j}}\rangle\!\rangle$ either
increases or decreases through evolution (Fig.\,\ref{fig:pase-diagram};
Methods under `Parameter-sweep diagram'). (Note that although the model
displays an unlimited increase or decrease of
$\langle\!\langle k_{\tilde{i}\tilde{j}}\rangle\!\rangle$ over time, the
model is intended to inform about short-term evolution as described
above; therefore, its result should be considered as providing
information about an instantaneous rate of evolution in a steady state
for given parameters.)

\begin{figure}[t]
\centerline{\includegraphics{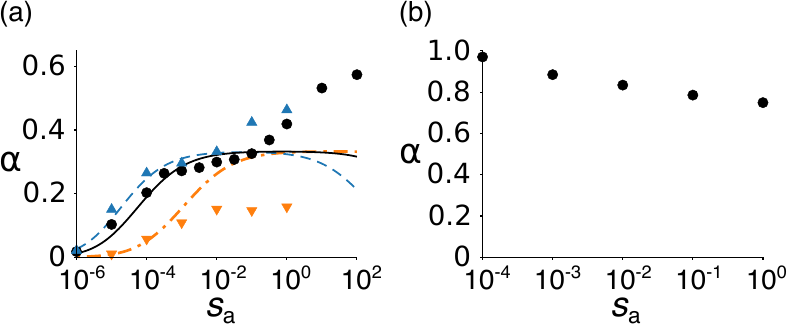}}
\caption{\label{fig:alpha-vs-s}Scaling exponent $\alpha$ of
  parameter-region boundaries where
  $\Delta\langle\!\langle k_{\tilde{i}\tilde{j}}\rangle\!\rangle$
  changes sign as a function of selection strength. (a)
  Quantitative-trait model ($M=5\times10^5$, and $\sigma=10^{-4}$). Data
  points are simulation results (see Methods under `Parameter-sweep
  diagram'). Lines are prediction by
  Eqs.\,(\ref{eq:diff-variance-closed-vw}) and
  (\ref{eq:diff-variance-closed-va}). $s_\mathrm{w}=s_\mathrm{a}$ (black
  circle and solid line), $s_\mathrm{w}=10s_\mathrm{a}$ (blue triangle
  up and dashed line), $10s_\mathrm{w}=s_\mathrm{a}$ (orange triangle
  down and dash-dotted line). (b) Binary-trait model
  ($s_\mathrm{w}=s_\mathrm{a}$ and $M=5\times10^5$).}
\end{figure}

The two parameter regions mentioned above are demarcated by scaling
relation $N\,\propto\, m^{-\alpha}$, where $\alpha\downarrow0$ as
$s\downarrow0$ (Fig.\,\ref{fig:alpha-vs-s}a)---i.e., the evolution of
$\langle\!\langle k_{\tilde{i}\tilde{j}}\rangle\!\rangle$ becomes
increasingly independent of $m$ as $s$ decreases. Similar scaling
relations hold also when $s_\mathrm{w}=10s_\mathrm{a}$, or
$s_\mathrm{w}=0.1 s_\mathrm{a}$ (Fig.\,\ref{fig:alpha-vs-s}a). These
results generalize those previously obtained with specific models of
protocells \cite{Takeuchi2016,Takeuchi2017}.

\subsection*{Mathematical analysis of the scaling relation}
Next, we present a theory that can account for
$N\,\propto\, m^{-\alpha}$ under the assumptions that $s_\mathrm{a}$ and
$s_\mathrm{w}$ are sufficiently small. Although such a theory could in
principle be built by calculating the dynamics of the frequency
distribution of $k_{ij}$, for simplicity, we instead calculate the
dynamics of the moments of this distribution. The expected change of
$\langle\!\langle k_{\tilde{i}\tilde{j}}\rangle\!\rangle$ per generation
is expressed by Price's equation as follows
\cite{Price1972,hamilton_innate_1975} (see Supplemental Text S1
``Derivation of Eq.\,(\ref{eq:price-mean})''):
\begin{equation}
  \label{eq:price-mean}
  \begin{split}
    \mathbb{E}\left[\Delta \langle\!\langle k_{\tilde{i}\tilde{j}}\rangle\!\rangle\right] =\,
    &\langle\!\langle
    w_{\tilde{i}\tilde{j}}\rangle\!\rangle^{-1}\left\{\mathrm{cov}_{\tilde{i}}
      \left[\langle w_{i\tilde{j}}\rangle, \langle k_{i\tilde{j}}\rangle\right]
     +\mathrm{ave}_{\tilde{i}}\left[\mathrm{cov}_{i\tilde{j}}\left[w_{ij}, k_{ij}\right]\right]\right\},
  \end{split}
\end{equation}
where $\mathbb{E}[x]$ is the expected value of $x$ after one iteration
of the Wright-Fisher process, $\mathrm{cov}_{\tilde{i}}[x_i,y_i]$ is the
covariance between $x_i$ and $y_i$ over collectives,
$\mathrm{cov}_{i\tilde{j}}[x_{ij},y_{ij}]$ is the covariance between
$x_{ij}$ and $y_{ij}$ over replicators in collective $i$, and
$\mathrm{ave}_{\tilde{i}}[x_i]$ is $x_i$ averaged over collectives (see
Table\,\ref{tbl:symbols} for precise definitions). Note that the
right-hand side of Eq.\,(\ref{eq:price-mean}) is divided by
$\langle\!\langle w_{\tilde{i}\tilde{j}}\rangle\!\rangle$, so that
$\mathbb{E}[\Delta \langle\!\langle
k_{\tilde{i}\tilde{j}}\rangle\!\rangle]$ depends on relative rather than
absolute fitness (note also that relative fitness is independent of the
absolute values of $k_{ij}$ and $\langle k_{i\tilde{j}}\rangle$, as
described in Model).

Expanding $\langle w_{i\tilde{j}}\rangle$ and $w_{ij}$ in
Eq.\,(\ref{eq:price-mean}) as a Taylor series around
$\langle k_{i\tilde{j}}\rangle=\langle\!\langle
k_{\tilde{i}\tilde{j}}\rangle\!\rangle$ and
$k_{ij}=\langle k_{i\tilde{j}}\rangle$ \cite{Iwasa1991}, we obtain (see
Supplemental Text S1 ``Derivation of Eq.\,(\ref{eq:price-mean-exp})'')
\begin{equation}
  \label{eq:price-mean-exp}
    \mathbb{E}\left[\Delta \langle\!\langle k_{\tilde{i}\tilde{j}}\rangle\!\rangle\right] =
    s_\mathrm{a}v_\mathrm{a} - s_\mathrm{w}v_\mathrm{w} +
    O(s_\mathrm{w}^2) +O(s_\mathrm{a}^2),
\end{equation}
where $v_\mathrm{a}$ is the variance of $\langle k_{i\tilde{j}}\rangle$
among collectives, and $v_\mathrm{w}$ is the average variance of
$k_{ij}$ among replicators within a collective
(Table\,\ref{tbl:symbols}). Equation\,(\ref{eq:price-mean-exp}) implies
that if $s_\mathrm{a}$ and $s_\mathrm{w}$ are sufficiently small, the
boundary of the parameter regions, on which
$\mathbb{E}[\Delta \langle\!\langle
k_{\tilde{i}\tilde{j}}\rangle\!\rangle]=0$, is given by the following
equation: $s_\mathrm{a}v_\mathrm{a} = s_\mathrm{w}v_\mathrm{w}$. Since
this equation is expected to imply scaling relation
$N\,\propto\, m^{-\alpha}$, we need to calculate $v_\mathrm{w}$ and
$v_\mathrm{a}$ to calculate $\alpha$.

To calculate $v_\mathrm{w}$ and $v_\mathrm{a}$, we first consider a
neutral case where $s_\mathrm{a}=s_\mathrm{w}=0$. Let the total variance
be $v_\mathrm{t} = v_\mathrm{a} + v_\mathrm{w}$. In each generation, $M$
replicators are randomly sampled from replicators of the previous
generation with mutation. Mutation increases $v_\mathrm{t}$ to the
variance of $k_{ij}+\epsilon$, which is $v_\mathrm{t} + m\sigma$ since
$k_{ij}$ and $\epsilon$ are uncorrelated (the variance of $\epsilon$ is
$m\sigma$). Moreover, the sampling decreases the variance by a factor of
$1-M^{-1}$ (in general, sample variance of sample size $M$ is smaller
than population variance by a factor of $1-M^{-1}$). Therefore, the
expected total variance of the next generation is
\begin{equation}
\label{eq:total-variance-neutral}
\mathbb{E}\left[v_\mathrm{t}'\right] = (1-M^{-1})(v_\mathrm{t} + m \sigma).
\end{equation}

Likewise, the expected within-collective variance of the next generation
can be calculated as follows. To enable this calculation, we assume that
all collectives always consist of $\beta^{-1}N$ replicators, where
$\beta$ is a constant (as will be described later, this approximation
becomes invalid for $s\gtrsim 1$; however, its validity for $s\ll1$ is
suggested by the fact that it enables us to calculate scaling exponent
$\alpha$ correctly). Randomly sampling $\beta^{-1}N$
replicators from a collective with mutation is expected to change
$v_\mathrm{w}$ to
\begin{equation}
  \label{eq:within-variance-neutral}
  \mathbb{E}\left[v_\mathrm{w}'\right] = (1-\beta N^{-1})(v_\mathrm{w} +m\sigma).
\end{equation}
Since $\mathbb{E}[v_\mathrm{a}'] = \mathbb{E}[v_\mathrm{t}'] - \mathbb{E}[v_\mathrm{w}']$, we obtain
\begin{equation}
\label{eq:among-variance-neutral}
\mathbb{E}\left[v_\mathrm{a}'\right] = (1-M^{-1})v_\mathrm{a} + (\beta N^{-1}-M^{-1})(v_\mathrm{w}+m \sigma),
\end{equation}
where the first term on the right-hand side indicates a decrease due to
random genetic drift, and the second term indicates an increase due to
random walks of $\langle k_{i\tilde{j}}\rangle$ through
within-collective neutral evolution. Note that
Eqs.\,(\ref{eq:within-variance-neutral}) and
(\ref{eq:among-variance-neutral}) partially incorporate the
collective-level division-removal process implicitly through the
assumption of a constant collective size.

Next, we incorporate the effect of selection on $v_\mathrm{w}$ and
$v_\mathrm{a}$. Allowing for the fact that replicators are sampled with
probabilities proportional to fitness $w_{ij}$, we can use Price's
equation to express the expected values of $v_\mathrm{w}$ and
$v_\mathrm{a}$ after one iteration of the Wright-Fisher process as
follows (see Supplemental Text S1 ``Derivation of
Eq.\,(\ref{eq:price-variance})'') \cite{Zhang2010}:
\begin{equation}
  \label{eq:price-variance}
  \begin{split}
    \mathbb{E}\left[v_\mathrm{w}'\right] &= \left(1-\beta N^{-1}\right)\left[v_\mathrm{w}+m\sigma-s_\mathrm{w}c_\mathrm{w}+O(s_\mathrm{w}^2)\right]\\
  \mathbb{E}\left[v_\mathrm{a}'\right] &= \left(1-M^{-1}\right)\left[v_\mathrm{a}+s_\mathrm{a}c_\mathrm{a}+O\left(s_\mathrm{a}^2\right)+O\left(s_\mathrm{w}^2\right)\right]\\
&\ +\left(\beta
  N^{-1}-M^{-1}\right)\left[v_\mathrm{w}+m\sigma-s_\mathrm{w}c_\mathrm{w}+O(s_\mathrm{w}^2)\right]
,
\end{split}
\end{equation}
where $c_\mathrm{w}$ is the average third central moments of $k_{ij}$
within a collective, and $c_\mathrm{a}$ is the third central moment of
$\langle k_{i\tilde{j}}\rangle$. Besides the assumption of a constant
collective size, the derivation of Eq.\,(\ref{eq:price-variance})
involves the additional assumption that the variance of $k_{ij}$ within
collective $i$ is statistically independent of
$\langle k_{i\tilde{j}}\rangle$ as $i$ varies.

\begin{figure}[t]
  \centerline{\includegraphics{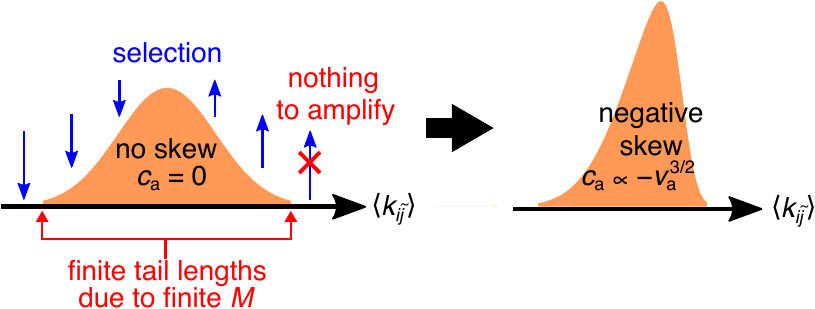}}
  \caption{\label{fig:building-asymmetry}Mechanism by which trait
    distribution becomes skewed owing to selection and finiteness of
    population. Drawing depicts frequency distributions of
    collective-level trait $\langle k_{i\tilde{j}}\rangle$ (orange) and
    effect of among-collective selection (blue arrows) (for simplicity,
    within-collective selection is not depicted). Distribution is
    initially assumed to be symmetric (left), so that its third central
    moment $c_\mathrm{a}$ is zero. Tails of distribution have finite
    lengths due to finiteness of total population size $M$ (red
    arrows). Because of finite lengths, left and right tails react
    differently to selection depending on whether they are amplified or
    reduced (red cross; see also main text). Consequently, distribution
    gets skewed (right), and $c_\mathrm{a}$ becomes negative. It is
    postulated based on dimension that
    $c_\mathrm{a}\,\propto\,-v_\mathrm{a}^{3/2}$ at steady state, where
    $v_\mathrm{a}$ is variance of $\langle k_{i\tilde{j}}\rangle$.}
\end{figure}

Given that the dimension of $c_\mathrm{w}$ and $c_\mathrm{a}$ is
equivalent to that of $v_\mathrm{w}^{3/2}$ and $v_\mathrm{a}^{3/2}$, we
make a postulate, which we verify later by simulations, that
\begin{equation}
  \label{eq:postulate}
  \begin{split}
    c_\mathrm{a}&=-\gamma_\mathrm{a} v_\mathrm{a}^{3/2},\\
    c_\mathrm{w}&=\gamma_\mathrm{w} v_\mathrm{w}^{3/2},
  \end{split}
\end{equation}
where $\gamma_\mathrm{a}$ and $\gamma_\mathrm{w}$ are positive
constants. An intuitive reason for postulating that $c_\mathrm{a}<0$ is
due to the finiteness of $M$, as follows
(Fig.\,\ref{fig:building-asymmetry}). The distribution of
$\langle k_{i\tilde{j}}\rangle$ has a finite range since $M$ is
finite. The right tail of this distribution, the one with greater
$\langle k_{i\tilde{j}}\rangle$, is exponentially amplified by selection
among collectives; however, the right tail cannot be extended because
its length is finite \cite{Tsimring1996,Hallatschek2011}. By contrast,
the left tail is contracted by among-collective selection, and this
contraction is unaffected by the finiteness of the tail length. Likewise,
the finiteness of tail lengths does not affect the rightward shift of
the mean of the distribution due to among-collective
selection. Consequently, asymmetry builds up such that the right tail
becomes shorter than the left tail, hence $c_\mathrm{a}<0$. The same
argument can be applied to $c_\mathrm{w}$, but the direction of
selection is opposite, hence the opposite sign: $c_\mathrm{w} >0$.

Combining Eqs.\,(\ref{eq:price-variance}) and (\ref{eq:postulate}), we
obtain
  \begin{align}
  \label{eq:diff-variance-closed-vw}
    \mathbb{E}\left[v_\mathrm{w}'\right] &= \left(1-\beta
                                           N^{-1}\right)\left[v_\mathrm{w}+m\sigma-\gamma_\mathrm{w}s_\mathrm{w}v_\mathrm{w}^{3/2}+O(s_\mathrm{w}^2)\right]\\
    \label{eq:diff-variance-closed-va}
  \mathbb{E}\left[v_\mathrm{a}'\right] &= \left(1-M^{-1}\right)\left[v_\mathrm{a}-\gamma_\mathrm{a}s_\mathrm{a}v_\mathrm{a}^{3/2}+O(s_\mathrm{w}^2)+O(s_\mathrm{a}^2)\right]\\
&\ +\left(\beta
  N^{-1}-M^{-1}\right)\left[v_\mathrm{w}+m\sigma-\gamma_\mathrm{w}s_\mathrm{w}v_\mathrm{w}^{3/2}+O(s_\mathrm{w}^2)\right]\nonumber
,
  \end{align}

Equations\,(\ref{eq:diff-variance-closed-vw}) and
(\ref{eq:diff-variance-closed-va}) enable us to calculate $v_\mathrm{w}$
and $v_\mathrm{a}$ at a steady state if $s_\mathrm{a}$ and
$s_\mathrm{w}$ are sufficiently small (a steady state is defined as
$\mathbb{E}\left[v_\mathrm{w}'\right]=v_\mathrm{w}$ and
$\mathbb{E}\left[v_\mathrm{a}'\right]=v_\mathrm{a}$). For illustration,
let us consider extreme conditions in which the expressions of
$v_\mathrm{w}$ and $v_\mathrm{a}$ become simple. Specifically, if
$\beta^{-1} N\gg 1$ and
$s_\mathrm{w}\ll[\gamma_\mathrm{w}
\sqrt{m\sigma}(\beta^{-1}N)^{3/2}]^{-1}$,
Eq.\,(\ref{eq:diff-variance-closed-vw}) implies that
\begin{equation}
\label{eq:variance-closed-vw}
v_\mathrm{w} \approx \beta^{-1} N m\sigma.
\end{equation}
Moreover,
Eq.\,(\ref{eq:diff-variance-closed-va}) implies that
\begin{equation}
  \label{eq:variance-implicit-va}
  M^{-1}v_\mathrm{a}+\gamma_\mathrm{a}s_\mathrm{a}v_\mathrm{a}^{3/2}\approx\left(\beta N^{-1}-M^{-1}\right)v_\mathrm{w},
\end{equation}
where the term involving $s_\mathrm{a}M^{-1}$ is ignored under the
assumption that both $s_\mathrm{a}$ and $M^{-1}$ are sufficiently small
(and the assumptions that $\beta^{-1} N\gg 1$ and
$s_\mathrm{w}\ll[\gamma_\mathrm{w}
\sqrt{m\sigma}(\beta^{-1}N)^{3/2}]^{-1}$ are used again). Substituting
Eq.\,(\ref{eq:variance-closed-vw}) into
Eq.\,(\ref{eq:variance-implicit-va}), we obtain
\begin{equation}
  \label{eq:variance-closed-va}
  v_\mathrm{a} \approx \begin{cases}
    Mm\sigma\left(1-\frac{\beta^{-1}N}{M}\right) &\text{if\ } s_\mathrm{a}\ll (\gamma_\mathrm{a} \sqrt{m\sigma}M^{3/2})^{-1}  \\
    \left[\frac{m\sigma}{\gamma_\mathrm{a} s_\mathrm{a}}\left(1-\frac{\beta^{-1}N}{M}\right)\right]^{2/3} &\text{if\ }
    s_\mathrm{a} \gg (\gamma_\mathrm{a} \sqrt{m\sigma}M^{3/2})^{-1}.
  \end{cases}
\end{equation}

Equation~(\ref{eq:variance-closed-va}) shows that $v_\mathrm{a}$ at a
steady state is independent of $N$ if $\beta^{-1}N \ll M$, a result that
might be contrary to one's intuition since by the law of large number,
increasing $N$ reduces random genetic drift within collectives and thus
decelerates the growth of $v_\mathrm{a}$. Indeed, the increase of
$v_\mathrm{a}$ per generation is approximately proportional to
$N^{-1}v_\mathrm{w}$ according to the second term of
Eq.\,(\ref{eq:diff-variance-closed-va}). However, since
$v_\mathrm{w}\,\propto\, Nm$ according to
Eq.\,(\ref{eq:variance-closed-vw}), $N$ cancels out, so that
$v_\mathrm{a}$ is independent of $N$ (see Supplemental
Fig.\,\sfigSingleLevel\ for simulation results). This cancellation
resembles that occurring in the rate of neutral molecular evolution,
which is also independent of population size \cite{Kimura1968}.

\begin{figure}[t]
  \centerline{\includegraphics{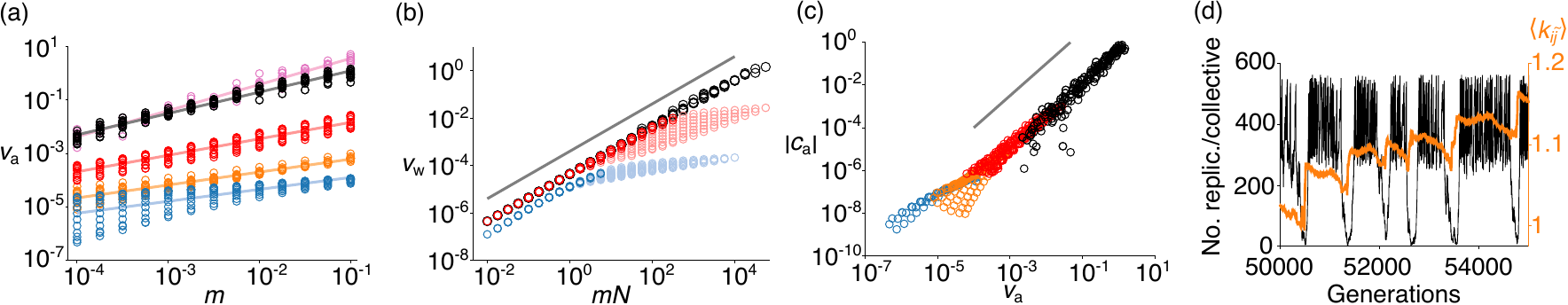}}
  \caption{\label{fig:moment-validation} Testing theory by simulations
    ($M=5\times 10^5$, $\sigma=10^{-4}$, and
    $s_\mathrm{w}=s_\mathrm{a}=s$). (a) Circles are simulation results:
    $s=10^{-6}$ (pink), $10^{-5}$ (black), $10^{-2}$ (red), $1$
    (orange), $10^{2}$ (blue). Lines are least squares regression:
    $v_\mathrm{a}\,\propto\, m^\eta$, where $\eta=0.98$ (pink), $0.8$
    (black), $0.62$ (red), $0.48$ (orange), $0.45$ (blue). (b) Circles
    are simulation results: $s=10^{-5}$ (black), $10^{-2}$ (red),
    $10^{2}$ (blue). For $s=10^{-2}$ or $10^{2}$, lighter (or darker)
    colours indicate
    $\Delta \langle\!\langle k_{\tilde{i}\tilde{j}}\rangle\!\rangle<0$
    (or $>0$, respectively). Line is $v_\mathrm{w}\,\propto\, mN$, as
    predicted by Eq.\,(\ref{eq:variance-closed-vw}). (c) Circles are simulation
    results: $s=10^{-5}$ (black), $10^{-2}$ (red), $10^0$ (orange), and
    $10^{2}$ (blue). Line is
    $|c_\mathrm{a}|\,\propto\, v_\mathrm{a}^{3/2}$, as postulated in
    Eq.\,(\ref{eq:postulate}). (d) Dynamics of common ancestors of
    collectives: number of replicators per collective (black) and
    $\langle k_{i\tilde{j}}\rangle$ (orange). $s=1$, $N=562$, and
    $m=0.01$. See also Methods under `Ancestor tracking.'}
\end{figure}

To examine the validity of Eqs.\,(\ref{eq:diff-variance-closed-vw}) and
(\ref{eq:diff-variance-closed-va}), we measured $v_\mathrm{a}$,
$v_\mathrm{w}$, $c_\mathrm{a}$, and $c_\mathrm{w}$ through simulations,
assuming $s_\mathrm{w}=s_\mathrm{a}=s$
(Fig.\,\ref{fig:moment-validation}). The results show that
$v_\mathrm{a}\,\propto\, m$ for a very small value of $s$ (viz.,
$10^{-6}$) in agreement with Eq.\,(\ref{eq:variance-closed-va})
(Fig.\,\ref{fig:moment-validation}a).  Moreover,
$v_\mathrm{w}\,\propto\, mN$ as predicted by
Eq.\,(\ref{eq:variance-closed-vw}) (Fig.\,\ref{fig:moment-validation}b),
except for cases where
$\Delta\langle\!\langle k_{\tilde{i}\tilde{j}}\rangle\!\rangle<0$ (this
deviation will be discussed later). Finally,
$c_\mathrm{a}\approx \gamma_\mathrm{a}v_\mathrm{a}^{3/2}$ if $s\ll1$
(Fig.\,\ref{fig:moment-validation}c), and
$c_\mathrm{w}\approx\gamma_\mathrm{w} v_\mathrm{w}^{3/2}$ if $s\ll1$ and
$\Delta\langle\!\langle
k_{\tilde{i}\tilde{j}}\rangle\!\rangle\lessapprox 0$ (Supplemental
Fig.\,\sfigWithinMoment), as postulated in
Eq.\,(\ref{eq:postulate}). Taken together, these results support the
validity of Eqs.\,(\ref{eq:diff-variance-closed-vw}) and
(\ref{eq:diff-variance-closed-va}) when $s_\mathrm{w}$ and
$s_\mathrm{a}$ are sufficiently small, and $m$ and $N$ are close to the
boundary of the parameter regions (i.e.,
$\Delta \langle\!\langle
k_{\tilde{i}\tilde{j}}\rangle\!\rangle\approx0$).

Using Eqs.\,(\ref{eq:diff-variance-closed-vw}) and
(\ref{eq:diff-variance-closed-va}), we can calculate the scaling
exponent ($\alpha$) of the boundary of the parameter regions for
sufficiently small $s_\mathrm{a}$ and $s_\mathrm{w}$. Since
$\mathbb{E}[\Delta\langle\!\langle
k_{\tilde{i}\tilde{j}}\rangle\!\rangle]=0$ on the parameter boundary,
Eq.\,(\ref{eq:price-mean-exp}) implies that
$v_\mathrm{a}/v_\mathrm{w}\approx s_\mathrm{w}/s_\mathrm{a}$. Thus, for
extreme parameter conditions (viz., $1\ll\beta^{-1}N\ll M$, and
$s_\mathrm{w}\ll[\gamma_\mathrm{w}
\sqrt{m\sigma}(\beta^{-1}N)^{3/2}]^{-1}$),
Eqs.\,(\ref{eq:variance-closed-vw}) and (\ref{eq:variance-closed-va})
imply that
\begin{equation}
\begin{aligned}
  &\alpha\approx 0 &&\text{if\ \ } s_\mathrm{a}\ll (\gamma_\mathrm{a} \sqrt{m\sigma}M^{3/2})^{-1}\\
  &\alpha\approx 1/3 &&\text{if\ \ } s_\mathrm{a} \gg (\gamma_\mathrm{a} \sqrt{m\sigma}M^{3/2})^{-1}.
    \end{aligned}
\end{equation}
For $s_\mathrm{a}\sim(\gamma_\mathrm{a} \sqrt{m\sigma}M^{3/2})^{-1}$,
Eqs.\,(\ref{eq:variance-closed-vw}) and (\ref{eq:variance-implicit-va})
imply that
\begin{equation}
  \label{eq:kaneko-crossover}
  rM^{-1}\beta^{-1}\left(N+\sqrt{r}s_\mathrm{a}\Gamma N^{3/2}m^{1/2}\right)\approx1,
\end{equation}
where $r=s_\mathrm{w}/s_\mathrm{a}$ and $\Gamma=\gamma_\mathrm{a}\sqrt{\sigma/\beta}M$,
and Eq.\,(\ref{eq:kaneko-crossover}) implies that $\alpha$ increases
from zero to one third as $\sqrt{r}s_\mathrm{a}$ increases from zero.

We also numerically obtained $\alpha$ by calculating the values of $N$
and $m$ ($m\in[10^{-4}, 10^{-1}]$) that satisfy
$v_\mathrm{a}/v_\mathrm{w}= s_\mathrm{w}/s_\mathrm{a}$ at a steady state
using Eqs.\,(\ref{eq:diff-variance-closed-vw}),
(\ref{eq:diff-variance-closed-va}), and the values of $\beta$,
$\gamma_\mathrm{a}$, and $\gamma_\mathrm{w}$ estimated from
Fig.\,\ref{fig:moment-validation}bc and Supplemental
Fig.\,\sfigWithinMoment, respectively (viz., $\beta^{-1}=0.45$ and
$\gamma_\mathrm{a}=0.26$ through least squares regression of
Eqs.\,(\ref{eq:variance-closed-vw}) and (\ref{eq:postulate}) for
$s=10^{-6}$ and $10^{-2}$, respectively; $\gamma_\mathrm{w}=0.25$
through least squares regression of Eq.\,(\ref{eq:postulate}) for
$\Delta \langle\!\langle
k_{\tilde{i}\tilde{j}}\rangle\!\rangle\lessapprox0$). The results agree
with the simulation results for $s_\mathrm{a}<1$ when $r=1$ and $10$,
and for $s_\mathrm{a}<10^{-3}$ when $r=0.1$
(Fig.\,\ref{fig:alpha-vs-s}a). We do not know why the validity of
analytical prediction is restricted when $r$ is small. Overall, the
above results support the validity of
Eqs.\,(\ref{eq:diff-variance-closed-vw}) and
(\ref{eq:diff-variance-closed-va}) for sufficiently small values of
$s_\mathrm{a}$ and $s_\mathrm{w}$.

In addition, we note that the postulate in Eq.\,(\ref{eq:postulate}) is
also supported by previous studies calculating the evolution of a
quantitative trait (viz., fitness) subject to single-level selection
\cite{Tsimring1996,Hallatschek2011}. These studies show that fitness
increases through evolution at a rate proportional to the two-third
power of the mutation rate. That result is consistent with
Eqs.\,(\ref{eq:diff-variance-closed-vw}) and
(\ref{eq:diff-variance-closed-va}) and, hence, also with
Eq.\,(\ref{eq:postulate}), as follows. Since Ref.\,\cite{Tsimring1996}
assumes single-level selection and a very large population, let us also
assume that $s_\mathrm{w}=0$ and $M\to\infty$, respectively, in our
model. Then, Eqs.\,(\ref{eq:price-mean-exp}) and
(\ref{eq:variance-closed-va}) imply that logarithmic fitness,
$\ln \langle\!\langle
w_{\tilde{i}\tilde{j}}\rangle\!\rangle\,\propto\,\langle
k_{\tilde{i}\tilde{j}}\rangle$, increases at a rate proportional to
$m^{2/3}$ (Supplemental Fig.\,\sfigSingleLevel). Reversing the argument,
we can also use the model of Ref.\,\cite{Tsimring1996} to estimate the
value of $\gamma_\mathrm{a}$ as about 0.25 (Supplemental Text S1 under
``Estimation of $\gamma_\mathrm{a}$''), which matches the value measured
in our model (viz.\ 0.26). Moreover, the model of
Ref.\,\cite{Tsimring1996} can also be applied to estimate
$\gamma_\mathrm{w}$, and the value of $\gamma_\mathrm{w}$ measured in
our model is about 0.25 (Supplemental Fig.\,\sfigWithinMoment). Taken
together, these agreements corroborate the validity of
Eq.\,(\ref{eq:postulate}).

Finally, to clarify why Eqs.\,(\ref{eq:diff-variance-closed-vw}) and
(\ref{eq:diff-variance-closed-va}) deviate from the simulation results
for $s\gtrsim1$ or
$\Delta \langle\!\langle k_{\tilde{i}\tilde{j}}\rangle\!\rangle<0$, we
tracked the genealogy of collectives backwards in time to observe the
common ancestors of all collectives (Methods under `Ancestor
tracking'). Figure~\ref{fig:moment-validation}d displays the dynamics of
$\langle k_{i\tilde{j}}\rangle$ and $n_i$ (the per-collective number of
replicators) in these ancestors along their single line of descent. The
results indicate that the model displays a phenomenon previously
described as \textit{evolutionarily stable disequilibrium} (ESD, for
short) \cite{Takeuchi2016}. Briefly, the collectives constantly
oscillate between growing and shrinking phases
(Fig.\,\ref{fig:moment-validation}d). During the growing phase, the
collectives continually grow and divide, and their
$\langle k_{i\tilde{j}}\rangle$ values gradually decline through
within-collective evolution, a decline that eventually puts the
collectives to a shrinking phase. In the shrinking phase, the
collectives steadily decrease in the number of constituent replicators;
however, their $\langle k_{i\tilde{j}}\rangle$ values abruptly jump at
the end of the shrinking phase, a transition that brings the collectives
back to the growing phase. This sudden increase of
$\langle k_{i\tilde{j}}\rangle$ is due to random genetic drift induced
by very severe within-collective population bottlenecks. Although such
an increase of $\langle k_{i\tilde{j}}\rangle$ is an extremely rare
event, it is always observed in the lineage of common ancestors because
these ancestors are the survivors of among-collective selection, which
favours high $\langle k_{i\tilde{j}}\rangle$ values
\cite{Takeuchi2016,Takeuchi2017}.

ESD breaks the assumption---involved in
Eqs.\,(\ref{eq:diff-variance-closed-vw}) and
(\ref{eq:diff-variance-closed-va})---that all collectives always consist
of $\beta^{-1}N$ replicators because ESD allows extremely small
collectives to regrow and contribute significantly to $v_\mathrm{w}$ and
$v_\mathrm{a}$ (note that the contributions of collectives to
$v_\mathrm{w}$ and $v_\mathrm{a}$ are proportional to the number of
replicators they contain, as defined in Table\,\ref{tbl:symbols}). We
found that ESD occurs for $s\gtrsim 1$
(Fig.\,\ref{fig:moment-validation}d) or for
$\Delta\langle k_{\tilde{i}\tilde{j}}\rangle<0$ (Supplemental
Fig.\,\sfigDisequilibrium). Thus, ESD might be responsible for the
failure of Eqs.\,(\ref{eq:diff-variance-closed-vw}) and
(\ref{eq:diff-variance-closed-va}) to predict $\alpha$ for $s\gtrsim 1$
(Fig.\,\ref{fig:alpha-vs-s}) as well as the fact that
$c_\mathrm{a}\neq\gamma_\mathrm{a}v_\mathrm{a}^{2/3}$ for $s\gtrsim 1$
(Fig.\,\ref{fig:moment-validation}c). In addition, ESD might also be
responsible for the fact that $v_\mathrm{w}$ is not proportional to $mN$
when $\Delta\langle k_{\tilde{i}\tilde{j}}\rangle<0$
(Fig.\,\ref{fig:moment-validation}b).

\begin{figure}[t]
  \centerline{\includegraphics{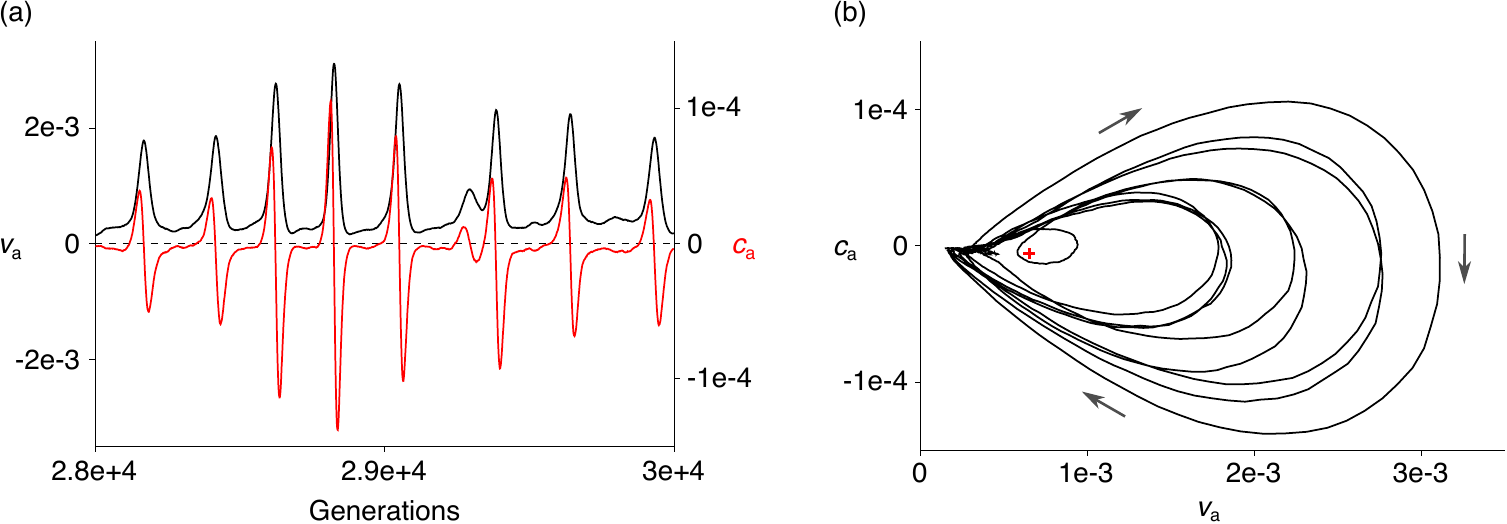}}
  \caption{\label{fig:oscillation}Oscillation of $v_\mathrm{a}$ and
    $c_\mathrm{a}$ observed in simulations ($N=5623$, $m=0.1$,
    $s_\mathrm{a}=s_\mathrm{w}=1$, $M=5\times 10^5$ and
    $\sigma=10^{-4}$). (a) $v_\mathrm{a}$ (black, left coordinate) and
    $c_\mathrm{a}$ (red, right coordinate) as functions of
    generations. (b) Phase-space trajectory of same data as shown in
    a. Cross indicates mean values of $v_\mathrm{a}$ and $c_\mathrm{a}$
    in this trajectory. Arrows indicate direction of trajectory.}
\end{figure}

Another potential reason for the failure of
Eqs.\,(\ref{eq:diff-variance-closed-vw}) and
(\ref{eq:diff-variance-closed-va}) for $s\geq 1$ is the fact that
$v_\mathrm{a}$ and $c_\mathrm{a}$ constantly oscillate with a periodic
sign change of $c_\mathrm{a}$ (Fig.\,\ref{fig:oscillation}). This
oscillation not only invalidates the assumption that
$c_\mathrm{a}=\gamma_\mathrm{a}v_\mathrm{a}^{2/3}$, but also makes it
questionable to consider a steady-state solution of
Eq.\,(\ref{eq:diff-variance-closed-vw}) and
(\ref{eq:diff-variance-closed-va}). Finally, we add that this
oscillation is distinct from ESD, in that it is observed in terms of
$v_\mathrm{a}$ and $c_\mathrm{a}$, which are properties of an entire
population of collectives, whereas ESD is observed in terms of the
properties of common ancestors of collectives.

\subsection*{Comparison to a binary-trait model}
To examine further the generality of the scaling relation described
above, we next consider a study by Kimura
\cite{Kimura1984,Kimura1986}. Kimura has investigated a binary-trait
(i.e., two-allele) model of multilevel evolution formulated based on a
diffusion equation. Using this model, Kimura has revealed the following
scaling relation that holds when within- and among-collective evolution
exactly balance each other out:
\begin{equation}
  \label{eq:kimura-our-notation}
  N = \frac{\beta s_\mathrm{a}}{4s_\mathrm{w}} m^{-1}
\end{equation}
(the notation has been converted to ours as described in Supplemental
Text S1 under ``Converting Kimura's notation into ours'')
\cite{Kimura1984,Kimura1986}. Equation\,(\ref{eq:kimura-our-notation})
is derived under the assumption that the steady-state frequency of the
altruistic allele is identical to that in the absence of selection, thus
involving a weak-selection approximation
\cite{Kimura1984,Kimura1986}. Therefore, the scaling exponent in
Kimura's model ($\alpha\approx1$) differs from that in ours
($\alpha\approx0$) for $s_\mathrm{a}\approx0$ and
$s_\mathrm{w}\approx0$.

To study how $\alpha$ depends on $s$ (where
$s=s_\mathrm{w}=s_\mathrm{a}$) if the trait is binary, we modified our
model into a binary-trait model by assuming that $k_{ij}$ switches
between zero and one at mutation rate $m$. By simulating the modified
model, we obtained a parameter-sweep diagram, where parameter regions
were defined by the sign of
$\langle\!\langle k_{\tilde{i}\tilde{j}}\rangle\!\rangle-1/2$ at steady
states (Supplemental Fig.\,\sfigPhaseDiagramDisc; this definition of
parameter regions is essentially equivalent to that used for the
quantitative-trait model, in that it can be rephrased in terms of the
sign of $\Delta \langle\!\langle k_{\tilde{i}\tilde{j}}\rangle\!\rangle$
at $\langle\!\langle k_{\tilde{i}\tilde{j}}\rangle\!\rangle=1/2$). The
results show that the parameter-region boundary constitutes scaling
relation $N\,\propto\, m^{-\alpha}$, where $\alpha\uparrow1$ as
$s\downarrow0$ (Fig.\,\ref{fig:alpha-vs-s}b)---i.e., the evolution of
$\langle\!\langle k_{\tilde{i}\tilde{j}}\rangle\!\rangle$ becomes
increasingly dependent on $m$ as $s$ decreases. Therefore, the way
$\alpha$ depends on $s$ is compatible with
Eq.\,(\ref{eq:kimura-our-notation}), but is opposite to that in the
quantitative-trait model, where $\alpha\downarrow0$ as $s\downarrow0$
(Fig.\,\ref{fig:alpha-vs-s}a).

To pinpoint why the two models yield such distinct predictions, we
re-derived Eq.\,(\ref{eq:kimura-our-notation}) using the method
developed in the previous section (for details, see Supplemental Text S1
under ``Derivation of Kimura's result through our method''). Briefly,
the most important difference from the quantitative-trait model is in
the definition of mutation: $\epsilon$ depends on $k_{ij}$ in the
binary-trait model (specifically, $\epsilon$ takes a value of
$1-2k_{ij_p}$ with a probability of $m$, where $k_{ij_p}$ is the trait
of a parental replicator). While this difference does not alter the
condition for a parameter-region boundary implied by
Eq.\,(\ref{eq:price-mean}), it significantly changes the calculation of
variances. Namely, Eqs.\,(\ref{eq:within-variance-neutral}) and
(\ref{eq:among-variance-neutral}) need to be modified to
\begin{align}
  \label{eq:within-variance-neutral-disc-boundary}
  \mathbb{E}\left[v'_\mathrm{w}\right] &\approx (1-\beta N^{-1})\left[v_\mathrm{w} + 4m(1-m)v_\mathrm{a}\right]\\
  \label{eq:among-variance-neutral-disc-boundary}
  \mathbb{E}\left[v'_\mathrm{a}\right] &\approx (1-M^{-1})v_\mathrm{a} + (\beta
    N^{-1}-M^{-1})v_\mathrm{w} - 4(1-\beta N^{-1})m(1-m)v_\mathrm{a},
\end{align}
respectively, where we have assumed that the parameters are on a
parameter-region boundary and, therefore, that
$\langle\!\langle
k_{\tilde{i}\tilde{j}}\rangle\!\rangle=1/2$. Equations\,(\ref{eq:within-variance-neutral-disc-boundary})
and (\ref{eq:among-variance-neutral-disc-boundary}) ignore the effect of
selection and are thus an approximation expected to be valid for
sufficiently weak selection. Dividing
Eq.\,(\ref{eq:within-variance-neutral-disc-boundary}) by
Eq.\,(\ref{eq:among-variance-neutral-disc-boundary}) on each side and
assuming a steady state (i.e.,
$\mathbb{E}\left[v'_\mathrm{w}\right]/\mathbb{E}\left[v'_\mathrm{a}\right]=v_\mathrm{w}/v_\mathrm{a}$),
we obtain
\begin{equation}
    \label{eq:variance-ratio-neutral-disc-boundary}
    \frac{v_\mathrm{w}}{v_\mathrm{a}} \approx \frac{4m(1-m)(1-\beta N^{-1})}{\beta N^{-1}-M^{-1}}.
\end{equation}
Imposing the condition for a parameter-region boundary,
$v_\mathrm{w}/v_\mathrm{a}\approx s_\mathrm{a}/s_\mathrm{w}$, we obtain
\begin{equation}
  \label{eq:kimura-rederived}
  \frac{4m(1-m)(1-\beta N^{-1})}{\beta N^{-1}-M^{-1}}\approx\frac{s_\mathrm{a}}{s_\mathrm{w}},
\end{equation}
which is approximately the same as Eq.\,(\ref{eq:kimura-our-notation})
if $m\ll1$ and $M^{-1} \ll \beta N^{-1} \ll 1$ as assumed by Kimura
\cite{Kimura1984,Kimura1986}.

Equations\,(\ref{eq:within-variance-neutral-disc-boundary}) and
(\ref{eq:among-variance-neutral-disc-boundary}) allow us to understand
why the two models display different scaling exponents. These equations
contain terms involving $\pm4m(1-m)v_\mathrm{a}$, which increase
$v_\mathrm{w}$ and commensurately decrease $v_\mathrm{a}$. This
`transfer' of variance occurs because mutation causes
$\langle k_{i\tilde{j}}\rangle$ to tend towards $1/2$, for which
$v_\mathrm{w}$ is maximized, in every collective. In other words,
mutation directly causes the convergent evolution of
$\langle k_{i\tilde{j}}\rangle$, raising the $v_\mathrm{w}/v_\mathrm{a}$
ratio. Consequently, the balance between within- and among-collective
evolution strongly depends on $m$. By contrast, the quantitative-trait
model assumes that mutation does not cause any directional evolutionary
change in $\langle k_{i\tilde{j}}\rangle$. Moreover, mutation equally
increases $v_\mathrm{w}$ and $v_\mathrm{a}$ according to
Eqs.\,(\ref{eq:diff-variance-closed-vw}),
(\ref{eq:diff-variance-closed-va}), and
(\ref{eq:variance-closed-vw}). Consequently, the balance between within-
and among-collective evolution does not much depend on $m$ if selection
is weak.

\section*{Discussion}
The results presented above suggest that scaling relation
$N\,\propto\,m^{-\alpha}$ is a general feature of conflicting multilevel
evolution. Scaling exponent $\alpha$, however, depends in a non-trivial
manner on the strength of selection and whether altruism is a
quantitative or binary trait.

Although we have assumed that the parameters involved in the scaling
relation---the mutation rate, selection strength, and the distinction
between quantitative and binary traits---are independent of each other,
these parameters are potentially correlated in reality. While such
correlations are not well understood \cite{Thompson2013,Kasper2017},
discussing them can illustrate the utility of the findings of this
study. For this illustration, we first note that whether altruism is a
quantitative or binary trait can be translated into the number of loci
involved in altruism: a quantitative trait involves many loci, whereas a
binary trait involves one. The number of loci is likely to be positively
correlated with the mutation rate of the trait, and it is possibly
negatively correlated with the effect size of mutation (e.g., a single
locus with large effects versus many loci with small effects). The
effect size of mutation, in turn, is possibly positively correlated with
the strength of selection. These correlations, which we assume here for
the sake of illustration, would imply a spectrum of altruism ranging
from a strongly-selected, binary trait with a low mutation rate to a
weakly-selected, quantitative trait with a high mutation rate (we are
ignoring the possibility that mutations have highly heterogeneous
effects). Such correlations would be conducive to the evolution of
altruism, an inference that is enabled by the following findings of this
study: binary-trait altruism is susceptible to the invasion by cheaters
for a high mutation rate, but this susceptibility decreases with
selection strength ($\alpha$ decreases with $s$); by contrast,
quantitative-trait altruism is relatively insensitive to mutation for
weak selection ($\alpha$ decreases to zero as $s$ decreases to zero).

Although the results of this study are phrased in the language of
multilevel selection
\cite{Wilson1975,Slatkin1978,Aoki1982,Crow1982,Leigh1983,Kimura1984,Kimura1986,Frank1994,Rispe2000,Goodnight2005,Traulsen2006,Bijma2007,Chuang2009,Leigh2010,Frank2012,Simon2013,Fontanari2014,Luo2014,Tarnita2013,Blokhuis2018,VanVliet2019,Cooney2019,Takeuchi2016,Takeuchi2017,Takeuchi2019b},
they can easily be rephrased, mutatis mutandis, in the language of kin
selection
\cite{Queller1985,Cheverud1985,Queller1992,Wolf1999,Frank1997,Gardner2007,Bijma2008,Chuang2010,McGlothlin2010,Queller2011,Akcay2012,McGlothlin2014}. To
do this, we define the relatedness of replicators as the regression
coefficient of $\langle k_{i\tilde{j}}\rangle$ on $k_{ij}$
\cite{hamilton_selfish_1970,Rice2004}, i.e.,
$R=v_\mathrm{a}/(v_\mathrm{a}+v_\mathrm{w})$, and express all the
results in terms of $R$ instead of
$v_\mathrm{w}/v_\mathrm{a}$. Therefore, our results are compatible with
the kin selection theory.

An important issue to address for future research is to test whether the
scaling relation is observed in reality. Such tests could in principle
be conducted through evolutionary experiments.

Capturing the essence of multilevel selection, the models and analyses
presented above are likely to have broad utility. They are generally
relevant for the evolution of altruism in replicators grouped into
reproducing collectives, e.g., symbionts, organelles, or genetic
elements grouped into cells \cite{burt_genes_2006}, cells grouped into
multicellular organisms \cite{buss_evolution_1987}, or other systems
that have emerged through major evolutionary transitions
\cite{maynard_smith_major_1995}.

\section*{Methods}

\subsection*{Parameter-sweep diagram}
In Fig.\,\ref{fig:pase-diagram}, the value of
$\Delta \langle\!\langle k_{\tilde{i}\tilde{j}}\rangle\!\rangle$ was
estimated from slopes of the least squares regression of
$\langle\!\langle k_{\tilde{i}\tilde{j}}\rangle\!\rangle$ against
time. The estimates were unreliable if
$|\Delta \langle\!\langle k_{\tilde{i}\tilde{j}}\rangle\!\rangle| <
3\times10^{-7}$ owing to a limitation of simulations, in which case the
RHS of Eq.\,(\ref{eq:price-mean}) was used as a proxy for
$\Delta\langle\!\langle k_{\tilde{i}\tilde{j}}\rangle\!\rangle$.

Parameter-region boundaries across which
$\Delta\langle\!\langle k_{\tilde{i}\tilde{j}}\rangle\!\rangle$ changes
sign were estimated as follows. The value of $N$ for which the RHS of
Eq.\,(\ref{eq:price-mean}) becomes zero was estimated for each selected
value of $m$ with linear interpolation from the values of the RHS of
Eq.\,(\ref{eq:price-mean}) (or
$\Delta \langle\!\langle k_{\tilde{i}\tilde{j}}\rangle\!\rangle$ if
$s\geq 10$) measured through simulations for two smallest values of $N$
for which the RHS of Eq.\,(\ref{eq:price-mean}) (or
$\Delta \langle\!\langle k_{\tilde{i}\tilde{j}}\rangle\!\rangle$ if
$s_\mathrm{a}\geq 10$) has different signs (the values of $m$ and $V$
for which simulations were run were selected as shown in
Fig.\,\ref{fig:pase-diagram}). The resulting estimates of $N$ were then
used to estimate the value of $\alpha$ through the least squares
regression of $N\,\propto\, m^{-\alpha}$.

\subsection*{Ancestor tracking}
Ancestor tracking is a method that provides novel information about
evolutionary dynamics by tracking the genealogy of individuals backwards
in time. In our study, individuals whose genealogy was tracked were
collectives. Since collectives undergo binary fission, their genealogy
can be pictured as a binary tree, where an event of binary fission is
represented by the coalescence of two branches of the tree. As the tree
is traversed from the tips to the root (i.e., from the present to the
past), all branches eventually coalesce to a single branch, the stem of
the tree, which represents the lineage of common ancestors of all
collectives present at a particular point in time. Information about
common ancestors can be visualized as time-series data along their line
of descent, i.e., along the stem of the tree. In
Fig.\,\ref{fig:moment-validation}d, $n_i$ and
$\langle k_{i\tilde{j}}\rangle$ of the common ancestors are plotted.

\section*{Competing interests}
We have no competing financial interests. 

\section*{Author Contributions}
N.T.\ conceived the study, designed, implemented and analysed the
models, and wrote the paper. K.K.\ discussed the design, results and
implications of the study, and revised the manuscript at all
stages. N.M.\ discussed the results of the study, revised the
manuscript, and contributed to the supplementary material.

\section*{Data accessibility}
Supplemental files are housed GSA's Figshare. Supplemental Texts and
Figures can be found in Sup File 1.
C\nolinebreak[4]\hspace{-.05em}\raisebox{.4ex}{\relsize{-3}{\textbf{++}}}
source code implementing the models can be found in Sup File 2.

\section*{Acknowledgements}
The authors thank Omri Barak, Kohtoh Yukawa, and Eita Nakamura for
discussion and acknowledge the contribution of NeSI to the results of
this research. This research was supported by the University of Auckland
computational biology theme fund and JSPS KAKENHI Grant Number 15H05746,
17H06386.

\section*{Supplemental Information captions}
\textbf{Text S1 Supplemental texts.} It consists of the following
sections:
\begin{enumerate}
\item Derivation of Eq.\,(\ref{eq:price-mean})
\item Derivation of Eq.\,(\ref{eq:price-mean-exp})
\item Derivation of Eq.\,(\ref{eq:price-variance})
\item Estimation of $\gamma_\mathrm{a}$
\item Derivation of Kimura's result through our method
\item Converting Kimura's notation into ours
\end{enumerate}

\noindent\textbf{Figure \sfigSingleLevel. Rate of logarithmic fitness increase
  as function of mutation rate measured through simulations.}

\noindent\textbf{Figure \sfigWithinMoment. Average third central moment
  of $k_{ij}$ within collective ($c_\text{w}$) measured through
  simulations.}

\noindent\textbf{Figure \sfigDisequilibrium. Dynamics of common ancestors of
  collectives.}

\noindent\textbf{Figure \sfigPhaseDiagramDisc. Parameter-sweep diagrams
  of binary-trait model.}

\bibliographystyle{vancouver}
\bibliography{reference}

\end{document}


\maketitle

\section{Derivation of Eq.\,(\eqPriceMean)}
\label{sec:derivation_of_price_eq}
In this section, we derive Eq.\,(\eqPriceMean) of the main text, which
is redisplayed below:
\begin{equation*}
  \begin{split}
    \mathbb{E}[\Delta \langle\!\langle k_{\tilde{i}\tilde{j}}\rangle\!\rangle] = \langle\!\langle w_{\tilde{i}\tilde{j}}\rangle\!\rangle^{-1}\left\{\mathrm{cov}_{\tilde{i}}\left[\langle w_{i\tilde{j}}\rangle, \langle k_{i\tilde{j}}\rangle\right]
     +\mathrm{ave}_{\tilde{i}}\left[\mathrm{cov}_{i\tilde{j}}\left[w_{ij}, k_{ij}\right]\right]\right\},
  \end{split}
\end{equation*}
where the symbols are defined as follows:
\begin{equation}
\langle\!\langle k_{\tilde{i}\tilde{j}}\rangle\!\rangle :=
\frac{1}{M}\sum_{i=1}^{L} n_i \langle k_{i\tilde{j}}\rangle,
\end{equation}
where $M$ is the total number of replicators, $n_i$ is the number of
replicators in collective $i$, $L$ is the number of collectives, and
\begin{equation}
  \begin{split}
    &\langle k_{i\tilde{j}}\rangle := \frac{1}{n_i}\sum_{j=1}^{n_i} k_{ij},\\
    &\langle\!\langle w_{\tilde{i}\tilde{j}}\rangle\!\rangle :=
    \frac{1}{M}\sum_{i=1}^{L} n_i \langle w_{i\tilde{j}}\rangle,\\
    &\langle w_{i\tilde{j}}\rangle := \frac{1}{n_i}\sum_{j=1}^{n_i} w_{ij},\\
    &\mathrm{cov}_{\tilde{i}}\left[\langle w_{i\tilde{j}}\rangle,
      \langle k_{i\tilde{j}}\rangle\right] := \frac{1}{M}\sum_{i=1}^{L} n_i
    \left(\langle w_{i\tilde{j}}\rangle-\langle\!\langle
      w_{\tilde{i}\tilde{j}}\rangle\!\rangle\right)\left(\langle
      k_{i\tilde{j}}\rangle-\langle\!\langle
      k_{\tilde{i}\tilde{j}}\rangle\!\rangle\right),\\
    &\mathrm{cov}_{i\tilde{j}}\left[w_{ij}, k_{ij}\right] :=
    \frac{1}{n_i}\sum_{j=1}^{n_i}\left(w_{ij}-\langle
      w_{i\tilde{j}}\rangle\right)\left(k_{ij}-\langle
      k_{i\tilde{j}}\rangle\right),\\
    &\mathrm{ave}_{\tilde{i}}\left[\mathrm{cov}_{i\tilde{j}}\left[w_{ij},
        k_{ij}\right]\right] := \frac{1}{M}\sum_{i=1}^{L} n_i\, \mathrm{cov}_{i\tilde{j}}\left[w_{ij}, k_{ij}\right].
\end{split}
\end{equation}

In each generation, a replicator is sampled $M$ times with replacement
from replicators of the previous generation with probabilities
proportional to fitness $w_{ij}$, as in the Wright-Fisher process (see
the main text under ``Model''). To express
$\langle\!\langle k_{{\tilde{i}}{\tilde{j}}}\rangle\!\rangle$ in the
next generation, we introduce the following symbols. Let $I_l$ be the
index of the collective to which the $l$th sampled replicator belongs
($l\in\{1, 2, \cdots, M\}$), $J_l$ be the index of the sampled
replicator within collective $I_l$, and $P(I_l=i, J_l=j)$ be the
probability that replicator $j$ in collective $i$ is sampled. By the
definition of the Wright-Fisher process,
\begin{equation}
  \label{eq:wright-fisher-prob}
  P(I_l=i,\ J_l=j) = \frac{w_{ij}}{M\langle\!\langle w_{\tilde{i}\tilde{j}}\rangle\!\rangle}.
\end{equation}
Moreover, let $\epsilon_{I_lJ_l}$ be the effect of mutation and
$P(\epsilon_{I_lJ_l})$ be its probability distribution function
($\epsilon_{I_lJ_l}$ takes a value of $0$ with a probability $1-m$ or a
value sampled from a Gaussian distribution with mean 0 and variance
$\sigma$ with a probability $m$). Finally, let $\mathbb{E}[x]$ denote
the expected value of $x$ after one iteration of the Wright-Fisher
process; e.g.,
\begin{equation}
  \mathbb{E}[k_{I_lJ_l}+\epsilon_{I_lJ_l}] = \sum_{i=1}^L\sum_{j=1}^{n_i}P(I_l=i, J_l=j) \int dP(\epsilon_{I_lJ_l})(k_{ij}+\epsilon_{I_lJ_l}).
\end{equation}
Using these definitions, we can express the expected change of
$\langle\!\langle k_{{\tilde{i}}{\tilde{j}}}\rangle\!\rangle$ per
generation, denoted by
$\mathbb{E}[\Delta\langle\!\langle
k_{{\tilde{i}}{\tilde{j}}}\rangle\!\rangle]$, as follows:
\begin{equation}
 \mathbb{E}[\Delta\langle\!\langle
 k_{{\tilde{i}}{\tilde{j}}}\rangle\!\rangle] =
 \mathbb{E}\left[M^{-1}\sum_{l=1}^M(k_{I_lJ_l}+\epsilon_{I_lJ_l})\right]
 - \langle\!\langle k_{{\tilde{i}}{\tilde{j}}}\rangle\!\rangle.
\end{equation}
Since $I_l$ and $J_l$ are independent and identically distributed for
different values of $l$, we can remove the summation in the above
equation to obtain
\begin{equation}
  \label{eq:expected_change_of_average_k}
  \begin{split}
    \mathbb{E}[\Delta\langle\!\langle k_{\tilde{i}\tilde{j}}\rangle\!\rangle] &=
\mathbb{E}\left[k_{IJ}+\epsilon_{IJ}\right] - \langle\!\langle
k_{{\tilde{i}}{\tilde{j}}}\rangle\!\rangle\\
&= \mathbb{E}\left[k_{IJ}\right] - \langle\!\langle
k_{{\tilde{i}}{\tilde{j}}}\rangle\!\rangle,\\
  \end{split}
\end{equation}
where we used the fact that $\mathbb{E}\left[\epsilon_{IJ}\right]=0$ and
omitted subscript $l$.

The first term on the RHS of
Eq.\,(\ref{eq:expected_change_of_average_k}) can be calculated as
follows:
\begin{equation}
  \label{eq:expected_average_k}
  \begin{split}
    \mathbb{E}\left[k_{IJ}\right] &= \sum_{i=1}^{L}\sum_{j=1}^{n_i}P(I=i, J=j)k_{ij}\\
    &= \sum_{i=1}^{L}\sum_{j=1}^{n_i} \frac{w_{ij}}{M\langle\!\langle
      w_{\tilde{i}\tilde{j}}\rangle\!\rangle}k_{ij}\\
    &= \langle\!\langle
      w_{\tilde{i}\tilde{j}}\rangle\!\rangle^{-1} \frac{1}{M}\sum_{i=1}^{L}\sum_{j=1}^{n_i}
        w_{ij}k_{ij}\\
    &= \langle\!\langle
      w_{\tilde{i}\tilde{j}}\rangle\!\rangle^{-1} \frac{1}{M}\sum_{i=1}^{L}n_i\frac{1}{n_i}\sum_{j=1}^{n_i}
        w_{ij}k_{ij}\\
    &=\langle\!\langle
      w_{\tilde{i}\tilde{j}}\rangle\!\rangle^{-1} \frac{1}{M}\sum_{i=1}^{L}n_i \left(\langle w_{i\tilde{j}}\rangle\langle
      k_{i\tilde{j}}\rangle + \mathrm{cov}_{i\tilde{j}}\left[w_{ij},
      k_{ij}\right] \right)\\
    &= \langle\!\langle
      w_{\tilde{i}\tilde{j}}\rangle\!\rangle^{-1}\bigg\{\frac{1}{M}\big(\sum_{i=1}^{L}n_i \langle w_{i\tilde{j}}\rangle\langle
      k_{i\tilde{j}}\rangle\big)
     +\mathrm{ave}_{\tilde{i}}\left[\mathrm{cov}_{i\tilde{j}}\left[w_{ij}, k_{ij}\right]\right]\bigg\}\\
    &= \langle\!\langle
      w_{\tilde{i}\tilde{j}}\rangle\!\rangle^{-1}\left(\mathrm{cov}_{\tilde{i}}\left[\langle w_{i\tilde{j}}\rangle,
      \langle k_{i\tilde{j}}\rangle\right] + \langle\!\langle
      w_{\tilde{i}\tilde{j}}\rangle\!\rangle\langle\!\langle k_{\tilde{i}\tilde{j}}\rangle\!\rangle
     +\mathrm{ave}_{\tilde{i}}\left[\mathrm{cov}_{i\tilde{j}}\left[w_{ij}, k_{ij}\right]\right]\right)\\
    &=  \langle\!\langle k_{\tilde{i}\tilde{j}}\rangle\!\rangle +
    \langle\!\langle w_{\tilde{i}\tilde{j}}\rangle\!\rangle^{-1}\big\{\mathrm{cov}_{\tilde{i}}\left[\langle w_{i\tilde{j}}\rangle,
      \langle k_{i\tilde{j}}\rangle\right]
     +\mathrm{ave}_{\tilde{i}}\left[\mathrm{cov}_{i\tilde{j}}\left[w_{ij}, k_{ij}\right]\right]\big\}.
  \end{split}
\end{equation}

Substituting Eq.\,(\ref{eq:expected_average_k}) into
Eq.\,(\ref{eq:expected_change_of_average_k}), we obtain
Eq.\,(\eqPriceMean).

\section{Derivation of Eq.\,(\eqPriceMeanExp)}
In this section, we derive Eq.\,(\eqPriceMeanExp) of the main text, which
is redisplayed below:
\begin{equation*}
  \begin{split}
    \mathbb{E}[\Delta \langle\!\langle k_{\tilde{i}\tilde{j}}\rangle\!\rangle] =
    s_\mathrm{a}v_\mathrm{a} - s_\mathrm{w}v_\mathrm{w} +
    O(s_\mathrm{w}^2) +O(s_\mathrm{a}^2),
  \end{split}
\end{equation*}
where the symbols are defined as follows:
\begin{equation}
  \begin{split}
    v_\mathrm{a} &:= \mathrm{cov}_{\tilde{i}}\left[\langle
      k_{i\tilde{j}}\rangle, \langle k_{i\tilde{j}}\rangle\right]=\frac{1}{M}\sum_{i=1}^{L} n_i
    \left(\langle
      k_{i\tilde{j}}\rangle-\langle\!\langle
      k_{\tilde{i}\tilde{j}}\rangle\!\rangle\right)^2\\
    v_{\mathrm{w}i} &:=
    \mathrm{cov}_{i\tilde{j}}\left[k_{ij},k_{ij}\right] = \frac{1}{n_i}\sum_{j=1}^{n_i}\left(k_{ij}-\langle
      k_{i\tilde{j}}\rangle\right)^2,\\
    v_\mathrm{w} &:=
    \mathrm{ave}_{\tilde{i}}\left[v_{\mathrm{w}i}\right]=\frac{1}{M}\sum_{i=1}^{L} n_i\,v_{\mathrm{w}i}.\\
  \end{split}
\end{equation}
Equation\,(\eqPriceMeanExp) is obtained by expanding
$\langle w_{i\tilde{j}}\rangle$ and $w_{ij}$ in
Eq.\,(\eqPriceMean), i.e.,
\begin{equation*}
  \begin{split}
    \Delta \langle\!\langle k_{\tilde{i}\tilde{j}}\rangle\!\rangle = \langle\!\langle w_{\tilde{i}\tilde{j}}\rangle\!\rangle^{-1}\mathrm{cov}_{\tilde{i}}\left[\langle w_{i\tilde{j}}\rangle, \langle k_{i\tilde{j}}\rangle\right]
     +\langle\!\langle w_{\tilde{i}\tilde{j}}\rangle\!\rangle^{-1}\mathrm{ave}_{\tilde{i}}\left[\mathrm{cov}_{i\tilde{j}}\left[w_{ij}, k_{ij}\right]\right],
  \end{split}
\end{equation*}
as Taylor series around
$\langle k_{i\tilde{j}}\rangle=\langle\!\langle
k_{\tilde{i}\tilde{j}}\rangle\!\rangle$ and
$k_{ij}=\langle k_{i\tilde{j}}\rangle$, respectively.

First, we obtain the first term of Eq.\,(\eqPriceMeanExp), which stems
from the first term of Eq.\,(\eqPriceMean). We assume that
$\langle w_{i\tilde{j}}\rangle$ is an analytic function of
$\langle k_{i\tilde{j}}\rangle$ and that
$\langle w_{i\tilde{j}}\rangle=\langle\!\langle
w_{\tilde{i}\tilde{j}}\rangle\!\rangle$ for
$\langle k_{i\tilde{j}}\rangle = \langle\!\langle
k_{\tilde{i}\tilde{j}}\rangle\!\rangle$. Expanding
$\langle w_{i\tilde{j}}\rangle$ around
$\langle k_{i\tilde{j}}\rangle = \langle\!\langle
k_{\tilde{i}\tilde{j}}\rangle\!\rangle$, we obtain
\begin{equation}
  \begin{split}
 \langle w_{i\tilde{j}}\rangle = \langle\!\langle
 w_{\tilde{i}\tilde{j}}\rangle\!\rangle + \frac{\partial \langle
   w_{i\tilde{j}}\rangle}{\partial\langle k_{i\tilde{j}}\rangle}\left(\langle k_{i\tilde{j}}\rangle-\langle\!\langle
k_{\tilde{i}\tilde{j}}\rangle\!\rangle\right) + \frac{\partial^2 \langle
   w_{i\tilde{j}}\rangle}{\partial\langle k_{i\tilde{j}}\rangle^2}\left(\langle k_{i\tilde{j}}\rangle-\langle\!\langle
k_{\tilde{i}\tilde{j}}\rangle\!\rangle\right)^2 + \cdots.
  \end{split}
\end{equation}
Dividing both sides by
$\langle\!\langle w_{\tilde{i}\tilde{j}}\rangle\!\rangle$, we obtain
\begin{equation}
  \label{eq:price-mean-taylor-cellular}
  \begin{split}
 \frac{\langle w_{i\tilde{j}}\rangle}{\langle\!\langle
w_{\tilde{i}\tilde{j}}\rangle\!\rangle} = 1 + \frac{1}{\langle\!\langle
w_{\tilde{i}\tilde{j}}\rangle\!\rangle}\frac{\partial \langle
   w_{i\tilde{j}}\rangle}{\partial\langle k_{i\tilde{j}}\rangle}\left(\langle k_{i\tilde{j}}\rangle-\langle\!\langle
k_{\tilde{i}\tilde{j}}\rangle\!\rangle\right) + \frac{1}{\langle\!\langle
w_{\tilde{i}\tilde{j}}\rangle\!\rangle}\frac{\partial^2 \langle
   w_{i\tilde{j}}\rangle}{\partial\langle k_{i\tilde{j}}\rangle^2}\left(\langle k_{i\tilde{j}}\rangle-\langle\!\langle
k_{\tilde{i}\tilde{j}}\rangle\!\rangle\right)^2 + \cdots.
  \end{split}
\end{equation}
By the definition of selection strength (see the main text under ``Model''),
\begin{equation}
  \begin{split}
\left.\frac{1}{\langle\!\langle
w_{\tilde{i}\tilde{j}}\rangle\!\rangle}\frac{\partial \langle
   w_{i\tilde{j}}\rangle}{\partial\langle k_{i\tilde{j}}\rangle}\right|_{\langle k_{i\tilde{j}}\rangle=\langle\!\langle
 k_{\tilde{i}\tilde{j}}\rangle\!\rangle} &= \left.\left(\frac{1}{\langle
   w_{i\tilde{j}}\rangle}\frac{\partial \langle
   w_{i\tilde{j}}\rangle}{\partial\langle k_{i\tilde{j}}\rangle}\right)\right|_{\langle k_{i\tilde{j}}\rangle=\langle\!\langle
 k_{\tilde{i}\tilde{j}}\rangle\!\rangle}\\
&:= s_\mathrm{a}
\end{split}
\end{equation}
By mathematical induction, it can be shown that 
\begin{equation}
\frac{1}{\langle w_{i\tilde{j}}\rangle} \frac{\partial^{l+1} \langle
   w_{i\tilde{j}}\rangle}{\partial\langle k_{i\tilde{j}}\rangle^{l+1}} =
 \left(\frac{\partial}{\partial \langle k_{i\tilde{j}}\rangle} + \frac{1}{\langle
   w_{i\tilde{j}}\rangle}\frac{\partial \langle
   w_{i\tilde{j}}\rangle}{\partial\langle k_{i\tilde{j}}\rangle}\right) \frac{1}{\langle w_{i\tilde{j}}\rangle} \frac{\partial^{l} \langle
   w_{i\tilde{j}}\rangle}{\partial\langle k_{i\tilde{j}}\rangle^{l}}
\end{equation}
for $l\in\{1, 2, 3, \cdots\}$.  Since it is assumed that
$\partial s_\mathrm{a}/\partial\langle k_{i\tilde{j}}\rangle=0$
(see the main text under ``Model''), the above equation implies that
\begin{equation}
\left.\left(\frac{1}{\langle w_{i\tilde{j}}\rangle} \frac{\partial^l \langle
   w_{i\tilde{j}}\rangle}{\partial\langle k_{i\tilde{j}}\rangle^l}\right)\right|_{\langle k_{i\tilde{j}}\rangle=\langle\!\langle
 k_{\tilde{i}\tilde{j}}\rangle\!\rangle}
 = s_\mathrm{a}^l.
\end{equation}
Given the above equation, Eq.\,(\ref{eq:price-mean-taylor-cellular})
implies that
\begin{equation}
  \begin{split}
 \frac{\langle w_{i\tilde{j}}\rangle}{\langle\!\langle
w_{\tilde{i}\tilde{j}}\rangle\!\rangle} = 1 + s_\mathrm{a}\left(\langle k_{i\tilde{j}}\rangle-\langle\!\langle
k_{\tilde{i}\tilde{j}}\rangle\!\rangle\right) + O(s_\mathrm{a}^2).
  \end{split}
\end{equation}
Therefore, 
\begin{equation}
  \label{eq:price-mean-taylor-cellular-final}
  \begin{split}
    \langle\!\langle
    w_{\tilde{i}\tilde{j}}\rangle\!\rangle^{-1}\mathrm{cov}_{\tilde{i}}\left[\langle
      w_{i\tilde{j}}\rangle, \langle k_{i\tilde{j}}\rangle\right] &= \mathrm{cov}_{\tilde{i}}\left[1+s_\mathrm{a} \left(\langle k_{i\tilde{j}}\rangle-\langle\!\langle
k_{\tilde{i}\tilde{j}}\rangle\!\rangle\right)+O(s_\mathrm{a}^2), \langle k_{i\tilde{j}}\rangle\right]\\
    &= s_\mathrm{a} v_\mathrm{a} +
    O(s_\mathrm{a}^2)
  \end{split}
\end{equation}

Second, we obtain the second term of Eq.\,(\eqPriceMeanExp), which stems
from the second term of Eq.\,(\eqPriceMean). Using the same method as
above, we can show that
\begin{equation}
  \label{eq:within-cov-linear}
  \begin{split}
\langle
w_{i\tilde{j}}\rangle^{-1}\mathrm{cov}_{i\tilde{j}}\left[w_{ij},
    k_{ij}\right] = -s_\mathrm{w} v_{\mathrm{w}i} + O(s_\mathrm{w}^2).
  \end{split}
\end{equation}
We assume that $\langle w_{i\tilde{j}}\rangle$ and $v_{\mathrm{w}i}$ are
statistically uncorrelated as $i$ varies (this is equivalent to assuming
that $\langle k_{i\tilde{j}}\rangle$ and $v_{\mathrm{w}i}$ are
uncorrelated). Under this assumption,
\begin{equation}
  \label{eq:price-mean-taylor-molecular-final}
  \begin{split}
\mathrm{ave}_{\tilde{i}}\left[\mathrm{cov}_{i\tilde{j}}\left[w_{ij},
    k_{ij}\right]\right] &= \mathrm{ave}_{\tilde{i}}\left[-\langle
  w_{i\tilde{j}}\rangle\, s_\mathrm{w}\, v_{\mathrm{w}i}
+ O(s_\mathrm{w}^2) \right]\\
&= -\mathrm{ave}_{\tilde{i}}\left[\langle
  w_{i\tilde{j}}\rangle\right]  s_\mathrm{w}\, \mathrm{ave}_{\tilde{i}}\left[v_{\mathrm{w}i}\right]
+ O(s_\mathrm{w}^2)\\
&= -\langle\!\langle w_{\tilde{i}\tilde{j}}\rangle\!\rangle s_\mathrm{w} v_\mathrm{w}
+ O(s_\mathrm{w}^2).
  \end{split}
\end{equation}

Substituting Eqs.\,(\ref{eq:price-mean-taylor-cellular-final}) and
(\ref{eq:price-mean-taylor-molecular-final}) into Eq.\,(\eqPriceMean),
we obtain Eq.\,(\eqPriceMeanExp).

\section{Derivation of Eq.\,(\eqPriceVariance)}
In this section, we derive Eq.\,(\eqPriceVariance), which is redisplayed
below:
\begin{equation*}
  \begin{split}
    \mathbb{E}\left[v_\mathrm{w}'\right] &= \left(1-\beta N^{-1}\right)\left[v_\mathrm{w}+m\sigma-s_\mathrm{w}c_\mathrm{w}+O(s_\mathrm{w}^2)\right]\\
  \mathbb{E}\left[v_\mathrm{a}'\right] &= \left(1-M^{-1}\right)\left[v_\mathrm{a}+s_\mathrm{a}c_\mathrm{a}+O\left((s_\mathrm{w}+s_\mathrm{a})^2\right)\right]\\
&\ +\left(\beta
  N^{-1}-M^{-1}\right)\left[v_\mathrm{w}+m\sigma-s_\mathrm{w}c_\mathrm{w}+O(s_\mathrm{w}^2)\right],
\end{split}
\end{equation*}

\subsection{Calculation of $\mathbb{E}[v_\mathrm{w}']$}
To calculate $\mathbb{E}[v_\mathrm{w}']$, we introduce the following
symbols. Let $n'_i$ be the number of replicators in collective $i$ after
one iteration of the Wright-Fisher process. Note that $n'_i$ is a random
variable and can be expressed as
\begin{equation}
  n'_i = \sum_{l=1}^{M}\delta_{I_li}
\end{equation}
where $\delta_{I_li}$ is the Kronecker delta (i.e., $\delta_{I_li}=1$ if
$I_l=i$, and $\delta_{I_li}=0$ otherwise). Moreover, let
$\langle k_{iJ_{\tilde{l}}}\rangle$ and
$\langle\epsilon_{iJ_{\tilde{l}}}\rangle$ be the sample mean of $k_{I_lJ_l}$
and $\epsilon_{I_lJ_l}$ within collective $i$:
\begin{equation}
  \begin{split}
    \langle k_{iJ_{\tilde{l}}}\rangle &:=
    \frac{1}{n_i'}\sum_{l=1}^{n'_i}k_{iJ_l}\\
    \langle\epsilon_{iJ_{\tilde{l}}}\rangle &:= \frac{1}{n_i'}\sum_{l=1}^{n'_i}\epsilon_{I_lJ_l},\\
\end{split}
\end{equation}
which are defined to be zero when $n'_i=0$.  The probability that
$J_l=j$ given $I_l=i$ is
\begin{equation}
  \label{eq:wright-fisher-prob-conditional}
  \begin{split}
    P(J=j|I=i) &= \frac{P(I=i,J=i)}{P(I=i)}\\
    &= \frac{P(I=i,J=i)}{\sum_{j=1}^{n_i}P(I=i,J=i)}\\
    &= \frac{\frac{w_{ij}}{M\langle\!\langle
        w_{\tilde{i}\tilde{j}}\rangle\!\rangle}}{\sum_{j=1}^{n_i}\frac{w_{ij}}{M\langle\!\langle
        w_{\tilde{i}\tilde{j}}\rangle\!\rangle}}\\
    &= \frac{\frac{w_{ij}}{M\langle\!\langle
        w_{\tilde{i}\tilde{j}}\rangle\!\rangle}}{\frac{n_i\langle w_{i\tilde{j}}\rangle}{M\langle\!\langle
        w_{\tilde{i}\tilde{j}}\rangle\!\rangle}}\\
    &=\frac{w_{ij}}{n_i\langle w_{i\tilde{j}}\rangle},
  \end{split}
\end{equation}
where Eq.\,(\ref{eq:wright-fisher-prob}) is used.

Using the symbols defined above, we can express the sample variance of
$k_{ij}$ within collective $i$ in the next generation as
\begin{equation}
  v'_{\mathrm{w}i} := \frac{1}{n_i'}\sum_{l=1}^{n'_i}\left(k_{iJ_l}+\epsilon_{iJ_l}-\langle
    k_{iJ_{\tilde{l}}}\rangle-\langle\epsilon_{iJ_{\tilde{l}}}\rangle\right)^2,
\end{equation}
which is defined to be zero when $n'_i=0$.  Using $v'_{\mathrm{w}i}$, we
can express $\mathbb{E}[v_\mathrm{w}']$ as follows:
\begin{equation}
  \mathbb{E}[v_\mathrm{w}'] =  \mathbb{E}\Bigg[\frac{1}{M}\sum_{i=1}^{L}n'_iv_{\mathrm{w}i}'\Bigg].
\end{equation}
In the last equation, we can separate $k_{iJ_l}$ and $\epsilon_{iJ_l}$ as follows:
\begin{equation}
  \begin{split}
    \label{eq:within-variance-separate-mut}
    \mathbb{E}[v_\mathrm{w}'] &=  \mathbb{E}\Bigg[\frac{1}{M}\sum_{i=1}^{L}n'_i\frac{1}{n_i'}\sum_{l=1}^{n'_i}\left(k_{iJ_l}+\epsilon_{iJ_l}-\langle
    k_{iJ_{\tilde{l}}}\rangle-\langle\epsilon_{iJ_{\tilde{l}}}\rangle\right)^2\Bigg]\\
  &=\mathbb{E}\Bigg[\frac{1}{M}\sum_{i=1}^{L}\sum_{l=1}^{n'_i}\left\{\left(k_{iJ_l}-\langle
    k_{iJ_{\tilde{l}}}\rangle\right)^2  +\left(\epsilon_{iJ_l} -\langle\epsilon_{iJ_{\tilde{l}}}\rangle\right)^2+2\left(k_{iJ_l}-\langle
    k_{iJ_{\tilde{l}}}\rangle\right)\left(\epsilon_{iJ_l}
    -\langle\epsilon_{iJ_{\tilde{l}}}\rangle\right)\right\}\Bigg]\\
  &=\mathbb{E}\Bigg[\frac{1}{M}\sum_{i=1}^{L}\sum_{l=1}^{n'_i}\left(k_{iJ_l}-\langle
      k_{iJ_{\tilde{l}}}\rangle\right)^2\Bigg] + \mathbb{E}\Bigg[\frac{1}{M}\sum_{i=1}^{L}\sum_{l=1}^{n'_i}\left(\epsilon_{iJ_l} -\langle\epsilon_{iJ_{\tilde{l}}}\rangle\right)^2\Bigg]\\
    &\ \ \ \ +2\mathbb{E}\Bigg[\frac{1}{M}\sum_{i=1}^{L}\sum_{l=1}^{n'_i}\left(k_{iJ_l}-\langle
    k_{iJ_{\tilde{l}}}\rangle\right)\left(\epsilon_{iJ_l}
    -\langle\epsilon_{iJ_{\tilde{l}}}\rangle\right)\Bigg]. \\
\end{split}
\end{equation}
The last term of the final line of
Eq.\,(\ref{eq:within-variance-separate-mut}) can be shown to be zero, as
follows. With the Kronecker delta $\delta_{I_li}$, this term can be
calculated as
\begin{equation}
  \begin{split}
    \label{eq:within-variance-mut-is-independent}
&\mathbb{E}\Bigg[\frac{1}{M}\sum_{i=1}^{L}\sum_{l=1}^{n'_i}\left(k_{iJ_l}-\langle
    k_{iJ_{\tilde{l}}}\rangle\right)\left(\epsilon_{iJ_l}
    -\langle\epsilon_{iJ_{\tilde{l}}}\rangle\right)\Bigg]\\
  &=\mathbb{E}\Bigg[\frac{1}{M}\sum_{i=1}^{L}\sum_{l=1}^{M}\delta_{I_li}\left(k_{I_lJ_l}-\langle
    k_{iJ_{\tilde{l}}}\rangle\right)\left(\epsilon_{iJ_l}
    -\langle\epsilon_{iJ_{\tilde{l}}}\rangle\right)\Bigg]\\
&=\frac{1}{M}\sum_{i=1}^{L}\sum_{l=1}^{M}\mathbb{E}\left[\delta_{I_li}\left(k_{I_lJ_l}-\langle
    k_{iJ_{\tilde{l}}}\rangle\right)\right]\mathbb{E}\left[\left(\epsilon_{iJ_l}
    -\langle\epsilon_{iJ_{\tilde{l}}}\rangle\right)\right]\\
&=0,
\end{split}
\end{equation}
where we used the fact that $k_{iJ_l}$ and $\epsilon_{iJ_l}$ are
independent of each other. Therefore,
\begin{equation}
  \label{eq:within-variance-k-delta-separated}
  \begin{split}
  \mathbb{E}[v_\mathrm{w}'] 
  &=\mathbb{E}\Bigg[\frac{1}{M}\sum_{i=1}^{L}\sum_{l=1}^{n'_i}\left(k_{iJ_l}-\langle
      k_{iJ_{\tilde{l}}}\rangle\right)^2\Bigg] + \mathbb{E}\Bigg[\frac{1}{M}\sum_{i=1}^{L}\sum_{l=1}^{n'_i}\left(\epsilon_{iJ_l} -\langle\epsilon_{iJ_{\tilde{l}}}\rangle\right)^2\Bigg].\\
\end{split}
\end{equation}

To calculate the first term of
Eq.\,(\ref{eq:within-variance-k-delta-separated}), we define
within-collective conditional expectation as follows:
\begin{equation}
  \label{eq:conditional-expectation}
  \mathbb{E}_{J|I=i}[x_{IJ}] := \sum_{j=1}^{n_i}P(J=j|I=i)x_{ij}.
\end{equation}
Using Eq.\,(\ref{eq:conditional-expectation}), we can transform the
first term of Eq.\,(\ref{eq:within-variance-k-delta-separated}) as follows:
\begin{equation}
  \label{eq:within-variance-separate-standard-error-of-mean}
  \begin{split}
    &\mathbb{E}\Bigg[\frac{1}{M}\sum_{i=1}^{L}\sum_{l=1}^{n'_i}\left(k_{iJ_l}-\langle
      k_{iJ_{\tilde{l}}}\rangle\right)^2\Bigg]\\
    &=
  \mathbb{E}\left[\frac{1}{M}\sum_{i=1}^{L}\sum_{l=1}^{n'_i}\left(k_{iJ_l}
    -\mathbb{E}_{J|I=i}[k_{IJ}] + \mathbb{E}_{J|I=i}[k_{IJ}] -\langle
    k_{iJ_{\tilde{l}}}\rangle\right)^2\right]\\
  &=\mathbb{E}\Bigg[\frac{1}{M}\sum_{i=1}^{L}\bigg\{\sum_{l=1}^{n'_i}\left(k_{iJ_l}
    -\mathbb{E}_{J|I=i}[k_{IJ}]\right)^2 + \sum_{l=1}^{n'_i}\left(\mathbb{E}_{J|I=i}[k_{IJ}] -\langle
    k_{iJ_{\tilde{l}}}\rangle\right)^2\\
  &\ \ \ \ \ \ +2\sum_{l=1}^{n'_i}\left(k_{iJ_l}
    -\mathbb{E}_{J|I=i}[k_{IJ}]\right)\left(\mathbb{E}_{J|I=i}[k_{IJ}] -\langle
    k_{iJ_{\tilde{l}}}\rangle\right)\bigg\}\Bigg]\\
  &=\mathbb{E}\Bigg[\frac{1}{M}\sum_{i=1}^{L}\bigg\{\sum_{l=1}^{n'_i}\left(k_{iJ_l}
    -\mathbb{E}_{J|I=i}[k_{IJ}]\right)^2 + n'_i\left(\mathbb{E}_{J|I=i}[k_{IJ}] -\langle
    k_{iJ_{\tilde{l}}}\rangle\right)^2\\
  &\ \ \ \ \ \ +2\left(\mathbb{E}_{J|I=i}[k_{IJ}] -\langle
    k_{iJ_{\tilde{l}}}\rangle\right)\sum_{l=1}^{n'_i}\left(k_{iJ_l}
    -\mathbb{E}_{J|I=i}[k_{IJ}]\right)\bigg\}\Bigg]\\
  &=\mathbb{E}\Bigg[\frac{1}{M}\sum_{i=1}^{L}\bigg\{\sum_{l=1}^{n'_i}\left(k_{iJ_l}
    -\mathbb{E}_{J|I=i}[k_{IJ}]\right)^2 + n'_i\left(\mathbb{E}_{J|I=i}[k_{IJ}] -\langle
    k_{iJ_{\tilde{l}}}\rangle\right)^2\\
  &\ \ \ \ \ \ +2\left(\mathbb{E}_{J|I=i}[k_{IJ}] -\langle
    k_{iJ_{\tilde{l}}}\rangle\right)n'_i\left(\langle k_{iJ_{\tilde{l}}}\rangle
    -\mathbb{E}_{J|I=i}[k_{IJ}]\right)\bigg\}\Bigg]\\
  &=\mathbb{E}\Bigg[\frac{1}{M}\sum_{i=1}^{L}\bigg\{\sum_{l=1}^{n'_i}\left(k_{iJ_l}
    -\mathbb{E}_{J|I=i}[k_{IJ}]\right)^2 - n'_i\left(\langle
    k_{iJ_{\tilde{l}}}\rangle-\mathbb{E}_{J|I=i}[k_{IJ}]\right)^2\bigg\}\Bigg]\\
  &=\frac{1}{M}\sum_{i=1}^{L}\mathbb{E}\Bigg[\sum_{l=1}^{n'_i}\left(k_{iJ_l}
    -\mathbb{E}_{J|I=i}[k_{IJ}]\right)^2\Bigg] - \frac{1}{M}\sum_{i=1}^{L}\mathbb{E}\left[n'_i\left(\langle
    k_{iJ_{\tilde{l}}}\rangle-\mathbb{E}_{J|I=i}[k_{IJ}]\right)^2\right].
\end{split}
\end{equation}

The first term in the last line of
Eq.\,(\ref{eq:within-variance-separate-standard-error-of-mean}) is
calculated as follows:
\begin{equation}
  \label{eq:within-variance-no-sampling}
  \begin{split}
    &\frac{1}{M}\sum_{i=1}^{L}\mathbb{E}\Bigg[\sum_{l=1}^{n'_i}\left(k_{iJ_l}
      -\mathbb{E}_{J|I=i}[k_{IJ}]\right)^2\Bigg] \\
    =& \frac{1}{M}\sum_{i=1}^{L}\mathbb{E}\Bigg[\sum_{l=1}^{n'_i}\left(k_{iJ_l}^2
      -2\mathbb{E}_{J|I=i}[k_{IJ}]k_{iJ_l} + \mathbb{E}_{J|I=i}[k_{IJ}]^2 \right)\Bigg]\\
    =& \frac{1}{M}\sum_{i=1}^{L}\mathbb{E}\Bigg[\sum_{l=1}^{n'_i}k_{iJ_l}^2
      -2\mathbb{E}_{J|I=i}[k_{IJ}]\sum_{l=1}^{n'_i}k_{iJ_l} + n'_i\mathbb{E}_{J|I=i}[k_{IJ}]^2\Bigg]\\
    =& \frac{1}{M}\sum_{i=1}^{L}\Bigg\{\mathbb{E}\bigg[\sum_{l=1}^{n'_i}k_{iJ_l}^2\bigg]
      -2\mathbb{E}_{J|I=i}[k_{IJ}]\mathbb{E}\bigg[\sum_{l=1}^{n'_i}k_{iJ_l}\bigg] + \mathbb{E}[n'_i]\mathbb{E}_{J|I=i}[k_{IJ}]^2\Bigg\}.
  \end{split}
\end{equation}
Using the Kronecker delta $\delta_{I_li}$, we can show that
\begin{equation}
  \begin{split}
    \mathbb{E}\bigg[\sum_{l=1}^{n'_i}k_{iJ_l}\bigg] &=
    \mathbb{E}\bigg[\sum_{l=1}^{M}\delta_{I_li}k_{I_lJ_l}\bigg]\\
    &=\sum_{l=1}^{M}\mathbb{E}\left[\delta_{I_li}k_{I_lJ_l}\right]\\
    &=M\mathbb{E}\left[\delta_{Ii}k_{IJ}\right]\\
    &=M\sum_{j=1}^{n_i}P(I=i,J=j)k_{ij}\\
    &=MP(I=i)\sum_{j=1}^{n_i}P(J=j|I=i)k_{ij}\\
    &=M\frac{n_i\langle w_{i\tilde{j}}\rangle}{M\langle\!\langle w_{\tilde{i}\tilde{j}}\rangle\!\rangle}\mathbb{E}_{J|I=i}[k_{IJ}]\\
    &=\frac{n_i\langle w_{i\tilde{j}}\rangle}{\langle\!\langle w_{\tilde{i}\tilde{j}}\rangle\!\rangle}\mathbb{E}_{J|I=i}[k_{IJ}].
  \end{split}
\end{equation}
Likewise, we can show that
\begin{equation}
  \begin{split}
    \mathbb{E}\bigg[\sum_{l=1}^{n'_i}k_{iJ_l}^2\bigg] &=
    \mathbb{E}\bigg[\sum_{l=1}^{M}\delta_{I_li}k_{I_lJ_l}^2\bigg]\\
    &=\sum_{l=1}^{M}\mathbb{E}\left[\delta_{I_li}k_{I_lJ_l}^2\right]\\
    &=M\mathbb{E}\left[\delta_{Ii}k_{IJ}^2\right]\\
    &=M\sum_{j=1}^{n_i}P(I=i,J=j)k_{ij}^2\\
    &=MP(I=i)\sum_{j=1}^{n_i}P(J=j|I=i)k_{ij}^2\\
    &=\frac{n_i\langle w_{i\tilde{j}}\rangle}{\langle\!\langle w_{\tilde{i}\tilde{j}}\rangle\!\rangle}\mathbb{E}_{J|I=i}[k_{IJ}^2].\\
  \end{split}
\end{equation}
Also, we can show that
\begin{equation}
  \begin{split}
    \mathbb{E}[n'_i] &= \mathbb{E}\bigg[\sum_{l=1}^{M}\delta_{I_li}\bigg]\\
    &=\sum_{l=1}^{M}\mathbb{E}[\delta_{I_li}]\\
    &=M\mathbb{E}[\delta_{Ii}]\\
    &=M\sum_{j=1}^{n_i}P(I=i,J=j)\\
    &=MP(I=i)\\
    &=\frac{n_i\langle w_{i\tilde{j}}\rangle}{\langle\!\langle w_{\tilde{i}\tilde{j}}\rangle\!\rangle}.\\
  \end{split}
\end{equation}
Using the above results, we can transform the last line of
Eq.\,(\ref{eq:within-variance-no-sampling}) as follows:
\begin{equation}
  \label{eq:within-variance-no-sampling-easier}
  \begin{split}
 &\frac{1}{M}\sum_{i=1}^{L}\bigg\{\mathbb{E}\bigg[\sum_{l=1}^{n'_i}k_{iJ_l}^2\bigg]
      -2\mathbb{E}_{J|I=i}[k_{IJ}]\mathbb{E}\bigg[\sum_{l=1}^{n'_i}k_{iJ_l}\bigg]
      + \mathbb{E}[n'_i]\mathbb{E}_{J|I=i}[k_{IJ}]^2\bigg\}\\
      = & \frac{1}{M}\sum_{i=1}^{L}\bigg\{\frac{n_i\langle w_{i\tilde{j}}\rangle}{\langle\!\langle w_{\tilde{i}\tilde{j}}\rangle\!\rangle}\mathbb{E}_{J|I=i}[k_{IJ}^2]
      -2\mathbb{E}_{J|I=i}[k_{IJ}]\frac{n_i\langle w_{i\tilde{j}}\rangle}{\langle\!\langle w_{\tilde{i}\tilde{j}}\rangle\!\rangle}\mathbb{E}_{J|I=i}[k_{IJ}]
      + \frac{n_i\langle w_{i\tilde{j}}\rangle}{\langle\!\langle w_{\tilde{i}\tilde{j}}\rangle\!\rangle}\mathbb{E}_{J|I=i}[k_{IJ}]^2\bigg\}\\
      = & \frac{1}{M}\sum_{i=1}^{L}\bigg\{\frac{n_i\langle w_{i\tilde{j}}\rangle}{\langle\!\langle w_{\tilde{i}\tilde{j}}\rangle\!\rangle}\mathbb{E}_{J|I=i}[k_{IJ}^2]
      -\frac{n_i\langle w_{i\tilde{j}}\rangle}{\langle\!\langle w_{\tilde{i}\tilde{j}}\rangle\!\rangle}\mathbb{E}_{J|I=i}[k_{IJ}]^2\bigg\}\\
      = & \frac{1}{M}\sum_{i=1}^{L}\frac{n_i\langle w_{i\tilde{j}}\rangle}{\langle\!\langle w_{\tilde{i}\tilde{j}}\rangle\!\rangle}\left\{\mathbb{E}_{J|I=i}[k_{IJ}^2]-\mathbb{E}_{J|I=i}[k_{IJ}]^2\right\}.\\
  \end{split}
\end{equation}

The conditional expectation in the last line of
Eq.\,(\ref{eq:within-variance-no-sampling-easier}) can be
calculated as follows:
\begin{equation}
  \label{eq:witin-variance-no-sampling-final}
  \begin{split}
    &\mathbb{E}_{J|I=i}[k_{IJ}^2]-\mathbb{E}_{J|I=i}[k_{IJ}]^2\\
    &=\sum_{j=1}^{n_i}P(J=j|i=i)k^2_{ij} - \bigg(\sum_{j=1}^{n_i}P(J=j|i=i)k_{ij}\bigg)^2\\
    &=\sum_{j=1}^{n_i}\frac{w_{ij}}{n_i\langle w_{i\tilde{j}}\rangle}k^2_{ij} - \bigg(\sum_{j=1}^{n_i}\frac{w_{ij}}{n_i\langle w_{i\tilde{j}}\rangle}k_{ij}\bigg)^2\\
    &=\langle w_{i\tilde{j}}\rangle^{-1}\frac{1}{n_i}\sum_{j=1}^{n_i}w_{ij}k^2_{ij} - \bigg(\langle w_{i\tilde{j}}\rangle^{-1}\frac{1}{n_i}\sum_{j=1}^{n_i}w_{ij}k_{ij}\bigg)^2\\
    &=\langle w_{i\tilde{j}}\rangle^{-1}\left(\mathrm{cov}_{i\tilde{j}}[w_{ij},k^2_{ij}]+\langle
      w_{i\tilde{j}}\rangle \langle
      k^2_{i\tilde{j}}\rangle
      \right)-  \left\{\langle w_{i\tilde{j}}\rangle^{-1} \left(\mathrm{cov}_{i\tilde{j}}[w_{ij},k_{ij}]+\langle
      w_{i\tilde{j}}\rangle \langle
      k_{i\tilde{j}}\rangle\right)\right\}^2\\
    &=\langle w_{i\tilde{j}}\rangle^{-1}\mathrm{cov}_{i\tilde{j}}[w_{ij},k^2_{ij}]+ \langle
      k^2_{i\tilde{j}}\rangle -  \left\{\langle w_{i\tilde{j}}\rangle^{-1} \mathrm{cov}_{i\tilde{j}}[w_{ij},k_{ij}]+
      \langle k_{i\tilde{j}}\rangle\right\}^2\\
    &=v_{\mathrm{w}i} + \langle w_{i\tilde{j}}\rangle^{-1}\mathrm{cov}_{i\tilde{j}}[w_{ij},k^2_{ij}] -  \left\{\langle
        w_{i\tilde{j}}\rangle^{-1}
        \mathrm{cov}_{i\tilde{j}}[w_{ij},k_{ij}]\right\}^2 - 2 \langle
        w_{i\tilde{j}}\rangle^{-1}
        \mathrm{cov}_{i\tilde{j}}[w_{ij},k_{ij}] \langle k_{i\tilde{j}}\rangle\\
    &=v_{\mathrm{w}i}+\langle w_{i\tilde{j}}\rangle^{-1}\mathrm{cov}_{i\tilde{j}}\left[w_{ij},(k_{ij}-\langle k_{i\tilde{j}}\rangle)^2\right] -  \left\{\langle
        w_{i\tilde{j}}\rangle^{-1}
        \mathrm{cov}_{i\tilde{j}}[w_{ij},k_{ij}]\right\}^2,\\
  \end{split}
\end{equation}
where we used the fact that
$\langle k^2_{i\tilde{j}}\rangle-\langle
k_{i\tilde{j}}\rangle^2=v_{\mathrm{w}i}$.  The last line of
Eq.\,(\ref{eq:witin-variance-no-sampling-final}) can be interpreted as
the expected variance of $k_{ij}$ within collective $i$ after one
iteration of the Wright-Fisher process excluding the effect of random
sampling. Thus, let us introduce the following symbol:
\begin{equation}
  \label{eq:within-variance-selection}
  \begin{split}
\Delta_\mathrm{s} v_{\mathrm{w}i} &:= 
\left(\mathbb{E}_{J|I=i}[k_{IJ}^2]-\mathbb{E}_{J|I=i}[k_{IJ}]^2\right) -
v_{\mathrm{w}i}\\
&=\langle w_{i\tilde{j}}\rangle^{-1}\mathrm{cov}_{i\tilde{j}}\left[w_{ij},(k_{ij}-\langle k_{i\tilde{j}}\rangle)^2\right] -  \left\{\langle
        w_{i\tilde{j}}\rangle^{-1}
        \mathrm{cov}_{i\tilde{j}}[w_{ij},k_{ij}]\right\}^2,\\
\end{split}
\end{equation}
which denotes the expected change of the variance of $k_{ij}$ within
collective $i$ due to within-collective selection.

Combining Eqs.\,(\ref{eq:within-variance-no-sampling}),
(\ref{eq:within-variance-no-sampling-easier}),
(\ref{eq:witin-variance-no-sampling-final}), and
(\ref{eq:within-variance-selection}), we can transform the first term in
the last line of
Eq.\,(\ref{eq:within-variance-separate-standard-error-of-mean}) as
follows
\begin{equation}
  \label{eq:within-variance-no-sampling-average}
  \begin{split}
    \frac{1}{M}\sum_{i=1}^{L}\mathbb{E}\Bigg[\sum_{l=1}^{n'_i}\left(k_{iJ_l}
      -\mathbb{E}_{J|I=i}[k_{IJ}]\right)^2\Bigg] 
    = \frac{1}{M}\sum_{i=1}^{L}\frac{n_i\langle w_{i\tilde{j}}\rangle}{\langle\!\langle w_{\tilde{i}\tilde{j}}\rangle\!\rangle}\left(v_{\mathrm{w}i}+\Delta_\mathrm{s} v_{\mathrm{w}i}\right)\\
  \end{split}
\end{equation}

Next, we calculate the second term in the last line of
Eq.\,(\ref{eq:within-variance-separate-standard-error-of-mean}) as follows:
\begin{equation}
  \label{eq:within-variance-error-of-mean}
  \begin{split}
    &\frac{1}{M}\sum_{i=1}^{L}\mathbb{E}\left[n'_i\left(\langle
        k_{iJ_{\tilde{l}}}\rangle-\mathbb{E}_{J|I=i}[k_{IJ}]\right)^2\right]\\
    &=\frac{1}{M}\sum_{i=1}^{L}\mathbb{E}\Bigg[n'_i\bigg(\frac{1}{n'_i}\sum_{l=1}^{n'_i}k_{iJ_{l}}-\frac{1}{n'_i}\sum_{l=1}^{n'_i}\mathbb{E}_{J|I=i}[k_{IJ}]\bigg)^2\Bigg]\\
    &=\frac{1}{M}\sum_{i=1}^{L}\mathbb{E}\Bigg[n'_i\bigg\{\frac{1}{n'_i}\sum_{l=1}^{n'_i}\left(k_{iJ_{l}}-\mathbb{E}_{J|I=i}[k_{IJ}]\right)\bigg\}^2\Bigg]\\
    &=\frac{1}{M}\sum_{i=1}^{L}\mathbb{E}\Bigg[\frac{1}{n'_i}\bigg\{\sum_{l=1}^{n'_i}\left(k_{iJ_{l}}-\mathbb{E}_{J|I=i}[k_{IJ}]\right)\bigg\}^2\Bigg]\\
    &=\frac{1}{M}\sum_{i=1}^{L}\mathbb{E}\Bigg[\frac{1}{n'_i}\bigg\{\sum_{l=1}^{n'_i}\left(k_{iJ_{l}}-\mathbb{E}_{J|I=i}[k_{IJ}]\right)^2\\
    & \ \ \ \ \ +2\sum_{m\neq
      l}\left(k_{iJ_{l}}-\mathbb{E}_{J|I=i}[k_{IJ}]\right)\left(k_{iJ_{m}}-\mathbb{E}_{J|I=i}[k_{IJ}]\right)\bigg\}\Bigg]\\
    &=\frac{1}{M}\sum_{i=1}^{L}\mathbb{E}\Bigg[\frac{1}{n'_i}\sum_{l=1}^{n'_i}\left(k_{iJ_{l}}-\mathbb{E}_{J|I=i}[k_{IJ}]\right)^2\Bigg],\\
\end{split}
\end{equation}
where we used the fact that $k_{iJ_{l}}$ and $k_{iJ_{m}}$ are
independent of each other for $l\neq m$ in the final step. Since $n'_i$
can be zero, the last line of
Eq.\,(\ref{eq:within-variance-error-of-mean}) needs to interpreted as
follows:
\begin{equation}
  \begin{split}
    \frac{1}{n'_i}\sum_{l=1}^{n'_i}\left(k_{iJ_{l}}-\mathbb{E}_{J|I=i}[k_{IJ}]\right)^2
    =\begin{cases}
      0, & \text{if } n'_i=0\\
      \frac{1}{n'_i}\sum_{l=1}^{n'_i}\left(k_{iJ_{l}}-\mathbb{E}_{J|I=i}[k_{IJ}]\right)^2,
      & \text{if } n'_i>0.
  \end{cases}
\end{split}
\end{equation}
Thus,
\begin{equation}
  \begin{split}
    \mathbb{E}\Bigg[\frac{1}{n'_i}\sum_{l=1}^{n'_i}\left(k_{iJ_{l}}-\mathbb{E}_{J|I=i}[k_{IJ}]\right)^2\Bigg]
    =\begin{cases}
      0, & \text{if } n'_i=0\\
      \mathbb{E}_{J|I=i}[k_{IJ}^2]-\mathbb{E}_{J|I=i}[k_{IJ}]^2,
      & \text{if } n'_i>0,
  \end{cases}
\end{split}
\end{equation}
where the case for $n'_i>0$ follows from a calculation similar to
Eqs.\,(\ref{eq:within-variance-no-sampling}) and
(\ref{eq:within-variance-no-sampling-easier}).

To calculate the last line of
Eq.\,(\ref{eq:within-variance-error-of-mean}), we separate the case
where $n'_i>0$ $\forall i\in\{1,2,3,\cdots,L\}$ and the case where
$n'_i=0$ for some $i\in \{1,2,3,\cdots,L\}$. Let $S$ be a proper subset
of $\{1,2,3,\cdots,L\}$, and $P(S)$ be the probability that $n'_i=0$
$\forall i\in S$ and $n'_i>0$ $\forall i\not\in S$. The value of $P(S)$ can
be estimated as follows:
\begin{equation}
  \label{eq:probability-of-collective-death}
  \begin{split}
    P(S)&\leq \left[1-\sum_{i\in S}P(I=i)\right]^M\\
     &=\Bigg[1-\sum_{i\in S} \frac{n_i\langle w_{i\tilde{j}}\rangle}{M\langle\!\langle w_{\tilde{i}\tilde{j}}\rangle\!\rangle}\Bigg]^M\\
      &\approx \exp\left(-\sum_{i\in S} \frac{n_i\langle
          w_{i\tilde{j}}\rangle}{\langle\!\langle
          w_{\tilde{i}\tilde{j}}\rangle\!\rangle}\right),\\
  \end{split}
\end{equation}
where the RHS of the first inequality is the probability that $n'_i=0$
$\forall i\in S$ and $n'_i\geq0$ $\forall i\not\in S$. Using these
symbols and Eq.\,(\ref{eq:within-variance-selection}), we can express
the last line of Eq.\,(\ref{eq:within-variance-error-of-mean}) as
follows:
\begin{equation}
  \label{eq:within-variance-error-of-mean-with-collective-death}
  \begin{split}
    &\frac{1}{M}\sum_{i=1}^{L}\mathbb{E}\Bigg[\frac{1}{n'_i}\sum_{l=1}^{n'_i}\left(k_{iJ_{l}}-\mathbb{E}_{J|I=i}[k_{IJ}]\right)^2\Bigg],\\
    &= \frac{1}{M}\sum_{S}P(S)\sum_{i\not\in
      S}\left(\mathbb{E}_{J|I=i}[k_{IJ}^2]-\mathbb{E}_{J|I=i}[k_{IJ}]^2\right)\\
    &= \frac{1}{M}\sum_{S}P(S)\sum_{i\not\in
      S}\left(v_{\mathrm{w}i}+\Delta_\mathrm{s} v_{\mathrm{w}i}\right)\\
    &=
    \frac{1}{M}\sum_{S}P(S)\Bigg[\sum_{i=1}^L\left(v_{\mathrm{w}i}+\Delta_\mathrm{s} v_{\mathrm{w}i}\right)-\sum_{i\in
    S}\left(v_{\mathrm{w}i}+\Delta_\mathrm{s} v_{\mathrm{w}i}\right)\Bigg]\\
    &= \frac{1}{M}
    \sum_{i=1}^{L}\left(v_{\mathrm{w}i}+\Delta_\mathrm{s} v_{\mathrm{w}i}\right)
    - \frac{1}{M}\sum_{S\neq\emptyset}P(S)\sum_{i\in
    S}\left(v_{\mathrm{w}i}+\Delta_\mathrm{s} v_{\mathrm{w}i}\right)\\
  \end{split}
\end{equation}
where $\sum_{S}$ is a summation over all possible
$S\subset\{1,2,\cdots,L\}$, and $\sum_{S\neq\emptyset}$ is the same
summation excluding the case where $S$ is empty.

We assume that the second term of the last line of
Eq.\,(\ref{eq:within-variance-error-of-mean-with-collective-death}) is
negligible for the following reasons. If $n_i\gg 1$ for some $i\in S$,
then Eq.\,(\ref{eq:probability-of-collective-death}) implies that
$P(S)\approx 0$. Contrariwise, if the statement that $n_i\gg1$ is false
for all $i\in S$, then the value of
$\sum_{i\in S}(v_{\mathrm{w}i}+\Delta_\mathrm{s} v_{\mathrm{w}i})$ is likely to be
small because $n_i$ is small. Under this assumption,
Eq.\,(\ref{eq:within-variance-error-of-mean-with-collective-death})
implies that
\begin{equation}
\label{eq:within-variance-error-of-mean-final}
  \begin{split}
    &\frac{1}{M}\sum_{i=1}^{L}\mathbb{E}\Bigg[\frac{1}{n'_i}\sum_{l=1}^{n'_i}\left(k_{iJ_{l}}-\mathbb{E}_{J|I=i}[k_{IJ}]\right)^2\Bigg]\approx \frac{1}{M}
    \sum_{i=1}^{L}\left(v_{\mathrm{w}i}+\Delta_\mathrm{s}
      v_{\mathrm{w}i}\right).\\
  \end{split}
\end{equation}

The second term of Eq.\,(\ref{eq:within-variance-k-delta-separated}) is
calculated as follows:
\begin{equation}
  \label{eq:within-variance-delta}
  \begin{split}
    &\mathbb{E}\Bigg[\frac{1}{M}\sum_{i=1}^{L}\sum_{l=1}^{n'_i}\left(\epsilon_{iJ_l}
      -\langle\epsilon_{iJ_{\tilde{l}}}\rangle\right)^2\Bigg]\\
    &=\mathbb{E}\Bigg[\frac{1}{M}\sum_{i=1}^{L}\sum_{l=1}^{n'_i}\left(\epsilon_{iJ_l}^2
      -2\langle\epsilon_{iJ_{\tilde{l}}}\rangle\epsilon_{iJ_l}+\langle\epsilon_{iJ_{\tilde{l}}}\rangle^2\right)\Bigg]\\
    &=\mathbb{E}\Bigg[\frac{1}{M}\sum_{i=1}^{L}\bigg(\sum_{l=1}^{n'_i}\epsilon_{iJ_l}^2
    -2\langle\epsilon_{iJ_{\tilde{l}}}\rangle\sum_{l=1}^{n'_i}\epsilon_{iJ_l}+\langle\epsilon_{iJ_{\tilde{l}}}\rangle^2\sum_{l=1}^{n'_i}\bigg)\Bigg]\\
    &=\mathbb{E}\Bigg[\frac{1}{M}\sum_{i=1}^{L}\bigg(\sum_{l=1}^{n'_i}\epsilon_{iJ_l}^2
    -2n'_i\langle\epsilon_{iJ_{\tilde{l}}}\rangle^2+n'_i\langle\epsilon_{iJ_{\tilde{l}}}\rangle^2\bigg)\Bigg]\\
    &=\mathbb{E}\Bigg[\frac{1}{M}\sum_{i=1}^{L}\bigg(\sum_{l=1}^{n'_i}\epsilon_{iJ_l}^2
    -n'_i\langle\epsilon_{iJ_{\tilde{l}}}\rangle^2\bigg)\Bigg]\\
    &=\mathbb{E}\Bigg[\frac{1}{M}\sum_{l=1}^{M}\epsilon_{I_lJ_l}^2
    -\frac{1}{M}\sum_{i=1}^{L}n'_i\langle\epsilon_{iJ_{\tilde{l}}}\rangle^2\Bigg]\\
    &=\mathbb{E}\Bigg[\frac{1}{M}\sum_{l=1}^{M}\epsilon_{I_lJ_l}^2
    -\frac{1}{M}\sum_{i=1}^{L}n'_i\bigg(\frac{1}{n_i'}\sum_{l=1}^{n'_i}\epsilon_{I_lJ_l}\bigg)^2\Bigg]\\
    &=m\sigma
    -\frac{1}{M}\mathbb{E}\Bigg[\sum_{i=1}^{L}\frac{1}{n_i'}\bigg(\sum_{l=1}^{n'_i}\epsilon_{I_lJ_l}\bigg)^2\Bigg]\\
    &=m\sigma
    -\frac{1}{M}\mathbb{E}\Bigg[\sum_{i=1}^{L}\frac{1}{n_i'}\bigg(\sum_{l=1}^{n'_i}\epsilon^2_{I_lJ_l}
    + 2\sum_{l\neq m}\epsilon_{I_lJ_l}\epsilon_{I_mJ_m}\bigg)\Bigg]\\
    &=m\sigma
    -\frac{1}{M}\mathbb{E}\Bigg[\sum_{i=1}^{L}\frac{1}{n_i'}\sum_{l=1}^{n'_i}\epsilon^2_{I_lJ_l}\Bigg]\\
    &=m\sigma
    -\frac{1}{M}\sum_{S}P(S)\sum_{i\not\in
      S}\mathbb{E}[\epsilon_{IJ}^2]\\
    &=m\sigma
    -\frac{1}{M}\sum_{S}P(S)\bigg(\sum_{i=1}^Lm\sigma- \sum_{i\in
      S}m\sigma\bigg)\\
    &=m\sigma
    -\frac{L}{M}m\sigma + \frac{1}{M}\sum_{S}P(S)\sum_{i\in
      S}m\sigma\\
    &= \bigg(1-\frac{L}{M}+\frac{1}{M}\sum_SP(S)|S|\bigg)m\sigma,\\
    &\approx \bigg(1-\frac{L}{M}\bigg)m\sigma,\\
    \end{split}
\end{equation}
where $|S|$ is the number of elements in $S$, and we have assumed that
$\sum_SP(S)|S|\ll L$ in the last step.

Combining Eqs.\, (\ref{eq:within-variance-k-delta-separated}),
(\ref{eq:within-variance-separate-standard-error-of-mean}),
(\ref{eq:within-variance-no-sampling-average}),
(\ref{eq:within-variance-error-of-mean}),
(\ref{eq:within-variance-error-of-mean-final}), and
(\ref{eq:within-variance-delta}),
we obtain
\begin{equation}
  \label{eq:within-variance-exact-final}
  \begin{split}
    \mathbb{E}[v_\mathrm{w}'] &\approx
    \frac{1}{M}\sum_{i=1}^{L}\frac{n_i\langle
      w_{i\tilde{j}}\rangle}{\langle\!\langle
      w_{\tilde{i}\tilde{j}}\rangle\!\rangle}\left\{v_{\mathrm{w}i}+\Delta_\mathrm{s}
      v_{\mathrm{w}i}\right\}\\
    & \ \ \ - \frac{1}{M}\sum_{i=1}^{L}\left\{v_{\mathrm{w}i}+\Delta_\mathrm{s} v_{\mathrm{w}i}\right\}+ \left(1-\frac{L}{M}\right)m\sigma \\
    &=\langle\!\langle
    w_{\tilde{i}\tilde{j}}\rangle\!\rangle^{-1}\mathrm{cov}_{\tilde{i}}\left[\langle
      w_{i\tilde{j}}\rangle, v_{\mathrm{w}i}+\Delta_\mathrm{s}
      v_{\mathrm{w}i}\right] + M^{-1}\sum_{i=1}^{L}n_i\left(v_{\mathrm{w}i}+\Delta_\mathrm{s} v_{\mathrm{w}i}\right)\\
    &\ \ \ -
    M^{-1}\sum_{i=1}^{L}n_in_i^{-1}\left\{v_{\mathrm{w}i}+\Delta_\mathrm{s}
      v_{\mathrm{w}i}\right\} + M^{-1}\sum_{i=1}^{L}n_i(1-n_i^{-1})m\sigma.\\
    &=\langle\!\langle
    w_{\tilde{i}\tilde{j}}\rangle\!\rangle^{-1}\mathrm{cov}_{\tilde{i}}\left[\langle
      w_{i\tilde{j}}\rangle, v_{\mathrm{w}i}+\Delta_\mathrm{s}
      v_{\mathrm{w}i}\right] +
    M^{-1}\sum_{i=1}^{L}n_i(1-n_i^{-1})\left(v_{\mathrm{w}i}+\Delta_\mathrm{s}
      v_{\mathrm{w}i}+m\sigma\right)\\
    &=\langle\!\langle
    w_{\tilde{i}\tilde{j}}\rangle\!\rangle^{-1}\mathrm{cov}_{\tilde{i}}\left[\langle
      w_{i\tilde{j}}\rangle, v_{\mathrm{w}i}+\Delta_\mathrm{s}
      v_{\mathrm{w}i}\right] +
    \mathrm{ave}_{\tilde{i}}\left[(1-n_i^{-1})\left(v_{\mathrm{w}i}+\Delta_\mathrm{s} v_{\mathrm{w}i}+m\sigma\right)\right]\\
    &= \mathrm{ave}_{\tilde{i}}\left[(1-n_i^{-1})\left(v_{\mathrm{w}i}+\Delta_\mathrm{s} v_{\mathrm{w}i}+m\sigma\right)\right],\\
  \end{split}
\end{equation}
where we have assumed in the last step that
$\langle w_{i\tilde{j}}\rangle$ and
$v_{\mathrm{w}i}+\Delta_\mathrm{s}v_{\mathrm{w}i}$ are statistically
uncorrelated as $i$ varies, as we have assumed in
Eq.\,(\ref{eq:price-mean-taylor-molecular-final}).

To enable the further calculation of
Eq.\,(\ref{eq:within-variance-exact-final}), we assume that
\begin{equation}
  \label{eq:constant-V-assumption}
  n_i = \beta N^{-1}.
\end{equation}
Under this assumption, we can transform
Eq.\,(\ref{eq:within-variance-exact-final}) as follows
\begin{equation}
  \label{eq:within-variance-approximate}
  \begin{split}
    \mathbb{E}[v_\mathrm{w}']  &\approx \mathrm{ave}_{\tilde{i}}\left[(1-n_i^{-1})\left(v_{\mathrm{w}i}+m\sigma+\Delta_\mathrm{s} v_{\mathrm{w}i}\right)\right]\\
    & \approx \left(1-\beta
  N^{-1}\right)\mathrm{ave}_{\tilde{i}}\Big[v_{\mathrm{w}i}+m\sigma+\langle
  w_{i\tilde{j}}\rangle^{-1}\mathrm{cov}_{i\tilde{j}}\left[w_{ij},(k_{ij}-\langle
    k_{i\tilde{j}}\rangle)^2\right] \\
  & \ \ \ \ \ \ \ \ \ \ \ \ \ \ \ \ \ \ \ \ \ \ \ \ \ \ -\left\{\langle w_{i\tilde{j}}\rangle^{-1}
        \mathrm{cov}_{i\tilde{j}}[w_{ij},k_{ij}]\right\}^2\Big].\\
    &= \left(1-\beta
      N^{-1}\right)\Big(v_{\mathrm{w}} +
    m\sigma+ 
      \mathrm{ave}_{\tilde{i}}\left[\langle w_{i\tilde{j}}\rangle^{-1}\mathrm{cov}_{i\tilde{j}}[w_{ij},(k_{ij}-\langle
        k_{i\tilde{j}}\rangle)^2]\right] +O(s^2_\mathrm{w}) \Big)  ,\\
  \end{split}
\end{equation}
where we used Eq.\,(\ref{eq:within-cov-linear}) in the last
step. Expanding $w_{ij}$ as a Taylor series around
$k_{ij}=\langle k_{i\tilde{j}}\rangle$, we can show that
\begin{equation}
  \label{eq:within-collective-ignore-4th-mom}
  \begin{split}
    \langle w_{i\tilde{j}}\rangle^{-1}\mathrm{cov}_{i\tilde{j}}[w_{ij},(k_{ij}-\langle
      k_{i\tilde{j}}\rangle)^2] = -s_{\mathrm{w}}c_{\mathrm{w}i} + O(s_{\mathrm{w}}^2),
  \end{split}
\end{equation}
where $c_{\mathrm{w}i}:=\langle (k_{i\tilde{j}}-\langle k_{i\tilde{j}}\rangle)^3\rangle$, and that
\begin{equation}
  \label{eq:within-collective-3rd-mom}
  \begin{split}
    \mathrm{ave}_{\tilde{i}}\left[\langle w_{i\tilde{j}}\rangle^{-1}\mathrm{cov}_{i\tilde{j}}[w_{ij},(k_{ij}-\langle
      k_{i\tilde{j}}\rangle)^2]\right] = -s_{\mathrm{w}}c_{\mathrm{w}} + O(s_{\mathrm{w}}^2),
  \end{split}
\end{equation}
where
$c_{\mathrm{w}}:=\mathrm{ave}_{\tilde{i}}[c_{\mathrm{w}i}]$. Substituting
Eq.\,(\ref{eq:within-collective-3rd-mom}) into
Eq.\,(\ref{eq:within-variance-approximate}), we obtain
\begin{equation}
  \label{eq:within-variance-approximate-final}
  \begin{split}
    \mathbb{E}[v_\mathrm{w}'] & \approx \left(1-\beta
      N^{-1}\right)\left(v_{\mathrm{w}} +
    m\sigma- s_\mathrm{w}c_\mathrm{w} +O(s^2_\mathrm{w}) \right).\\
  \end{split}
\end{equation}

\subsection{Calculation of $\mathbb{E}\left[v_\mathrm{a}'\right]$}
To calculate $\mathbb{E}\left[v_\mathrm{a}'\right]$, we first calculate
$\mathbb{E}\left[v_\mathrm{t}'\right]$ and then use the fact that
$\mathbb{E}\left[v_\mathrm{a}'\right]=\mathbb{E}\left[v_\mathrm{t}'\right]-\mathbb{E}\left[v_\mathrm{w}'\right]$.

We can express $\mathbb{E}\left[v_\mathrm{t}'\right]$ as follows:
\begin{equation}
  \label{total-v-definition}
  \begin{split}
    \mathbb{E}\left[v_\mathrm{t}'\right]
    &=
    \mathbb{E}\Bigg[\frac{1}{M}\sum_{l=1}^{M}\bigg\{k_{I_lJ_l}+\epsilon_{iJ_l}-\frac{1}{M}\sum_{l=1}^{M}(k_{I_lJ_l}+\epsilon_{iJ_l})\bigg\}^2\Bigg]\\
    &=
    \left(1-\frac{1}{M}\right)\mathbb{E}\left[\left(k_{IJ}-\mathbb{E}[k_{IJ}]\right)^2\right]
    + \left(1-\frac{1}{M}\right)m\sigma,\\
  \end{split}
\end{equation}
where we used the fact that $k_{I_lJ_l}$ and $\epsilon_{iJ_l}$ are
independent of each other, as we did in
Eqs.\,(\ref{eq:within-variance-separate-mut}),
(\ref{eq:within-variance-mut-is-independent}), and
(\ref{eq:within-variance-k-delta-separated}).

The first term of the last line of Eq.\,(\ref{total-v-definition}) is
calculated as follows:
\begin{equation}
  \label{eq:total-v-no-sampling}
  \begin{split}
    & \mathbb{E}\left[\left(k_{IJ}-\mathbb{E}[k_{IJ}]\right)^2\right]\\
    &= \mathbb{E}[\Big[\left(k_{ij}-\mathbb{E}_{J|I=i}[k_{IJ}]+\mathbb{E}_{J|I=i}[k_{IJ}]-\mathbb{E}[k_{IJ}]\right)^2\Big]\\
    &= \mathbb{E}\Big[\left(k_{ij}-\mathbb{E}_{J|I=i}[k_{IJ}]\right)^2\Big]
      +\mathbb{E}\Big[\left(\mathbb{E}_{J|I=i}[k_{IJ}]-\mathbb{E}[k_{IJ}]\right)^2\Big]\\
      & \ \ \ - 2 \mathbb{E}\Big[\left(k_{ij}-\mathbb{E}_{J|I=i}[k_{IJ}]\right)\left(\mathbb{E}_{J|I=i}[k_{IJ}]-\mathbb{E}[k_{IJ}]\right)\Big]\\
    &=\mathbb{E}\Big[\left(k_{ij}-\mathbb{E}_{J|I=i}[k_{IJ}]\right)^2\Big]
      +\mathbb{E}\Big[\left(\mathbb{E}_{J|I=i}[k_{IJ}]-\mathbb{E}[k_{IJ}]\right)^2\Big]\\
      & \ \ \ - 2 \sum_{i=1}^L\sum_{j=1}^{n_i}P(I=i,J=j)\left(\mathbb{E}_{J|I=i}[k_{IJ}]-\mathbb{E}[k_{IJ}]\right)\left(k_{ij}-\mathbb{E}_{J|I=i}[k_{IJ}]\right)\\
    &=\mathbb{E}\Big[\left(k_{ij}-\mathbb{E}_{J|I=i}[k_{IJ}]\right)^2\Big]
      +\mathbb{E}\Big[\left(\mathbb{E}_{J|I=i}[k_{IJ}]-\mathbb{E}[k_{IJ}]\right)^2\Big]\\
      & \ \ \ - 2 \sum_{i=1}^LP(I=i)\left(\mathbb{E}_{J|I=i}[k_{IJ}]-\mathbb{E}[k_{IJ}]\right)\sum_{j=1}^{n_i}P(J=j|I=i)\left(k_{ij}-\mathbb{E}_{J|I=i}[k_{IJ}]\right)\\
    &= \mathbb{E}\Big[\left(k_{ij}-\mathbb{E}_{J|I=i}[k_{IJ}]\right)^2\Big]
      +\mathbb{E}\Big[\left(\mathbb{E}_{J|I=i}[k_{IJ}]-\mathbb{E}[k_{IJ}]\right)^2\Big]\\
\end{split}
\end{equation}

The first term of the last line of Eq.\,(\ref{eq:total-v-no-sampling}) is
calculated as follows:
\begin{equation}
  \label{eq:within-v-no-sampling}
  \begin{split}
    &\mathbb{E}\Big[\left(k_{ij}-\mathbb{E}_{J|I=i}[k_{IJ}]\right)^2\Big]\\
    &=\mathbb{E}\Big[k^2_{ij}-2\mathbb{E}_{J|I=i}[k_{IJ}]k_{ij}+\mathbb{E}_{J|I=i}[k_{IJ}]^2)\Big]\\
    &=\sum_{i=1}^L\sum_{j=1}^{n_i}P(I=i,J=j)\left(k^2_{ij}-2\mathbb{E}_{J|I=i}[k_{IJ}]k_{ij}+\mathbb{E}_{J|I=i}[k_{IJ}]^2\right)\\
    &=\sum_{i=1}^LP(I=i)\bigg(\sum_{j=1}^{n_i}P(J=j|I=i)k^2_{ij}\\
    & \ \ \ -2\mathbb{E}_{J|I=i}[k_{IJ}]\sum_{j=1}^{n_i}P(J=j|I=i)k_{ij} +\mathbb{E}_{J|I=i}[k_{IJ}]^2\bigg)\\
    &=\sum_{i=1}^LP(I=i)\left(\mathbb{E}_{J|I=i}[k^2_{IJ}]-\mathbb{E}_{J|I=i}[k_{IJ}]^2\right)\\
    &=\sum_{i=1}^L\frac{n_i\langle w_{i\tilde{j}}\rangle}{M\langle\!\langle w_{\tilde{i}\tilde{j}}\rangle\!\rangle}\left(\mathbb{E}_{J|I=i}[k^2_{IJ}]-\mathbb{E}_{J|I=i}[k_{IJ}]^2\right)\\
    &=\sum_{i=1}^L\frac{n_i\langle
      w_{i\tilde{j}}\rangle}{M\langle\!\langle
      w_{\tilde{i}\tilde{j}}\rangle\!\rangle}\left(v_{\mathrm{w}i}+\Delta_\mathrm{s}
      v_{\mathrm{w}i}\right),\\
    &=\langle\!\langle
    w_{\tilde{i}\tilde{j}}\rangle\!\rangle^{-1}\mathrm{cov}_{\tilde{i}}\left[\langle
      w_{i\tilde{j}}\rangle, v_{\mathrm{w}i}+\Delta_\mathrm{s}
      v_{\mathrm{w}i}\right] + M^{-1}\sum_{i=1}^{L}n_i\left(v_{\mathrm{w}i}+\Delta_\mathrm{s} v_{\mathrm{w}i}\right)\\
    &=\mathrm{ave}_{\tilde{i}}\left[v_{\mathrm{w}i}+\Delta_\mathrm{s} v_{\mathrm{w}i}\right],\\
  \end{split}
\end{equation}
where we used Eq.\,(\ref{eq:within-variance-selection}) and the
assumption that $\langle w_{i\tilde{j}}\rangle$ and
$v_{\mathrm{w}i}+\Delta_\mathrm{s}v_{\mathrm{w}i}$ are statistically
uncorrelated as $i$ varies, which has already been made in
Eq.\,(\ref{eq:price-mean-taylor-molecular-final}). Using
Eq.\,(\ref{eq:within-cov-linear}) and
(\ref{eq:within-collective-3rd-mom}), we can transform the last line of
Eq.\,(\ref{eq:within-v-no-sampling}) as follows:
\begin{equation}
\mathrm{ave}_{\tilde{i}}\left[v_{\mathrm{w}i}+\Delta_\mathrm{s} v_{\mathrm{w}i}\right]
= v_{\mathrm{w}} -s_\mathrm{w}c_\mathrm{w} +O(s^2_\mathrm{w}).
\end{equation}
Therefore,
\begin{equation}
  \label{eq:within-v-no-sampling-final}
  \begin{split}
\mathbb{E}\Big[\left(k_{ij}-\mathbb{E}_{J|I=i}[k_{IJ}]\right)^2\Big]
= v_{\mathrm{w}} -s_\mathrm{w}c_\mathrm{w} +O(s^2_\mathrm{w}).
  \end{split}
\end{equation}

The second term of the last line of Eq.\,(\ref{eq:total-v-no-sampling}) is
calculated as follows:
\begin{equation}
  \label{eq:among-v-no-sampling}
  \begin{split}
    &\mathbb{E}\left[\left(\mathbb{E}_{J|I=i}[k_{IJ}]-\mathbb{E}[k_{IJ}]\right)^2\right]\\
    &=\mathbb{E}\left[\mathbb{E}_{J|I=i}[k_{IJ}]^2-2\mathbb{E}[k_{IJ}]\mathbb{E}_{J|I=i}[k_{IJ}]+\mathbb{E}[k_{IJ}]^2\right]\\
    &=\sum_{i=1}^L\sum_{j=1}^{n_i}P(I=i,J=j)\left(\mathbb{E}_{J|I=i}[k_{IJ}]^2-2\mathbb{E}[k_{IJ}]\mathbb{E}_{J|I=i}[k_{IJ}]+\mathbb{E}[k_{IJ}]^2\right)\\
    &=\sum_{i=1}^LP(I=i)\mathbb{E}_{J|I=i}[k_{IJ}]^2-2\mathbb{E}[k_{IJ}]\sum_{i=1}^LP(I=i)\mathbb{E}_{J|I=i}[k_{IJ}]+\mathbb{E}[k_{IJ}]^2\\
    &=\sum_{i=1}^LP(I=i)\mathbb{E}_{J|I=i}[k_{IJ}]^2-\mathbb{E}[k_{IJ}]^2\\
  \end{split}
\end{equation}

The first term of the last line of Eq.\,(\ref{eq:among-v-no-sampling})
is calculated as follows:
\begin{equation}
  \label{eq:among-v-mean-of-square}
  \begin{split}
    &\sum_{i=1}^LP(I=i)\mathbb{E}_{J|I=i}[k_{IJ}]^2\\
    &=\sum_{i=1}^LP(I=i)\bigg[\sum_{j=1}^{n_i}P(J=j|I=i)k_{ij}\bigg]^2\\
    &=\sum_{i=1}^LP(I=i)\bigg[\sum_{j=1}^{n_i}\frac{w_{ij}}{n_i\langle w_{i\tilde{j}}\rangle}k_{ij}\bigg]^2\\
    &=\sum_{i=1}^LP(I=i)\bigg[\frac{1}{\langle w_{i\tilde{j}}\rangle n_i}\sum_{j=1}^{n_i}w_{ij}k_{ij}\bigg]^2\\
    &=\sum_{i=1}^LP(I=i)\left[\langle w_{i\tilde{j}}\rangle^{-1}\mathrm{cov}_{i\tilde{j}}[w_{ij},k_{ij}]+\langle k_{i\tilde{j}}\rangle\right]^2\\
    &=\sum_{i=1}^L\frac{n_i\langle w_{i\tilde{j}}\rangle}{M\langle\!\langle
  w_{\tilde{i}\tilde{j}}\rangle\!\rangle}\left[\langle w_{i\tilde{j}}\rangle^{-1}\mathrm{cov}_{i\tilde{j}}[w_{ij},k_{ij}]+\langle
k_{i\tilde{j}}\rangle\right]^2\\
    &=\langle\!\langle
  w_{\tilde{i}\tilde{j}}\rangle\!\rangle^{-1}\sum_{i=1}^L\frac{n_i}{M}\langle w_{i\tilde{j}}\rangle\left[\langle w_{i\tilde{j}}\rangle^{-1}\mathrm{cov}_{i\tilde{j}}[w_{ij},k_{ij}]+\langle k_{i\tilde{j}}\rangle\right]^2\\
    &=\langle\!\langle
  w_{\tilde{i}\tilde{j}}\rangle\!\rangle^{-1}\,\mathrm{cov}_{\tilde{i}}\left[\langle
    w_{i\tilde{j}}\rangle,\left\{\langle w_{i\tilde{j}}\rangle^{-1}\mathrm{cov}_{i\tilde{j}}[w_{ij},k_{ij}]+\langle
    k_{i\tilde{j}}\rangle\right\}^2\right]\\
  & \ \ \ +\mathrm{ave}_{\tilde{i}}\left[\left\{\langle w_{i\tilde{j}}\rangle^{-1}\mathrm{cov}_{i\tilde{j}}[w_{ij},k_{ij}]+\langle k_{i\tilde{j}}\rangle\right\}^2\right]\\
    &=\langle\!\langle w_{\tilde{i}\tilde{j}}\rangle\!\rangle^{-1}\,\mathrm{cov}_{\tilde{i}}\left[\langle
    w_{i\tilde{j}}\rangle,\left(\langle
    k_{i\tilde{j}}\rangle+\Delta_\mathrm{s}\langle
    k_{i\tilde{j}}\rangle\right)^2\right] +\mathrm{ave}_{\tilde{i}}\left[\left(\langle k_{i\tilde{j}}\rangle+\Delta_\mathrm{s}\langle k_{i\tilde{j}}\rangle\right)^2\right],
  \end{split}
\end{equation}
where we have used the following notation in the final step:
\begin{equation}
  \label{eq:within-k-change}
  \Delta_\mathrm{s}\langle k_{i\tilde{j}}\rangle := \langle w_{i\tilde{j}}\rangle^{-1}\mathrm{cov}_{i\tilde{j}}[w_{ij},k_{ij}].
\end{equation}

The second term of the last line of Eq.\,(\ref{eq:among-v-no-sampling})
is calculated as follows:
\begin{equation}
  \label{eq:among-v-square-of-mean}
  \begin{split}
    \mathbb{E}[k_{IJ}]^2 &= \Bigg(\sum_{i=1}^L\sum_{j=1}^{n_i}P(I=i,J=j)k_{ij}\Bigg)^2\\
    &= \Bigg(\sum_{i=1}^LP(I=i)\sum_{j=1}^{n_i}P(J=j|I=i)k_{ij}\Bigg)^2\\
    &= \Bigg(\sum_{i=1}^LP(I=i)\mathbb{E}_{J|I=i}[k_{IJ}]\Bigg)^2\\
    &= \Bigg[\sum_{i=1}^L\frac{n_i\langle w_{i\tilde{j}}\rangle}{M\langle\!\langle
  w_{\tilde{i}\tilde{j}}\rangle\!\rangle}\mathbb{E}_{J|I=i}[k_{IJ}]\Bigg]^2.\\
  \end{split}
\end{equation}
Doing the same calculation as in Eq.\,(\ref{eq:among-v-mean-of-square}),
we can transform  Eq.\,(\ref{eq:among-v-square-of-mean}) as follows:
\begin{equation}
  \label{eq:among-v-square-of-mean-final}
  \begin{split}
    &\Bigg[\sum_{i=1}^L\frac{n_i\langle w_{i\tilde{j}}\rangle}{M\langle\!\langle
      w_{\tilde{i}\tilde{j}}\rangle\!\rangle}\mathbb{E}_{J|I=i}[k_{IJ}]\Bigg]^2\\
    &=\Bigg[\sum_{i=1}^L\frac{n_i\langle w_{i\tilde{j}}\rangle}{M\langle\!\langle
      w_{\tilde{i}\tilde{j}}\rangle\!\rangle}\left(\langle
      k_{i\tilde{j}}\rangle+\Delta_\mathrm{s}\langle
      k_{i\tilde{j}}\rangle\right)\Bigg]^2\\
    &=\left(\langle\!\langle
      w_{\tilde{i}\tilde{j}}\rangle\!\rangle^{-1}\mathrm{cov}_{\tilde{i}}\left[\langle w_{i\tilde{j}}\rangle,\langle
      k_{i\tilde{j}}\rangle+\Delta_\mathrm{s}\langle
      k_{i\tilde{j}}\rangle\right] + \mathrm{ave}_{\tilde{i}}\left[\langle
      k_{i\tilde{j}}\rangle+\Delta_\mathrm{s}\langle
      k_{i\tilde{j}}\rangle\right]\right)^2\\
  \end{split}
\end{equation}

Combining Eqs.\,(\ref{eq:among-v-no-sampling}),
(\ref{eq:among-v-mean-of-square}), and
(\ref{eq:among-v-square-of-mean-final}), we obtain
\begin{equation}
  \label{eq:among-v-exact}
  \begin{split}
    &\mathbb{E}\left[\left(\mathbb{E}_{J|I=i}[k_{IJ}]-\mathbb{E}[k_{IJ}]\right)^2\right]\\
    &= \langle\!\langle w_{\tilde{i}\tilde{j}}\rangle\!\rangle^{-1}\,\mathrm{cov}_{\tilde{i}}\left[\langle
    w_{i\tilde{j}}\rangle,\left(\langle
    k_{i\tilde{j}}\rangle+\Delta_\mathrm{s}\langle
    k_{i\tilde{j}}\rangle\right)^2\right]
+\mathrm{ave}_{\tilde{i}}\left[\left(\langle
    k_{i\tilde{j}}\rangle+\Delta_\mathrm{s}\langle
    k_{i\tilde{j}}\rangle\right)^2\right]\\
&\ \ \ -\left(\langle\!\langle
      w_{\tilde{i}\tilde{j}}\rangle\!\rangle^{-1}\mathrm{cov}_{\tilde{i}}\left[\langle w_{i\tilde{j}}\rangle,\langle
      k_{i\tilde{j}}\rangle+\Delta_\mathrm{s}\langle
      k_{i\tilde{j}}\rangle\right] + \mathrm{ave}_{\tilde{i}}\left[\langle
      k_{i\tilde{j}}\rangle+\Delta_\mathrm{s}\langle
      k_{i\tilde{j}}\rangle\right]\right)^2\\
    &= \langle\!\langle w_{\tilde{i}\tilde{j}}\rangle\!\rangle^{-1}\,\mathrm{cov}_{\tilde{i}}\left[\langle
    w_{i\tilde{j}}\rangle,\left(\langle
    k_{i\tilde{j}}\rangle+\Delta_\mathrm{s}\langle
    k_{i\tilde{j}}\rangle\right)^2\right]
+\mathrm{ave}_{\tilde{i}}\left[\left(\langle
    k_{i\tilde{j}}\rangle+\Delta_\mathrm{s}\langle
    k_{i\tilde{j}}\rangle\right)^2\right]\\
&\ \ \ -\left(\langle\!\langle
      w_{\tilde{i}\tilde{j}}\rangle\!\rangle^{-1}\mathrm{cov}_{\tilde{i}}\left[\langle w_{i\tilde{j}}\rangle,\langle
      k_{i\tilde{j}}\rangle+\Delta_\mathrm{s}\langle
      k_{i\tilde{j}}\rangle\right]\right)^2 - \left(\mathrm{ave}_{\tilde{i}}\left[\langle
      k_{i\tilde{j}}\rangle+\Delta_\mathrm{s}\langle
      k_{i\tilde{j}}\rangle\right]\right)^2\\
 & \ \ \ -2 \langle\!\langle
      w_{\tilde{i}\tilde{j}}\rangle\!\rangle^{-1}\mathrm{cov}_{\tilde{i}}\left[\langle w_{i\tilde{j}}\rangle,\langle
      k_{i\tilde{j}}\rangle+\Delta_\mathrm{s}\langle
      k_{i\tilde{j}}\rangle\right]\mathrm{ave}_{\tilde{i}}\left[\langle
      k_{i\tilde{j}}\rangle+\Delta_\mathrm{s}\langle
      k_{i\tilde{j}}\rangle\right]\\
    &= \langle\!\langle w_{\tilde{i}\tilde{j}}\rangle\!\rangle^{-1}\,\mathrm{cov}_{\tilde{i}}\left[\langle
      w_{i\tilde{j}}\rangle,\left(\langle
        k_{i\tilde{j}}\rangle+\Delta_\mathrm{s}\langle
        k_{i\tilde{j}}\rangle-\mathrm{ave}_{\tilde{i}}\left[\langle
          k_{i\tilde{j}}\rangle+\Delta_\mathrm{s}\langle
          k_{i\tilde{j}}\rangle\right]\right)^2\right]\\
    &\ \ \ -\left(\langle\!\langle
      w_{\tilde{i}\tilde{j}}\rangle\!\rangle^{-1}\mathrm{cov}_{\tilde{i}}\left[\langle w_{i\tilde{j}}\rangle,\langle
        k_{i\tilde{j}}\rangle+\Delta_\mathrm{s}\langle
      k_{i\tilde{j}}\rangle\right]\right)^2\\
  &\ \ \ +\mathrm{ave}_{\tilde{i}}\left[\left(\langle
      k_{i\tilde{j}}\rangle+\Delta_\mathrm{s}\langle
      k_{i\tilde{j}}\rangle\right)^2\right]- \left(\mathrm{ave}_{\tilde{i}}\left[\langle
      k_{i\tilde{j}}\rangle+\Delta_\mathrm{s}\langle
      k_{i\tilde{j}}\rangle\right]\right)^2\\
    &= \langle\!\langle w_{\tilde{i}\tilde{j}}\rangle\!\rangle^{-1}\,\mathrm{cov}_{\tilde{i}}\left[\langle
      w_{i\tilde{j}}\rangle,\left(\langle
        k_{i\tilde{j}}\rangle+\Delta_\mathrm{s}\langle
        k_{i\tilde{j}}\rangle-\mathrm{ave}_{\tilde{i}}\left[\langle
          k_{i\tilde{j}}\rangle+\Delta_\mathrm{s}\langle
          k_{i\tilde{j}}\rangle\right]\right)^2\right]\\
    &\ \ \ -\left(\langle\!\langle
      w_{\tilde{i}\tilde{j}}\rangle\!\rangle^{-1}\mathrm{cov}_{\tilde{i}}\left[\langle w_{i\tilde{j}}\rangle,\langle
        k_{i\tilde{j}}\rangle+\Delta_\mathrm{s}\langle
      k_{i\tilde{j}}\rangle\right]\right)^2\\
  &\ \ \ +\mathrm{ave}_{\tilde{i}}\left[\left(\langle
      k_{i\tilde{j}}\rangle+\Delta_\mathrm{s}\langle
      k_{i\tilde{j}}\rangle-\mathrm{ave}_{\tilde{i}}\left[\langle
      k_{i\tilde{j}}\rangle+\Delta_\mathrm{s}\langle
      k_{i\tilde{j}}\rangle\right]\right)^2\right]\\
  \end{split}
\end{equation}

We consider each term in the last line of Eq.\,(\ref{eq:among-v-exact})
in terms of the order of $s_\mathrm{a}$ and $s_\mathrm{w}$. We begin
with the first term.
\begin{equation}
  \label{eq:among-v-approx-main}
  \begin{split}
    &\langle\!\langle w_{\tilde{i}\tilde{j}}\rangle\!\rangle^{-1}\,\mathrm{cov}_{\tilde{i}}\left[\langle
      w_{i\tilde{j}}\rangle,\left(\langle
        k_{i\tilde{j}}\rangle+\Delta_\mathrm{s}\langle
        k_{i\tilde{j}}\rangle-\mathrm{ave}_{\tilde{i}}\left[\langle
          k_{i\tilde{j}}\rangle+\Delta_\mathrm{s}\langle
          k_{i\tilde{j}}\rangle\right]\right)^2\right]\\
    &\langle\!\langle w_{\tilde{i}\tilde{j}}\rangle\!\rangle^{-1}\,\mathrm{cov}_{\tilde{i}}\left[\langle
      w_{i\tilde{j}}\rangle,\left(\langle
        k_{i\tilde{j}}\rangle - \langle\!\langle
        k_{\tilde{i}\tilde{j}}\rangle\!\rangle+\Delta_\mathrm{s}\langle
        k_{i\tilde{j}}\rangle-\mathrm{ave}_{\tilde{i}}\left[\Delta_\mathrm{s}\langle
          k_{i\tilde{j}}\rangle\right]\right)^2\right]\\
    &=\langle\!\langle w_{\tilde{i}\tilde{j}}\rangle\!\rangle^{-1}\,\mathrm{cov}_{\tilde{i}}\Big[\langle
    w_{i\tilde{j}}\rangle,\left(\langle
      k_{i\tilde{j}}\rangle -\langle\!\langle
      k_{\tilde{i}\tilde{j}}\rangle\!\rangle\right)^2+\left(\Delta_\mathrm{s}\langle
      k_{i\tilde{j}}\rangle-\mathrm{ave}_{\tilde{i}}\left[\Delta_\mathrm{s}\langle
        k_{i\tilde{j}}\rangle\right]\right)^2\\
    &\ \ \ \ \ \ \ \ \ \ \ \ \ \ \ \ \ \ \ \ -2\left(\langle
      k_{i\tilde{j}}\rangle -\langle\!\langle
      k_{\tilde{i}\tilde{j}}\rangle\!\rangle\right)\left(\Delta_\mathrm{s}\langle
      k_{i\tilde{j}}\rangle-\mathrm{ave}_{\tilde{i}}\left[\Delta_\mathrm{s}\langle
        k_{i\tilde{j}}\rangle\right]\right)\Big]\\
    &=\langle\!\langle w_{\tilde{i}\tilde{j}}\rangle\!\rangle^{-1}\,\mathrm{cov}_{\tilde{i}}\Big[\langle
    w_{i\tilde{j}}\rangle,\left(\langle
      k_{i\tilde{j}}\rangle -\langle\!\langle
      k_{\tilde{i}\tilde{j}}\rangle\!\rangle\right)^2\Big]+ O(s_\mathrm{w}^2)\\
    &\ \ \ -2\langle\!\langle w_{\tilde{i}\tilde{j}}\rangle\!\rangle^{-1}\,\mathrm{cov}_{\tilde{i}}\Big[\langle
    w_{i\tilde{j}}\rangle,\left(\langle
      k_{i\tilde{j}}\rangle -\langle\!\langle
      k_{\tilde{i}\tilde{j}}\rangle\!\rangle\right)\left(\Delta_\mathrm{s}\langle
      k_{i\tilde{j}}\rangle-\mathrm{ave}_{\tilde{i}}\left[\Delta_\mathrm{s}\langle
        k_{i\tilde{j}}\rangle\right]\right)\Big],\\
  \end{split}
\end{equation}
where we used Eqs.\,(\ref{eq:within-cov-linear}) and
(\ref{eq:price-mean-taylor-molecular-final}) in the last step. The last
term of the last line of Eq.\,(\ref{eq:among-v-approx-main}) is zero
because $\Delta_\mathrm{s}\langle k_{i\tilde{j}}\rangle$ is independent
of $\langle k_{i\tilde{j}}\rangle$ and $\langle w_{i\tilde{j}}\rangle$,
a fact that stems from the assumptions that
$\partial s_\mathrm{w}/\partial\langle k_{ij}\rangle=0$ (see the main
text under ``Model'') and  that $v_{\mathrm{w}i}$ is statistically
uncorrelated with $\langle w_{i\tilde{j}}\rangle$ as $i$ varies [see
Eq.\,(\ref{eq:price-mean-taylor-molecular-final})]. Thus, expanding
$\langle w_{i\tilde{j}}\rangle$ as a Taylor series around
$\langle k_{i\tilde{j}}\rangle =\langle\!\langle
k_{\tilde{i}\tilde{j}}\rangle\!\rangle$, we can transform the last line of
Eq.\,(\ref{eq:among-v-approx-main}) as follows:
\begin{equation}
  \label{eq:among-v-approx-main-final}
  \begin{split}
    &\langle\!\langle w_{\tilde{i}\tilde{j}}\rangle\!\rangle^{-1}\,\mathrm{cov}_{\tilde{i}}\Big[\langle
    w_{i\tilde{j}}\rangle,\left(\langle
      k_{i\tilde{j}}\rangle -\langle\!\langle
      k_{\tilde{i}\tilde{j}}\rangle\!\rangle\right)^2\Big]+ O(s_\mathrm{w}^2)
= s_\mathrm{a}c_\mathrm{a} + O(s_\mathrm{w}^2)+O(s_\mathrm{a}^2),
  \end{split}
\end{equation}
where we introduced the following symbol:
\begin{equation}
c_\mathrm{a}:=\mathrm{ave}_{\tilde{i}}\left[\left(\langle
      k_{i\tilde{j}}\rangle -\langle\!\langle
      k_{\tilde{i}\tilde{j}}\rangle\!\rangle\right)^3\right].
\end{equation}

Next, we consider the second term of the last line of
Eq.\,(\ref{eq:among-v-exact}). Equation\,(\ref{eq:price-mean-taylor-cellular-final})
implies that
\begin{equation}
  \label{eq:among-v-approx-cov-squared}
  \begin{split}
    \left(\langle\!\langle
      w_{\tilde{i}\tilde{j}}\rangle\!\rangle^{-1}\mathrm{cov}_{\tilde{i}}\left[\langle w_{i\tilde{j}}\rangle,\langle
        k_{i\tilde{j}}\rangle+\Delta_\mathrm{s}\langle
        k_{i\tilde{j}}\rangle\right]\right)^2 = O(s_\mathrm{a}^2).
\end{split}
\end{equation}

Finally, we consider the third term of the last line of
Eq.\,(\ref{eq:among-v-exact}) as follows:
\begin{equation}
  \label{eq:among-v-approx-ave-squared}
  \begin{split}
    &\mathrm{ave}_{\tilde{i}}\left[\left(\langle
        k_{i\tilde{j}}\rangle+\Delta_\mathrm{s}\langle
        k_{i\tilde{j}}\rangle-\mathrm{ave}_{\tilde{i}}\left[\langle
          k_{i\tilde{j}}\rangle+\Delta_\mathrm{s}\langle
          k_{i\tilde{j}}\rangle\right]\right)^2\right]\\
    &=\mathrm{ave}_{\tilde{i}}\left[\left(\langle
        k_{i\tilde{j}}\rangle-\langle\!\langle
        k_{\tilde{i}\tilde{j}}\rangle\!\rangle
        +\Delta_\mathrm{s}\langle
        k_{i\tilde{j}}\rangle-\mathrm{ave}_{\tilde{i}}\left[\Delta_\mathrm{s}\langle
          k_{i\tilde{j}}\rangle\right]\right)^2\right]\\
    &=\mathrm{ave}_{\tilde{i}}\Big[\left(\langle
      k_{i\tilde{j}}\rangle-\langle\!\langle
      k_{\tilde{i}\tilde{j}}\rangle\!\rangle\right)^2
    +\left(\Delta_\mathrm{s}\langle
      k_{i\tilde{j}}\rangle-\mathrm{ave}_{\tilde{i}}\left[\Delta_\mathrm{s}\langle
        k_{i\tilde{j}}\rangle\right]\right)^2\\
    &\ \ \  +2 \left(\langle
      k_{i\tilde{j}}\rangle-\langle\!\langle
      k_{\tilde{i}\tilde{j}}\rangle\!\rangle\right)\left(\Delta_\mathrm{s}\langle
      k_{i\tilde{j}}\rangle-\mathrm{ave}_{\tilde{i}}\left[\Delta_\mathrm{s}\langle
        k_{i\tilde{j}}\rangle\right]\right) \Big]\\
    &=\mathrm{ave}_{\tilde{i}}\Big[\left(\langle
      k_{i\tilde{j}}\rangle-\langle\!\langle
      k_{\tilde{i}\tilde{j}}\rangle\!\rangle\right)^2\Big]
    +\mathrm{ave}_{\tilde{i}}\Big[\left(\Delta_\mathrm{s}\langle
      k_{i\tilde{j}}\rangle-\mathrm{ave}_{\tilde{i}}\left[\Delta_\mathrm{s}\langle
        k_{i\tilde{j}}\rangle\right]\right)^2\Big]\\
    &\ \ \  +2 \mathrm{ave}_{\tilde{i}}\Big[\left(\langle
      k_{i\tilde{j}}\rangle-\langle\!\langle
      k_{\tilde{i}\tilde{j}}\rangle\!\rangle\right)\left(\Delta_\mathrm{s}\langle
      k_{i\tilde{j}}\rangle-\mathrm{ave}_{\tilde{i}}\left[\Delta_\mathrm{s}\langle
        k_{i\tilde{j}}\rangle\right]\right) \Big]\\
    &=\mathrm{ave}_{\tilde{i}}\Big[\left(\langle
      k_{i\tilde{j}}\rangle-\langle\!\langle
      k_{\tilde{i}\tilde{j}}\rangle\!\rangle\right)^2\Big]
    +\mathrm{ave}_{\tilde{i}}\Big[\left(\Delta_\mathrm{s}\langle
      k_{i\tilde{j}}\rangle-\mathrm{ave}_{\tilde{i}}\left[\Delta_\mathrm{s}\langle
        k_{i\tilde{j}}\rangle\right]\right)^2\Big]\\
    &= \mathrm{ave}_{\tilde{i}}\Big[\left(\langle
      k_{i\tilde{j}}\rangle-\langle\!\langle
      k_{\tilde{i}\tilde{j}}\rangle\!\rangle\right)^2\Big] +O(s_\mathrm{w}^2),\\
    &= v_\mathrm{a} + O(s_\mathrm{w}^2),\\
\end{split}
\end{equation}
where we have assumed that $\langle k_{i\tilde{j}}\rangle$ and
$\Delta_\mathrm{s}\langle k_{i\tilde{j}}\rangle$ are statistically
uncorrelated, an assumption that is essentially the same as the
assumption made in Eq.\,(\ref{eq:price-mean-taylor-molecular-final})
that $v_{\mathrm{w}i}$ and $\langle w_{i\tilde{j}}\rangle$ are
statistically uncorrelated.

Combining Eq.\,(\ref{eq:among-v-exact}), (\ref{eq:among-v-approx-main}),
(\ref{eq:among-v-approx-main-final}),
(\ref{eq:among-v-approx-cov-squared}), and
(\ref{eq:among-v-approx-ave-squared}), we obtain
\begin{equation}
  \label{eq:among-v-approx-final}
\mathbb{E}\left[\left(\mathbb{E}_{J|I=i}[k_{IJ}]-\mathbb{E}[k_{IJ}]\right)^2\right]
= v_\mathrm{a} + s_\mathrm{a}c_\mathrm{a} +
O(s_\mathrm{a}^2)+O(s_\mathrm{w}^2).
\end{equation}

Combining Eq.\,(\ref{total-v-definition}),
(\ref{eq:total-v-no-sampling}), (\ref{eq:within-v-no-sampling-final}),
and (\ref{eq:among-v-approx-final}), we obtain
\begin{equation}
  \label{total-v-approx}
  \mathbb{E}\left[v_\mathrm{t}'\right] =
  \left(1-M^{-1}\right)\left(
    v_\mathrm{a} + s_\mathrm{a}c_\mathrm{a} + v_{\mathrm{w}}
    -s_\mathrm{w}c_\mathrm{w} + m\sigma +
O(s_\mathrm{a}^2)+O(s_\mathrm{w}^2) \right).
\end{equation}

Substituting Eqs.\,(\ref{eq:within-variance-approximate-final}) and
(\ref{total-v-approx}) into
$\mathbb{E}\left[v_\mathrm{a}'\right]=\mathbb{E}\left[v_\mathrm{t}'\right]-\mathbb{E}\left[v_\mathrm{w}'\right]$,
we obtain
\begin{equation}
  \begin{split}
  \label{among-v-approx}
  \mathbb{E}\left[v_\mathrm{a}'\right] &\approx
  \left(1-M^{-1}\right)\left(
    v_\mathrm{a} + s_\mathrm{a}c_\mathrm{a} + v_{\mathrm{w}}
    -s_\mathrm{w}c_\mathrm{w} + m\sigma + O(s_\mathrm{a}^2)+O(s_\mathrm{w}^2) \right) \\
  &\ \ \ - \left(1-\beta
      N^{-1}\right)\left(v_{\mathrm{w}} +
    m\sigma- s_\mathrm{w}c_\mathrm{w}+ O(s_\mathrm{w}^2)\right)\\
  &=  \left(1-M^{-1}\right)\left(
    v_\mathrm{a} + s_\mathrm{a}c_\mathrm{a} + O(s_\mathrm{a}^2)+O(s_\mathrm{w}^2)\right)\\
& \ \ \ +\left(\beta N^{-1}-M^{-1}\right)\left(v_{\mathrm{w}} +
    m\sigma- s_\mathrm{w}c_\mathrm{w} + O(s_\mathrm{w}^2) \right) \\
\end{split}
\end{equation}

\section{Estimation of $\gamma_\mathrm{a}$}

Tsimring et al.~\cite{Tsimring1996} have investigated the time evolution
of the probability density $p(r,t)$ of fitness $r$ subject to mutation
and selection. In this section, we show that the results of Tsimring et
al.~\cite{Tsimring1996} imply $C \approx -0.25 V^{3/2}$, where $V$ and
$C$ are the variance and the third central moment of $p(r,t)$,
respectively. This implication is consistent with our postulate
$c_\mathrm{a}=-\gamma_\mathrm{a} v_\mathrm{a}^{3/2}$ made in Eq.\,(\eqPostulate) of
the main text, where $\gamma_\mathrm{a}$ was measured to be about 0.25 through
simulations.

Tsimring et al.~\cite{Tsimring1996} have considered the following
equation, which describes the time evolution of $p(r,t)$:
\begin{equation}
\frac{\partial}{\partial t}p(r,t)=\theta(p-p_c)\left(r-\langle r\rangle \right)p(r,t)+D\frac{\partial^2}{\partial r^2}p(r,t),\label{eq:FP}
\end{equation}
where $\theta(x)$ is the Heaviside step function, and $\langle r\rangle$
is the average fitness defined as
\[
\langle f(r) \rangle=\int_{-\infty}^\infty f(r)p(r,t)dr,
\]
and $D$ is a diffusion constant. The first term on the RHS of
Eq.\,(\ref{eq:FP}) describes the effect of selection; the second term,
that of mutation. The Heaviside step function accounts for the fact that
the probability density $p(r,t)$ must exceed a small threshold density
$p_c$ to grow because the size of a population is not infinite in
reality. Tsimring et al.\ have shown that Eq.\,(\ref{eq:FP}) allows a
travelling-wave solution, in which the peak of the density travels toward
higher values of $r$, while maintaining a pulse-like shape, at a
steady-state speed (denoted by $v$)
\begin{equation}
  \label{eq:vc}
  v=cD^{2/3},
\end{equation} 
where the value of $c$ depends weakly on $p_c$ and is around $4$ in a
wide range of $p_c$ \cite{Tsimring1996}.

Multiplying both sides of Eq.\,(\ref{eq:FP}) with $r$ or $(r-\langle
r\rangle)^2$ and integrating over the whole range, we get
\begin{eqnarray}
\frac{d}{dt}\langle r\rangle &=&V-\epsilon_1,
\label{eq:mean}\\
\frac{d}{dt} V &=&C-\epsilon_2+2D, \label{eq:variance}
\end{eqnarray}
where $V$,  $C$, $\epsilon_1$, and $\epsilon_2$ are defined as follows:
\begin{eqnarray}
V&=&\langle \left(r-\langle r\rangle\right)^2\rangle,\\
C&=&\langle \left(r-\langle r\rangle\right)^3\rangle, \\
\epsilon_1&=&\left\langle \theta(p_c-p) \left(r-\langle r\rangle\right)^2\right\rangle,\\
\epsilon_2&=&\left\langle \theta(p_c-p) \left(r-\langle r\rangle\right)^3\right\rangle.
\end{eqnarray}
In obtaining Eqs.\,(\ref{eq:mean}) and (\ref{eq:variance}), we have
assumed that the surface terms go to zero as $r\to\pm\infty$; i.e.,
$\lim_{r\to \pm \infty}P(r,t)= 0$, $\lim_{r\to \pm \infty}rP(r,t)= 0$,
$\lim_{r\to \pm \infty}r\frac{\partial p}{\partial r}= 0$, and
$\lim_{r\to \pm \infty}r^2\frac{\partial p}{\partial r}= 0$.

For a travelling-wave solution of Eq.\,(\ref{eq:FP}) with a constant
speed $v$ and shape, Eq.\,(\ref{eq:mean}) implies
\begin{equation}
  \label{eq:vVa}
  v= V-\epsilon_1.
\end{equation}
From Eqs.\,(\ref{eq:vc}) and (\ref{eq:vVa}), we get
\begin{equation}
 \label{eq:Dva}
D=\left(\frac{V-\epsilon_1}{c}\right)^{3/2}.
\end{equation}
Since $v$ and $\epsilon_1$ are constant, Eq.\,(\ref{eq:vVa}) implies
$dV/dt=0$. Thus, Eq.\,(\ref{eq:variance}) implies
\begin{equation}
C=-2D+\epsilon_2.
\label{eq:CaD}
\end{equation}
Equations~(\ref{eq:Dva}) and (\ref{eq:CaD}) imply
\[
C=-2\left(\frac{V-\epsilon_1}{c}\right)^{3/2}+\epsilon_2\approx -2c^{-3/2}V^{3/2},
\]
where we have assumed $\epsilon_1\ll V$ and $\epsilon_2\ll V$ to obtain
the last term. Since $c$ is about 4 according to Tsimring et
al.\ \cite{Tsimring1996}, we get
\[
C\approx -0.25V^{3/2}.
\]

\section{Converting Kimura's notation into ours}
Kimura \cite{Kimura1986} has investigated a binary-trait model of
multilevel selection and shown that within-collective selection exactly
balances out among-collective selection if
\begin{equation}
  \label{eq:kimura-original}
  \frac{c}{v+v'+m}-4Ns' = 0,
\end{equation}
where the symbols are as described in Table~\ref{supp:table:kimura} (see
also the next paragraph). Equation\,(\ref{eq:kimura-original}) includes
Eq.\,(\eqKimuraOurNotation) of the main text as a special
case. Equation\,(\ref{eq:kimura-original}) appears as Eq.\,(27) of
Ref.\,\cite{Kimura1986} or Eq.\,(4.8) of Ref.\,\cite{Kimura1984} as and
is derived therein under the assumption that the steady-state frequency
of the altruistic allele is identical to that in the absence of
selection, an approximation that is expected to be valid in the limit of
weak selection.

\begin{table}[tb]
  \centering
  \caption{\label{supp:table:kimura}{\bf Correspondence between Kimura's
      notation \cite{Kimura1986} and ours.}}
  \begin{tabularx}{\textwidth}{ccX}
    \hline
    Kimura's &  ours & description\\\hline
    $c$ & $s_\mathrm{a}$& among-collective selection coefficient \\
    $v$ & $m$ & mutation rate per generation from non-altruistic to altruistic allele \\
    $v'$ & $m$ & reverse mutation rate; we assumed $v=v'$ \\
    $s'$ & $s_\mathrm{w}$ & within-collective selection coefficient \\
    $m$ & 0 & among-collective migration rate\\
    $2N$ & $\beta^{-1}N$ & number of alleles per collective; Kimura
    considers diploid \\
    $\infty$ & $M$ & total number of alleles\\
    \hline
  \end{tabularx}
\end{table}

To convert Kimura's notation into ours, we assumed that the rate of
mutation from a non-altruistic to an altruistic allele is identical to
the rate of mutation from the altruistic to the non-altruistic allele,
so that the direction of mutation is unbiased as in our quantitative-trait
model. Moreover, we assumed that the migration rate among collectives is
zero since our model does not consider migration. Finally, we took
account of the fact that Kimura's model considers diploid as follows. In
Kimura's model, each collective consists of $N$ diploid individuals,
i.e., $2N$ alleles. The number of alleles per collective can be
considered as the average number of replicators per collective in our
model (i.e., $\beta^{-1}N$) because Kimura's model assumes no dominance.

\section{Derivation of Kimura's result through our method}

In this section, we derive Eq.\,(\eqKimuraOurNotation) of the main text,
which gives parameter-region boundaries of the binary-trait model, using
the method developed in the main text. The most important difference
between the binary-trait and quantitative-trait models resides in the
definition of $\epsilon_{IJ}$. Thus, we consider only terms involving
$\epsilon$ or $m\sigma$ as described below.

First, we show that the change in the definition of $\epsilon_{IJ}$ does
not affect the condition for the parameter-region boundary given by
Eq.\,(\eqPriceMean). In the binary-trait model, $\epsilon_{IJ}=0$ with
probability $1-m$ and $\epsilon_{IJ}=1-2k_{IJ}$ with probability $m$
($I$ and $J$ are random variables taking the indices of a sampled
replicator, as defined in
Section\,\ref{sec:derivation_of_price_eq}). Thus,
\begin{equation}
  \begin{split}
    \mathbb{E}\left[\epsilon_{IJ}\right]
    &=\sum_{i=1}^L\sum_{j=1}^{n_i}P(I=i, J=j) \int
    dP(\epsilon_{IJ})\epsilon_{IJ}\\
    &=\sum_{i=1}^{L}\sum_{j=1}^{n_i} \frac{w_{ij}}{M\langle\!\langle
      w_{\tilde{i}\tilde{j}}\rangle\!\rangle}m(1-2k_{ij})\\
    &=
    m(1-2\langle\!\langle k_{\tilde{i}\tilde{j}}\rangle\!\rangle) +
    O(s_\mathrm{w}^2 + s_\mathrm{a}^2 + ms_\mathrm{w}+ms_\mathrm{a}).\\
\end{split}
\end{equation}
Therefore, Eq.\,(\eqPriceMeanExp) needs to be modified as follows:
\begin{equation}
    \mathbb{E}\left[\Delta \langle\!\langle k_{\tilde{i}\tilde{j}}\rangle\!\rangle\right]
    = s_\mathrm{a}v_\mathrm{a} - s_\mathrm{w}v_\mathrm{w} +
    m(1-2\langle\!\langle k_{\tilde{i}\tilde{j}}\rangle\!\rangle) + O(s_\mathrm{w}^2 + s_\mathrm{a}^2 + ms_\mathrm{w}+ms_\mathrm{a}).
\end{equation}
This equation, however, becomes almost identical to
Eq.\,(\eqPriceMeanExp) if the parameters are on the parameter-region
boundary, on which
$\langle\!\langle k_{\tilde{i}\tilde{j}}\rangle\!\rangle=1/2$. Thus, the
condition for the parameter-region boundary when $s_\mathrm{a}$,
$s_\mathrm{w}$, and $m$ are sufficiently small is the same as in the
quantitative-trait model.

Next, we consider Eq.\,(\eqTotalVarianceNeutral) and show that the
change in the definition of $\epsilon_{IJ}$ makes a significant
difference, which explains the difference between the binary- and
quantitative-trait models. In Eq.\,(\eqTotalVarianceNeutral), $m\sigma$
represents the difference between $v_\textrm{t}$ and the variance of
$k_{IJ}+\epsilon_{IJ}$. In the quantitative-trait model, this difference
is simply the variance of $\epsilon_{IJ}$ because $\epsilon_{IJ}$ and
$k_{IJ}$ are independent of each other. In the binary-trait model,
however, $\epsilon_{IJ}$ and $k_{IJ}$ are not independent, and this fact
affects Eq.\,(\eqTotalVarianceNeutral), as follows. Under the assumption
that $s_\mathrm{a}=s_\mathrm{w}=0$, the variance of
$k_{IJ}+\epsilon_{IJ}$ in the binary-trait model is
\begin{equation}
  \begin{split}
    \mathbb{E}\left[\left(k_{IJ}+\epsilon_{IJ}-\mathbb{E}\left[k_{IJ}+\epsilon_{IJ}\right]\right)^2\right] &=
    (1-\mathbb{E}\left[k_{IJ}+\epsilon_{IJ}\right])^2P\left(k_{IJ}+\epsilon_{IJ}=1\right)\\
    &+ (0-\mathbb{E}\left[k_{IJ}+\epsilon_{IJ}\right])^2P\left(k_{IJ}+\epsilon_{IJ}=0\right),\\
  \end{split}
\end{equation}
where 
\begin{equation}
  \begin{split}
P\left(k_{IJ}+\epsilon_{IJ}=1\right) &= \langle\!\langle
k_{\tilde{i}\tilde{j}}\rangle\!\rangle \, (1-m) + \left(1-\langle\!\langle k_{\tilde{i}\tilde{j}}\rangle\!\rangle\right)m\\
P\left(k_{IJ}+\epsilon_{IJ}=0\right) &= \left(1-\langle\!\langle
k_{\tilde{i}\tilde{j}}\rangle\!\rangle\right)(1-m) + \langle\!\langle
k_{\tilde{i}\tilde{j}}\rangle\!\rangle\, m\\
\mathbb{E}\left[k_{IJ}+\epsilon_{IJ}\right]&=P\left(k_{IJ}+\epsilon_{IJ}=1\right).\\
  \end{split}
\end{equation}
Thus,
\begin{equation}
  \begin{split}
    \mathbb{E}\left[\left(k_{IJ}+\epsilon_{IJ}-\mathbb{E}\left[k_{IJ}+\epsilon_{IJ}\right]\right)^2\right] = v_\mathrm{t} + m(1-m)\left(1-2\langle\!\langle k_{\tilde{i}\tilde{j}}\rangle\!\rangle\right)^2,
  \end{split}
\end{equation}
where we used the fact that
$v_\mathrm{t}=\langle\!\langle
k_{\tilde{i}\tilde{j}}\rangle\!\rangle(1-\langle\!\langle
k_{\tilde{i}\tilde{j}}\rangle\!\rangle)$.  Therefore, the expected
sample variance of the next generation is
\begin{equation}
\label{eq:total-variance-neutral-disc}
\mathbb{E}\left[v_\mathrm{t}'\right] = (1-M^{-1})\left[v_\mathrm{t} + m(1-m)(1-2\langle\!\langle k_{\tilde{i}\tilde{j}}\rangle\!\rangle)^2\right].
\end{equation}
Likewise, under the assumption that all collectives always consist of
$\beta^{-1}N$ replicators, the expected sample variance within a
collective of the next generation is
\begin{equation}
\mathbb{E}\left[v_{\mathrm{w}i}'\right] = (1-\beta N^{-1})\left[v_{\mathrm{w}i} + m(1-m)(1-2\langle k_{i\tilde{j}}\rangle)^2\right],
\end{equation}
where the index of collectives $i$ needs to be kept because
$\langle k_{i\tilde{j}}\rangle$ depends on $i$. Averaging
$\mathbb{E}\left[v_{\mathrm{w}i}'\right]$ over $i$, we obtain
\begin{equation}
  \label{eq:within-variance-neutral-disc}
  \begin{split}
    \mathbb{E}\left[v_\mathrm{w}'\right] &\approx
    \mathrm{ave}_{\tilde{i}}\left[\mathbb{E}\left[v_{\mathrm{w}i}'\right]\right]\\
    &=(1-\beta N^{-1})\left\{v_\mathrm{w} +
      m(1-m)\left[(1-2\langle\!\langle
        k_{\tilde{i}\tilde{j}}\rangle\!\rangle)^2+4v_\mathrm{a}\right]\right\},
  \end{split}
\end{equation}
where we used the fact that $v_\mathrm{a}=\mathrm{ave}_{\tilde{i}}[\langle
  k_{i\tilde{j}}\rangle^2] - \langle\!\langle
k_{\tilde{i}\tilde{j}}\rangle\!\rangle^2$. Since
$\mathbb{E}[v_\mathrm{a}']=\mathbb{E}[v_\mathrm{t}']-\mathbb{E}[v_\mathrm{w}']$, we obtain
\begin{equation}
  \label{eq:among-variance-neutral-disc}
  \begin{split}
    \mathbb{E}\left[v_\mathrm{a}'\right] =&\ \left(1-M^{-1}\right)v_\mathrm{a} + \left(\beta
          N^{-1}-M^{-1}\right)\left[v_\mathrm{w}+m(1-m)
          \left(1-2\langle\!\langle
    k_{\tilde{i}\tilde{j}}\rangle\!\rangle\right)^2\right]\\& - 4\left(1-\beta N^{-1}\right)m(1-m)v_\mathrm{a}.
  \end{split}
\end{equation}
If the systems is on a parameter-region boundary,
$\langle\!\langle k_{\tilde{i}\tilde{j}}\rangle\!\rangle=1/2$. Thus,
setting $\langle\!\langle k_{\tilde{i}\tilde{j}}\rangle\!\rangle=1/2$,
we obtain
\begin{align}
  \label{eq:within-variance-neutral-disc-boundary}
    \mathbb{E}\left[v'_\mathrm{w}\right] &= (1-\beta N^{-1})\left[v_\mathrm{w} + 4m(1-m)v_\mathrm{a}\right]\\
  \label{eq:among-variance-neutral-disc-boundary}
  \mathbb{E}\left[v'_\mathrm{a}\right] &= \left(1-M^{-1}\right)v_\mathrm{a} + \left(\beta
    N^{-1}-M^{-1}\right)v_\mathrm{w} - 4\left(1-\beta N^{-1}\right)m\left(1-m\right)v_\mathrm{a}.
\end{align}
To apply the condition for the parameter-region boundary
$v_\mathrm{w}/v_\mathrm{a}\approx s_\mathrm{a}/s_\mathrm{w}$, we need to
calculate the ratio $v_\mathrm{w}/v_\mathrm{a}$. To this end, dividing
Eq.\,(\ref{eq:within-variance-neutral-disc-boundary}) by
Eq.\,(\ref{eq:among-variance-neutral-disc-boundary}) on each side, we
obtain
\begin{equation}
  \frac{\mathbb{E}\left[v'_\mathrm{w}\right]}{\mathbb{E}\left[v'_\mathrm{a}\right]}=\frac{(1-\beta N^{-1})\left[\frac{v_\mathrm{w}}{v_\mathrm{a}} + 4m(1-m)\right]}{(1-M^{-1}) + (\beta
    N^{-1}-M^{-1})\frac{v_\mathrm{w}}{v_\mathrm{a}} - 4(1-\beta N^{-1})m(1-m)}.
\end{equation}
Assuming a steady state (i.e.,
$\mathbb{E}[v'_\mathrm{w}]/\mathbb{E}[v'_\mathrm{a}]=v_\mathrm{w}/v_\mathrm{a}$), we obtain
\begin{equation}
    \label{eq:variance-ratio-neutral-disc-boundary}
    \frac{v_\mathrm{w}}{v_\mathrm{a}} = \frac{4m(1-m)(1-\beta N^{-1})}{\beta N^{-1}-M^{-1}}.
\end{equation}
Using the condition $v_\mathrm{w}/v_\mathrm{a}\approx
s_\mathrm{a}/s_\mathrm{w}$, we obtain
\begin{equation}
  \label{eq:kimura-rederived}
  \frac{4m(1-m)(1-\beta N^{-1})}{\beta N^{-1}-M^{-1}}\approx\frac{s_\mathrm{a}}{s_\mathrm{w}}.
\end{equation}
If $\beta N^{-1} \ll 1$, $m\ll1$, and $M\to\infty$ (Kimura's model
assumes that $M=\infty$), we obtain
\begin{equation}
  \frac{s_\mathrm{a}}{s_\mathrm{w}}\approx 4m\beta^{-1}N,
\end{equation}
which is the same as Eq.\,(\eqKimuraOurNotation).

\newpage

\section{Supplementary Figures}

\bigskip
\bigskip
\bigskip
\bigskip

\begin{figure}[h]
  \centerline{\includegraphics{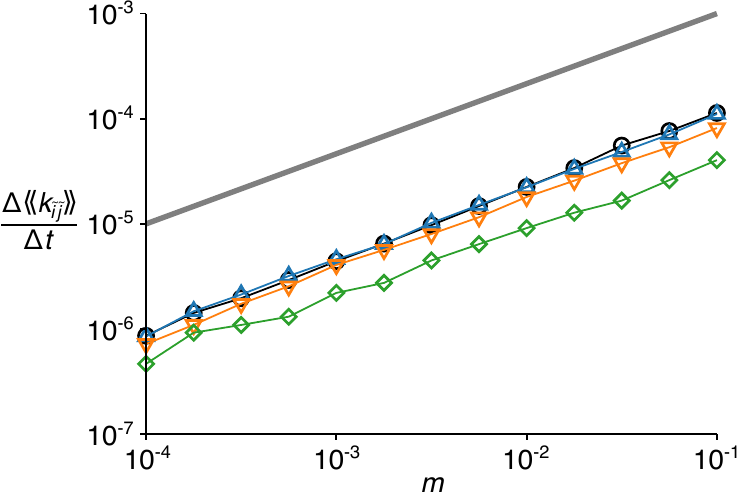}}
  \caption{\label{sfig:single-level}Rate of logarithmic fitness increase
    as function of mutation rate measured through simulations with no
    within-collective selection ($s_\mathrm{w}=0$,
    $s_\mathrm{a}=10^{-3}$, $M=5\times 10^5$, and
    $\sigma=10^{-4}$). Fitness is defined as
    $w_{ij} = e^{s_\mathrm{a}\langle k_{i\tilde{j}}\rangle}$. Symbols
    have following meaning: $N=10^2$ (black circles); $N=10^3$ (blue
    triangle up); $N=10^4$ (orange triangle down); $N=10^5$ (green
    diamond). Line is
    $\Delta\langle\!\langle
    k_{\tilde{i}\tilde{j}}\rangle\!\rangle/\Delta t\,\propto\, m^{2/3}$,
    as predicted by Eqs.\,(\eqPriceMeanExp) and (\eqVarianceClosedVa) in
    main text. This figure confirms that
    $\Delta\langle\!\langle
    k_{\tilde{i}\tilde{j}}\rangle\!\rangle/\Delta t\,\propto\, m^{2/3}$
    in agreement with Ref.\,\cite{Tsimring1996}. Note also that
    $\Delta\langle\!\langle k_{\tilde{i}\tilde{j}}\rangle\!\rangle$ is
    roughly independent of $N$ if $N\ll M$, which is consistent with
    prediction of Eq.\,(\eqVarianceClosedVa) in main text.}
\end{figure}

\bigskip
\bigskip
\bigskip
\bigskip
\bigskip

\begin{figure}[h]
  \centerline{\includegraphics{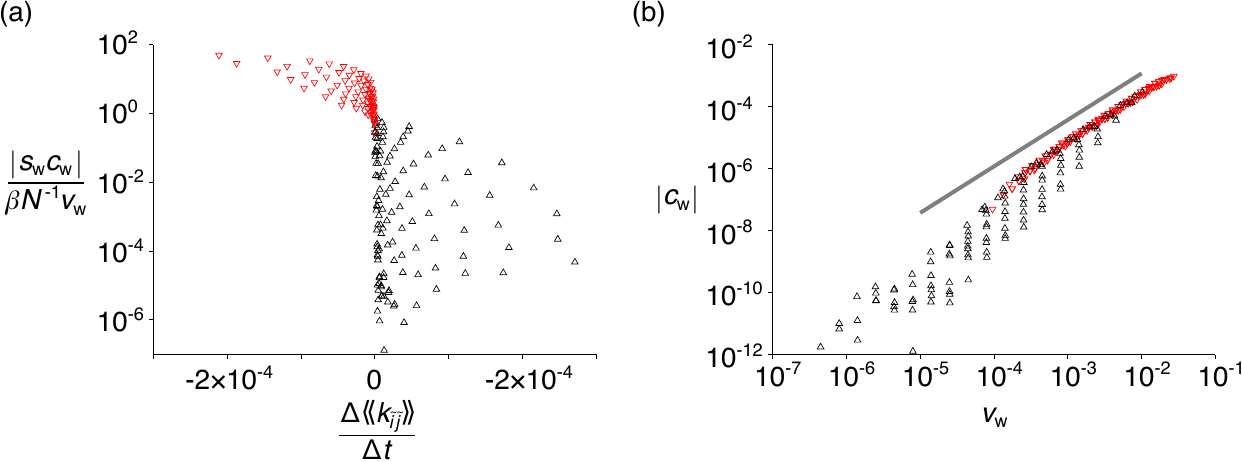}}
  \caption{\label{sfig:within-3rd-moment}Average third central moment of
    $k_{ij}$ within collective ($c_\text{w}$) measured through
    simulations ($M=5\times 10^5$, $\sigma=10^{-4}$, and
    $s_\mathrm{w}=s_\mathrm{a}=10^{-2}$). (a) Ratio between effect of
    selection and that of random genetic drift on $\Delta v_\text{w}$ as
    function of
    $\Delta\langle\!\langle
    k_{\tilde{i}\tilde{j}}\rangle\!\rangle/\Delta t$:
    $\Delta\langle\!\langle
    k_{\tilde{i}\tilde{j}}\rangle\!\rangle/\Delta t>0$ (black triangle
    up);
    $\Delta\langle\!\langle
    k_{\tilde{i}\tilde{j}}\rangle\!\rangle/\Delta t<0$ (red triangle
    down). (b) $|c_\text{w}|$ as function of $v_\text{w}$. Triangles are
    simulation results:
    $\Delta\langle\!\langle
    k_{\tilde{i}\tilde{j}}\rangle\!\rangle/\Delta t>3\times10^{-7}$
    (black triangle up);
    $\Delta\langle\!\langle
    k_{\tilde{i}\tilde{j}}\rangle\!\rangle/\Delta t<3\times10^{-7}$ (red
    triangle down). Line is $|c_\text{w}|\,\propto\,v_\text{w}^{3/2}$,
    as postulated in Eq.\,({\eqPostulate}). Least squares fitting of
    $|c_\text{w}|=\gamma_\text{w} v_\text{w}^{3/2}$ to data for
    $\Delta\langle\!\langle
    k_{\tilde{i}\tilde{j}}\rangle\!\rangle/\Delta t<3\times10^{-7}$
    yielded $\gamma_\text{w}\approx 0.25$.}
\end{figure}

\newpage
\ %
\bigskip
\bigskip
\bigskip
\bigskip

\begin{figure}[h]
  \centerline{\includegraphics{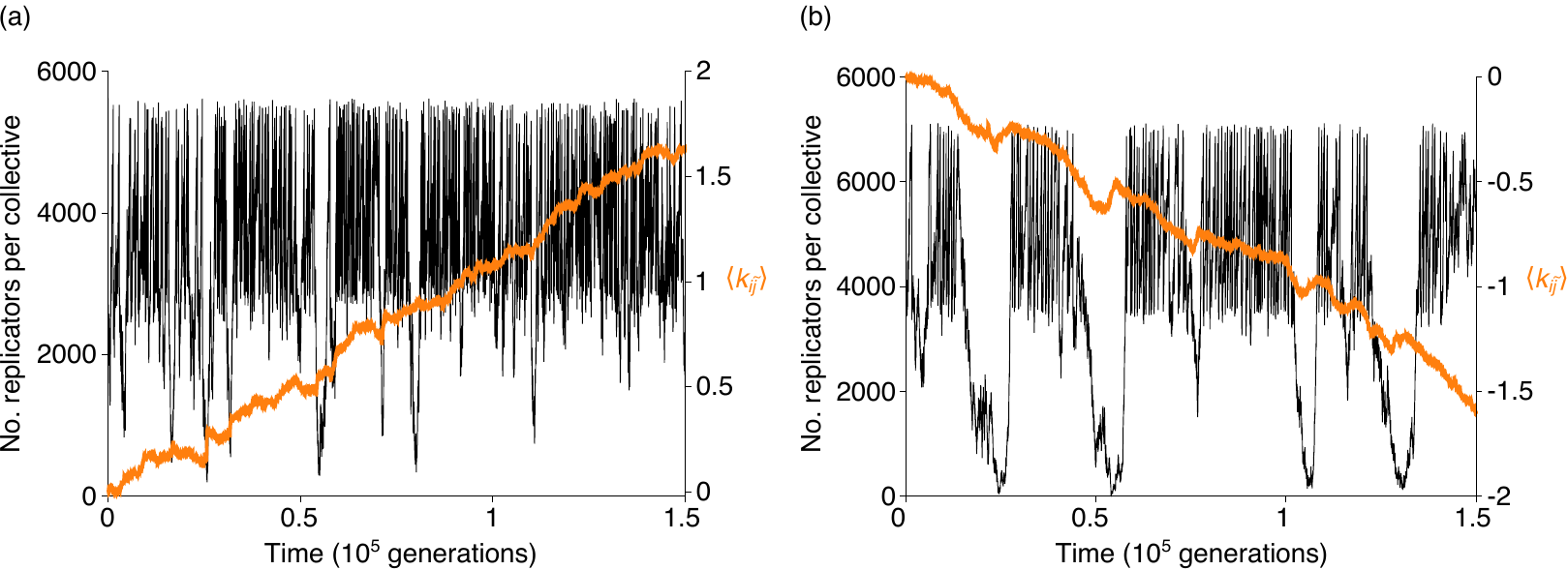}}
  \caption{\label{sfig:disequilibrium}Dynamics of common ancestors of
    collectives ($m=0.01$, $M=5\times10^5$
    $\sigma=10^{-4}$, and $s_\mathrm{a}=s_\mathrm{w}=0.01$). Plotted are
    number of replicators per collective (black; left coordinate) and
    $\langle k_{i\tilde{j}}\rangle$ (orange; right coordinate). (a)
    $N=5623$. In this case, $\Delta\langle
    k_{\tilde{i}\tilde{j}}\rangle>0$, and evolutionarily stable
    disequilibrium is not clearly observed. (b) $N=17783$. In this case,
    $\Delta\langle k_{\tilde{i}\tilde{j}}\rangle<0$, and evolutionarily
    stable disequilibrium is clearly observed.}
\end{figure}



\newpage
\ %
\bigskip
\bigskip
\bigskip
\bigskip

\begin{figure}[h]
  \centerline{\includegraphics[width=\textwidth]{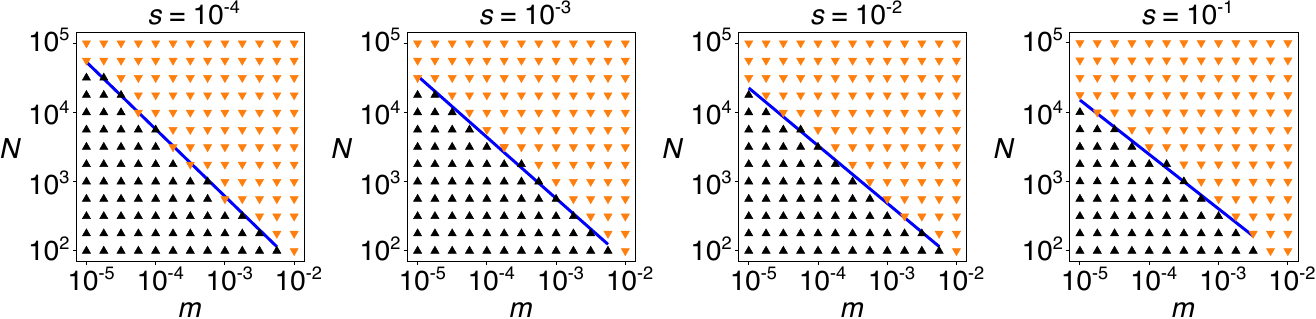}}
  \caption{\label{sfig:phase-diagram-disc}Parameter-sweep diagrams of
    binary-trait model ($s_\mathrm{w}=s_\mathrm{a}=s$ and
    $M=5\times10^5$). Symbols have following meaning:
    $\langle\!\langle k_{\tilde{i}\tilde{j}}\rangle\!\rangle>1/2$
    (triangle up);
    $\langle\!\langle k_{\tilde{i}\tilde{j}}\rangle\!\rangle<1/2$
    (triangle down). Lines are estimated parameter-region
    boundaries. Parameter-region boundaries were estimated as
    follows. Zeros of
    $\langle\!\langle k_{\tilde{i}\tilde{j}}\rangle\!\rangle-1/2$ were
    estimated with linear interpolation with respect to $N$ from two
    simulation points around parameter-region boundary for various $m$
    values between $10^{-5}$ and $10^{-1}$. Estimated zeros were used to
    obtain parameter-region boundary through least squares regression of
    $N\,\propto\, m^{-\alpha}$.}
\end{figure}

\bibliographystyle{vancouver}
\bibliography{reference}